\pdfoutput=1
\documentclass[]{article}
\usepackage[a4paper,top=3cm,bottom=3cm,left=2.5cm,right=2.5cm,marginparwidth=1.75cm]{geometry}
\usepackage{authblk}
\usepackage[numbers, sort&compress]{natbib}
\usepackage[english]{babel}
\usepackage{url,microtype}
\usepackage[colorlinks=true,linkcolor=black,citecolor=black,urlcolor=black]{hyperref}
\usepackage{doi}
\usepackage{amsmath}
\usepackage{accents}
\usepackage{amssymb}
\usepackage{mathtools}
\usepackage{graphicx}
\usepackage{epstopdf}
\usepackage[dvipsnames,table]{xcolor}
\usepackage{booktabs}
\usepackage{arydshln}
\usepackage{pdfpages}
\usepackage{soul}
\usepackage{units}
\usepackage[shortlabels]{enumitem}
\setlist[enumerate,1]{label=(\roman*)}
\usepackage{bbm}
\usepackage{subcaption}
\usepackage{placeins}
\usepackage[permil]{overpic}
\usepackage{algorithm}
\usepackage[noEnd=true,endLComment=]{algpseudocodex}
\usepackage{tikz}
\usetikzlibrary{shapes.geometric, arrows}
\usepackage{multirow}
\usepackage{multicol}
\usepackage{rotating}
\usepackage{pgfplots}

\newcommand{\ve}[1]{\text{\boldmath${#1}$}} 

\newcommand{\grads}{\nabla^\mathrm{S}}
\renewcommand{\top}{\mathrm{T}} 

\newtheorem{thm}{Theorem}[section]

\newtheorem{remark}[thm]{Remark}

\AfterEndEnvironment{prob}{\noindent\ignorespaces}

\newcommand{\draftHeavy}{false}

\title{Synthetic Dataset Generation, Physical Consistency and Active Learning for Shape-Informed Surrogate Modeling of Hemodynamics}
\author[1]{F. Romor$^{\star,}$}
\author[2,3]{R. Schussnig$^{\star,}$}
\author[3]{C.A. Taylor}
\author[2,4]{G.A. Holzapfel}
\author[1]{A. Caiazzo}
\affil[$^\star$]{these authors contributed equally to this work}
\affil[1]{Weierstra{\ss} Institute, Germany}
\affil[2]{Institute of Biomechanics, Graz University of Technology, Austria, schussnig@tugraz.at}
\affil[3]{Oden Institute for Computational Engineering and Sciences, University of Texas at Austin, United States}
\affil[4]{Department of Structural Engineering, Norwegian University of Science and Technology, Norway}

\begin{document}

    \maketitle

    \begin{abstract}
        Model order reduction has the potential to enable real-time
        patient-specific assessment of cardiovascular flows.
        However, it still faces significant challenges stemming from the 
        geometric variability of vascular anatomies.
        Building upon the SynthAorta dataset (Bo\v{s}njak et al., IEEE Trans.
        Med. Imag. 45(1), 2026,
        \href{https://doi.org/10.1109/TMI.2025.3599937}
             {doi:10.1109/TMI.2025.3599937})
        of physiological aortic geometries, we generate an extensive dataset of
        physiologically consistent computational fluid dynamics solutions
        through an automatic boundary-condition adjustment procedure.
        A common reference geometry allows all patient-specific anatomies
        to be represented by displacement fields, providing point-to-point
        correspondence without the need for geometric registration.

        Based on this dataset, we develop Graph U-Net and Transformer U-Net
        surrogate models to predict velocity and pressure fields at
        peak systole. We further investigate two strategies to enhance
        data efficiency and prediction accuracy: active learning and
        differentiable physical consistency constraints based on the
        finite element residual of the incompressible Navier--Stokes
        equations. The results show that active learning is strongly
        regime-dependent: while it offers little benefit over random sampling
        during early training, it improves performance during fine-tuning
        through gradient- and subspace-based acquisition strategies.
        Similarly, physical consistency constraints prove to be
        substantially more effective as a fine-tuning objective than as a
        regularizer applied throughout training. Finally, out-of-distribution
        studies---from both patients with coarctation and healthy 
        subjects---demonstrate that
        the primary limitation of surrogate models lies in the geometric
        support of the underlying parametrization rather than in the neural
        network architectures themselves.
        The proposed framework combines synthetic dataset generation,
        geometry-aware surrogate modeling, physics-informed optimization,
        and active learning into a scalable workflow for inter-patient
        hemodynamic prediction.
        \newline

    \end{abstract}

    \section{Introduction}
    Computational fluid dynamics (CFD), applied to blood flow modeling and
    simulation, can support non-invasive diagnosis and assessment of
    cardiovascular diseases, as well as treatment planning and patient
    monitoring (see, e.g., \cite{Schwarz2023review, Fumagalli2024review}).
    The development of numerical solvers, together with the constantly
    growing potential of computing resources, allows tackling applications
    of extremely high complexity.
    However, especially in the case of aortic blood flow, where fluid dynamics
    is characterized by highly nonlinear regimes, strongly dependent on patient
    geometry, the physical and geometrical complexity still requires considerable
    computational resources for physiological patient-specific simulations.
    This aspect is particularly relevant for \textit{outer-loop} applications
    concerning, e.g., inverse problems, optimization, or
    uncertainty quantification, in which a large number of (forward)
    simulations is required at each iteration step.
    Reduced-order models and scientific machine learning (ML) can address this
    challenge by considering approximate dynamics on lower-dimensional spaces.
    These techniques rely on existing simulation (training) data to define
    high-fidelity reduced representations of target applications, and their
    capabilities are strongly dependent on data availability.

    Compared with other applications of model order reduction, biomedical
    applications are still constrained by limited data availability and
    substantial inter-patient variability in both anatomy and physiological
    conditions, concerning anatomical shape, cardiac output, peripheral pressure,
    and mechanical parameters, among others.
    Since \textit{full-order} CFD simulations are computationally expensive,
    the predictive accuracy of inter-patient surrogate models is essential for
    clinically relevant applications.

    This work addresses different challenges concerning inter-patient
    variability in the context of surrogate models of aortic blood flow.
    We study the approximation properties and the performance of ML
    surrogate models trained on synthetic data, i.e., data generated
    through high-fidelity numerical blood flow simulations (full-order models)
    on a dataset of variable patient anatomies sharing a common reference shape.
    In particular, we investigate the
    accuracy in terms of relevant biomarkers, namely, peak velocity, pressure,
    and wall shear stress (WSS), when predicting these quantities across shapes.

    Handling the geometrical variability in the context of scientific ML
    requires tailored strategies for developing reduced spaces and
    deep learning architectures that can scale better during training and also
    generalize accurately to unseen geometries.
    Previous studies explored the use of the reduced basis method combined with
    domain decomposition strategies to attach archetypal
    subdomains~\cite{Pegolotti2021}, and combined with registration techniques
    to transport a reduced basis computed on a template shape onto patient
    geometries~\cite{Guibert2014}. Recent works focusing on ML
    models have addressed the development of graph neural networks (GNNs)
    approximating the averaged quantities of interest (QoIs) from measurement data on
    the vessel centerline~\cite{Yevtushenko2022}, the learning of equivalent
    one-dimensional (1D) models from measurement data, to be applied to variable
    three-dimensional (3D) anatomies~\cite{Pegolotti2024}, common GNN architectures
    applied to the 3D mesh~\cite{NANNINI2025110477},
    \emph{shape-informed} GNNs, combined with registration strategies and
    trained on registered simulation data~\cite{Romor2025Assimilation}, and
    GNNs trained on image data of carotid arteries, which leverage first
    principles as soft constraints and equivariant layers~\cite{Suk2026}.
    In this work, we focus on graph- and transformer-based
    architectures~\cite{bodnar2024foundationmodelearth}, as they have proven
    effective for large-scale applications, and apply them to a database of
    physiological aortic geometries~\cite{Bosnjak2025SynthAorta} generated
    from parametric deformations.

    The accuracy of surrogate models and their capability of generalizing to different patients strongly depend on the variability of the available training data. This aspect is particularly important
    in the context of blood flow in large vessels, where geometrical details and patient-specific boundary conditions strongly influence the fluid dynamics. Flexible registration techniques and
    parameter-dependent shape databases, such as the one considered in this work, provide useful tools to increase the size of the training set with additional simulations.
  Typical methods to perform data augmentation, increasing the number of different anatomical shapes, include steady-state free precession (SSFP) magnetic resonance imaging (MRI) to directly acquire them from patients, statistical shape models~\cite{Yevtushenko2022}, parametric deformations~\cite{Bosnjak2025SynthAorta}, and generative models~\cite{Katz2026}. Although the availability of large datasets keeps increasing, the cost of generating high-fidelity CFD solutions remains high. The exploration of the space of all possible anatomical shapes must thus be optimized to reduce the computational costs. In this regard, active learning (AL)~\cite{rygiel2025activelearningdeeplearningbased} and reinforcement learning provide promising strategies to select the most informative samples to be added to the training dataset.
  We test different strategies for the selection of additional training samples, comparing \textit{label-dependent} (supervised) techniques, i.e., techniques requiring the full CFD solution computed on the candidate geometries, with \textit{label-free} (unsupervised) ones, which only rely on surrogate approximations and are therefore suitable for practical applications.

  A further aspect investigated in this work concerns the introduction of \textit{physically consistent} soft constraints in the training loss,
  besides purely data-dependent residuals computed from the solutions on the different shapes. The enforcement of these constraints
  has been tested extensively in the literature. Among others, physics-informed neural networks~\cite{Cuomo2022} and neural emulators~\cite{NEURIPS2020_43e4e6a6} use ML models as ansatz spaces while minimizing equation-based loss terms or residuals of numerical schemes via automatic differentiation. We are interested instead in physics constraints used as regularizers for surrogate models, with the aim of modeling a whole joint distribution of input parameters and output solutions, where the parameters represent complex boundary conditions or different computational domains. The efficiency of such regularizers with respect to state-of-the-art ML methods is not often benchmarked, making it difficult to discern whether purely data-informed or physics-informed losses yield the highest gain in generalization error relative to computational cost.

  This work provides the following main contributions.
  \begin{enumerate}
  \item We present a detailed comparison of two state-of-the-art GNN and transformer-based architectures in terms of accuracy and computational cost. In particular, we show that the partial differential equation (PDE) residuals of the predicted solutions can be used as \emph{a posteriori} error estimators.
  \item Concerning AL strategies, our results
  show that physics-based strategies, often considered the most effective (e.g.,~\cite{rygiel2025activelearningdeeplearningbased}),
  are not necessarily the best choices for the considered applications.
  \item Physical consistency soft constraints are more effective as a fine-tuning objective than as a regularizer applied throughout training.
  \item We analyze how the considered surrogate models perform for out-of-distribution samples, based on shape databases published in different contexts, discussing how to improve the generalization capabilities of the surrogate models.
  \end{enumerate}

    The rest of this paper is organized as follows: Section~\ref{sec:dataset_generation} summarizes the main aspects of the dataset
    generation and curation including the underlying modeling assumptions and their implications for the designed surrogate models. Section~\ref{sec:surrogate_modeling} then focuses on the validation of state-of-the-art deep learning architectures for surrogate models of aortic flows (Graph U-Nets and Transformer U-Nets), including the analysis of inter-patient AL methods, the application of physics-based soft constraints during training and in fine-tuning, and the evaluation of the models' generalization capabilities. Section~\ref{sec:conclusion} discusses the conclusions, limitations, and possible future directions.

    \section{Dataset generation}
    \label{sec:dataset_generation}
    This section describes the dataset generation, discussing the shape variability and parametrization of physiological, healthy aortas (Section~\ref{sec:geometric_modeling}), the
    relevant aspects of the vascular flow modeling (Section~\ref{sec:vascular_flow}), and the details of the CFD solver (Section~\ref{sec:cfd_solver_and_results}).

    \subsection{Geometric modeling}
    \label{sec:geometric_modeling}
    The shape database used in this work considers a geometric parametrization
    of healthy aortas based on the SynthAorta
    dataset~\cite{Bosnjak2025SynthAorta}, generated by mapping a single common
    reference centerline model. Equipped with a parametrized transformation and
    radii at each point of the centerline, a broad range of
    healthy physiological aortas can be captured, albeit in a simplified manner.
    \citet{Bosnjak2025SynthAorta} proposed a set of hyperparameters defining healthy
    aortic geometries, which have a clear physiological meaning, are directly
    measurable, and are mutually independent.
    The stochastically sampled values for these hyperparameters are sourced from
    patient-specific data whenever possible.
    For the sources of the geometric hyperparameters, their definitions, and
    related distributions, we refer the reader to \cite{Bosnjak2025SynthAorta}.

    \begin{remark}
    Naturally, the pre-existing parametrization introduces assumptions and
    limitations regarding the aorta's shape, and hence might not cover the entire
    healthy range, let alone pathological configurations such as those resulting from
    abdominal aortic aneurysms~\cite{Doyle2014Aneurysm} or aortic
    dissection~\cite{Rolf2025MechanismsOfAorticDissection}.
    On the other hand, the parametrization allows generating vast high-quality
    datasets, which are crucial for ML purposes. To address this
    issue, out-of-distribution generalization is explored in
    Section~\ref{sec:results_ood}, investigating the potential of the presented
    methods in such scenarios. An alternative approach, not considered in this
    work, would be to consider arbitrary, non-parametric shape datasets, combined
    with additional registration methods~\cite{Romor2025Assimilation}, for which,
    however, the performance of the surrogate might depend on the properties of
    the available data~\cite{Romor2025Assimilation}.
    \end{remark}

    Based on the available collection of centerlines with corresponding radial
    data, smooth aortic surfaces of the SynthAorta dataset are reconstructed via
    convolution surfaces~\cite{Bloomenthal_1997a, FuentesSuarez_2019a}.
    Structured meshing methods based on convolution surfaces have been employed
    for tube-like structures~\cite{Bosnjak2024BlockStructuredMeshing},
    such as healthy aortas, but have also been extended to
    describe more general biological shapes~\cite{Pepe2025Deep}, pathological
    vessel configurations such as aortic
    dissection~\cite{Bosnjak2024SemiAutomatic}, or vessel wall
    geometries~\cite{Schussnig2025Hyperelasticity}.

    Since each aorta in the SynthAorta parametrization is represented through a
    limited set of hyperparameters, it inevitably cannot reproduce the full
    geometric detail of any individual patient's anatomy: local, high-fidelity
    features not captured by the chosen hyperparameters are necessarily lost.
    For aortic flow, however, \citet{Bosnjak2024GeometricUncertainty} have shown
    that such local geometric variations play only a secondary role for global
    hemodynamic QoIs, such as volumetric flow rates,
    pressures, and averaged WSS. Whenever local QoIs must be
    resolved with high accuracy, e.g., at elevated Reynolds numbers,
    patient-specific CFD remains the gold standard; global QoIs, in
    contrast, can be inferred reliably given a sufficiently rich training
    dataset. The methods presented in this work therefore necessarily trade
    some geometric fidelity to individual patients for the ability to generalize
    across the broader anatomical variability captured by the parametrization.

\begin{figure}[!ht]
        \centering
        \includegraphics[width=0.10\linewidth, draft=\draftHeavy
        ]{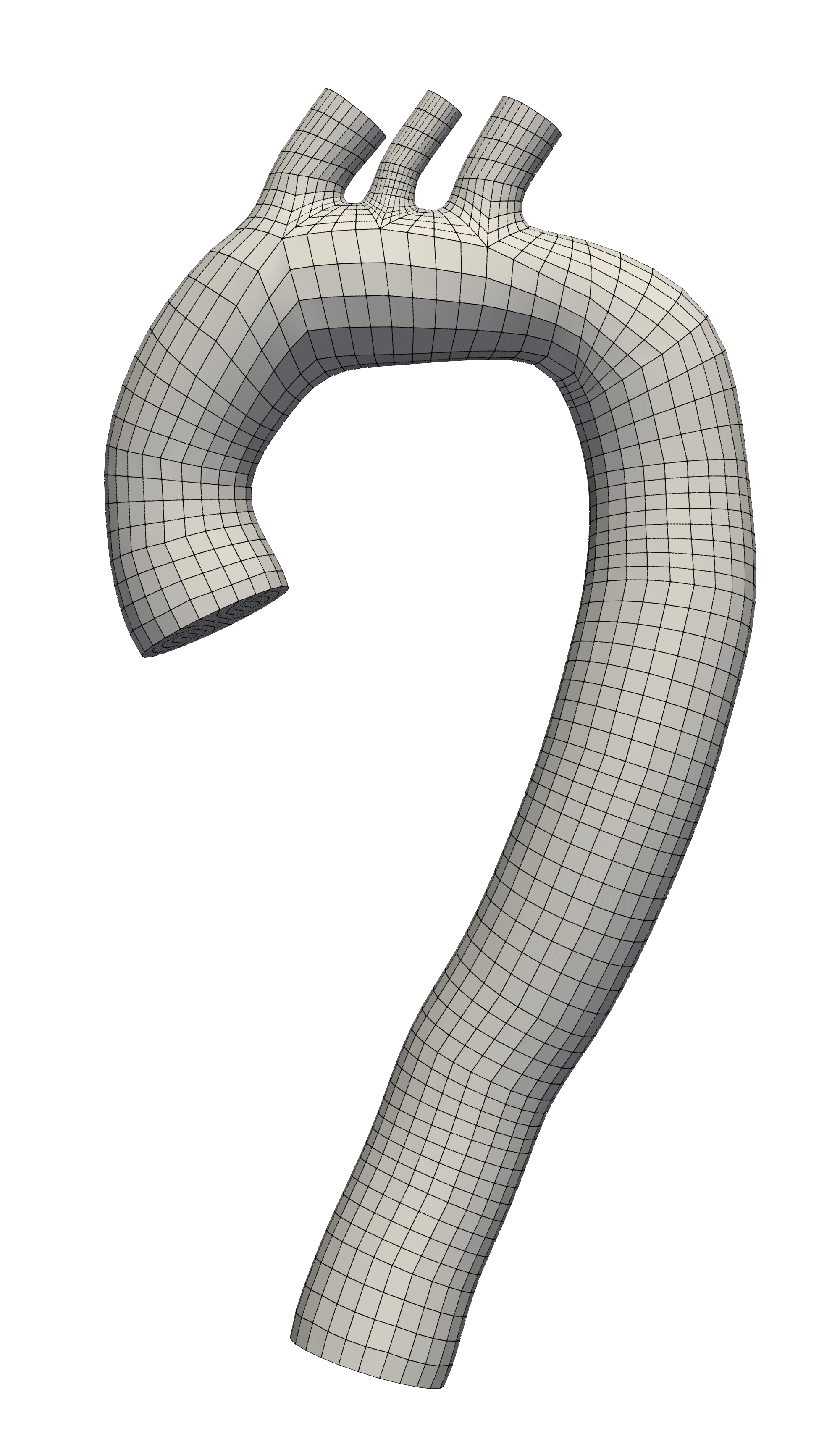}
        \hspace{0.00\linewidth}
        \includegraphics[width=0.10\linewidth, draft=\draftHeavy
        ]{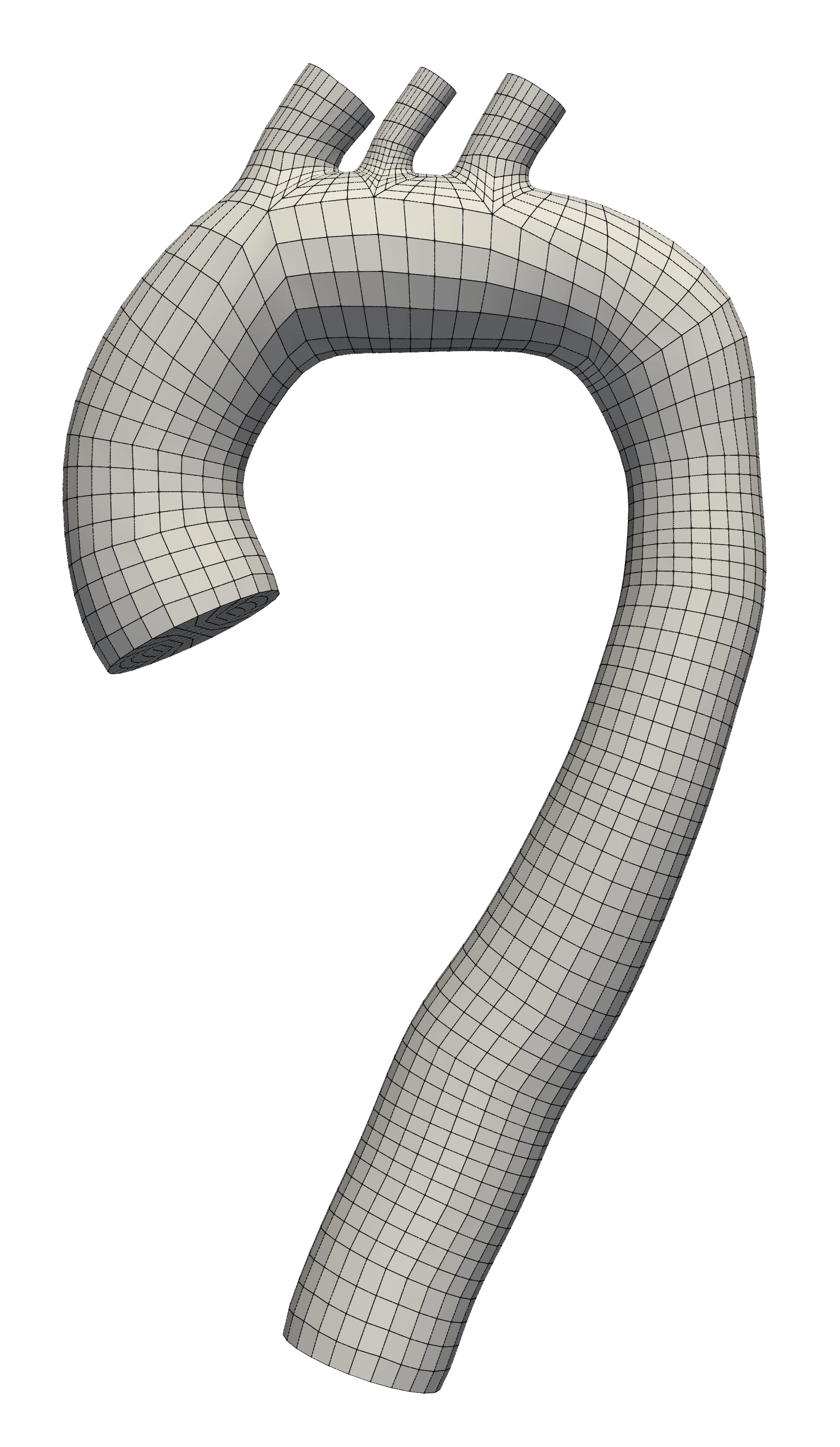}
        \hspace{0.00\linewidth}
        \includegraphics[width=0.10\linewidth, draft=\draftHeavy
        ]{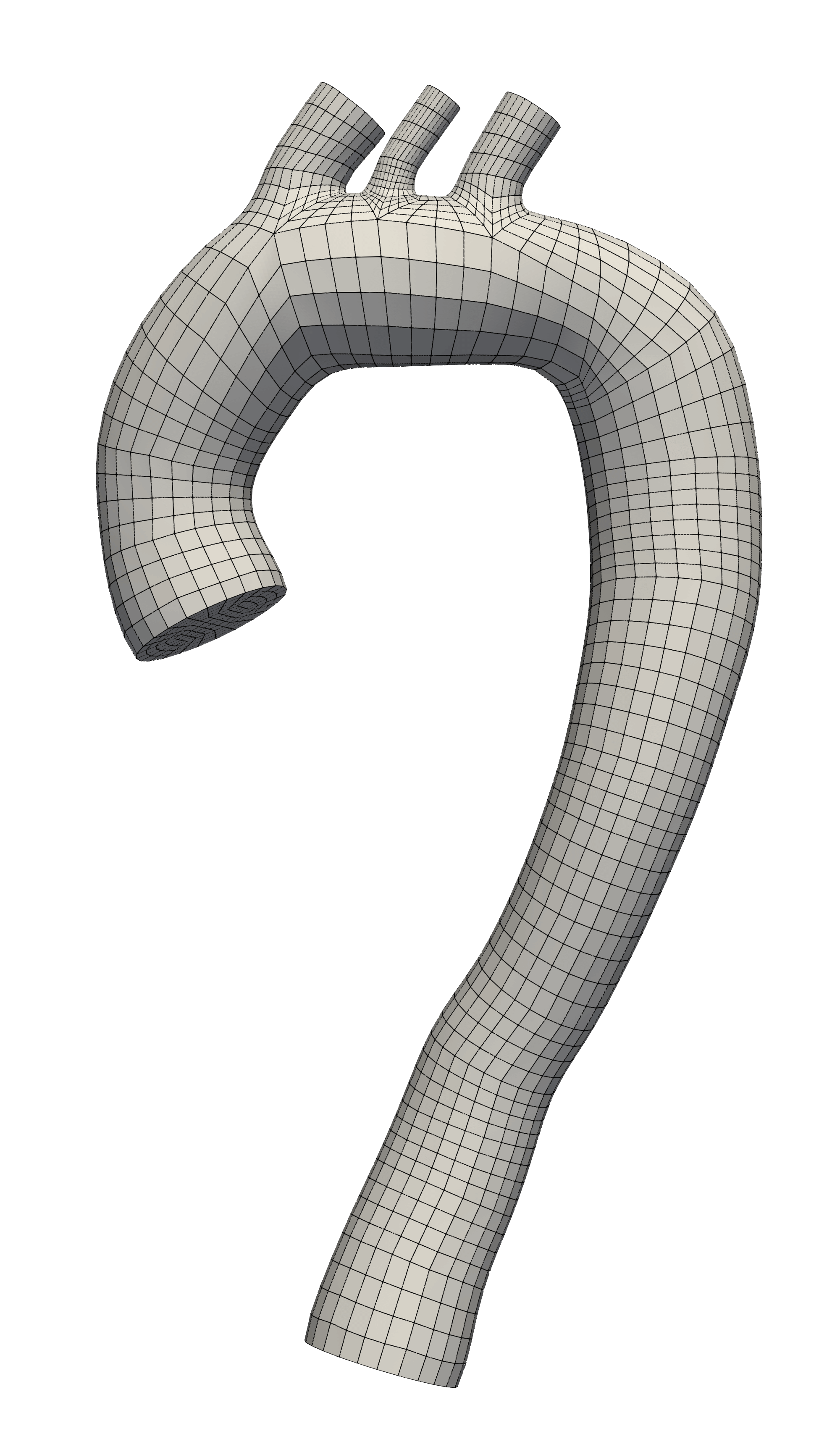}
        \hspace{0.00\linewidth}
        \includegraphics[width=0.10\linewidth, draft=\draftHeavy
        ]{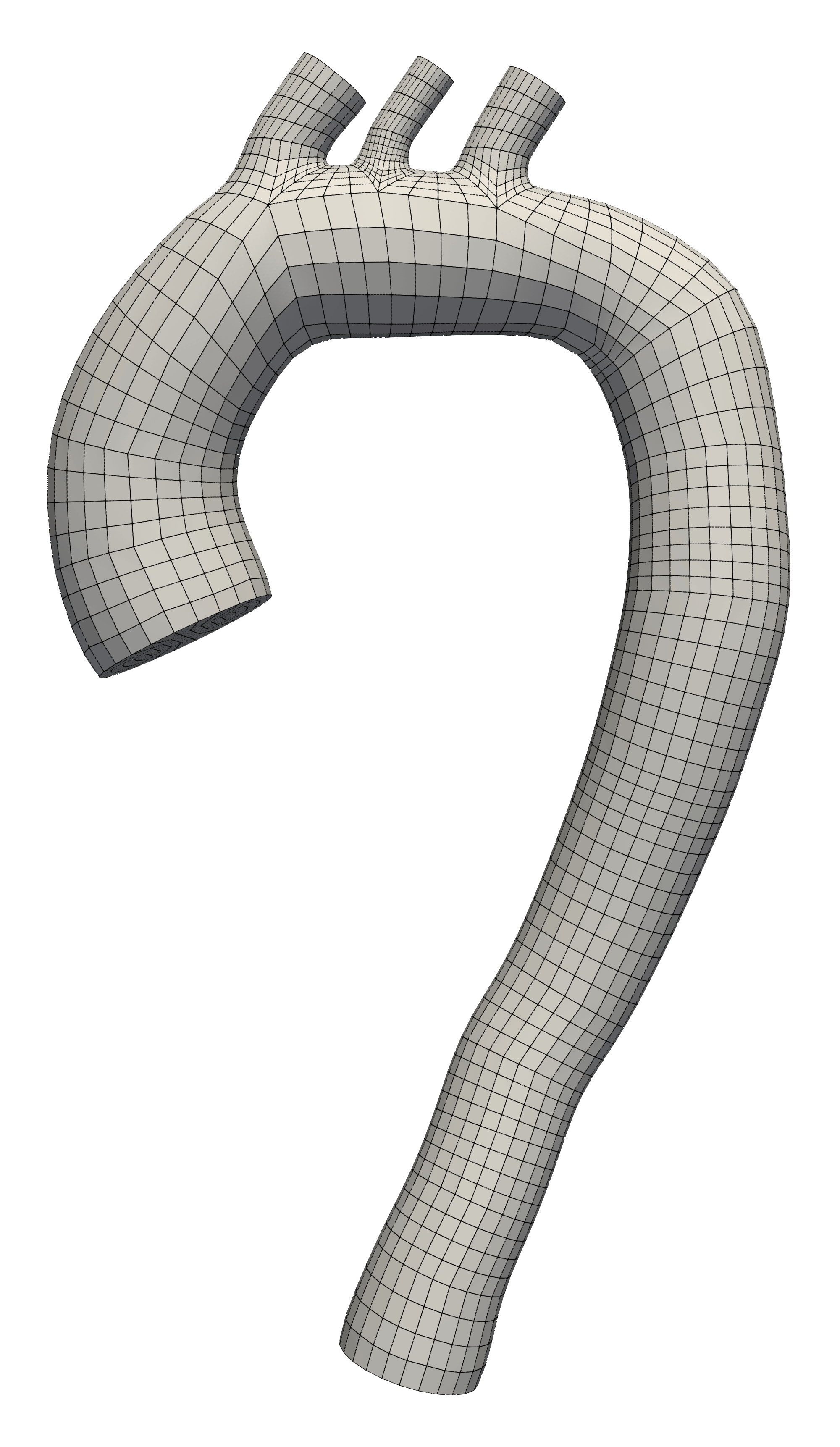}
        \hspace{0.00\linewidth}
        \includegraphics[width=0.10\linewidth, draft=\draftHeavy
        ]{case_5_frontal_small}
        \hspace{0.00\linewidth}
        \includegraphics[width=0.10\linewidth, draft=\draftHeavy
        ]{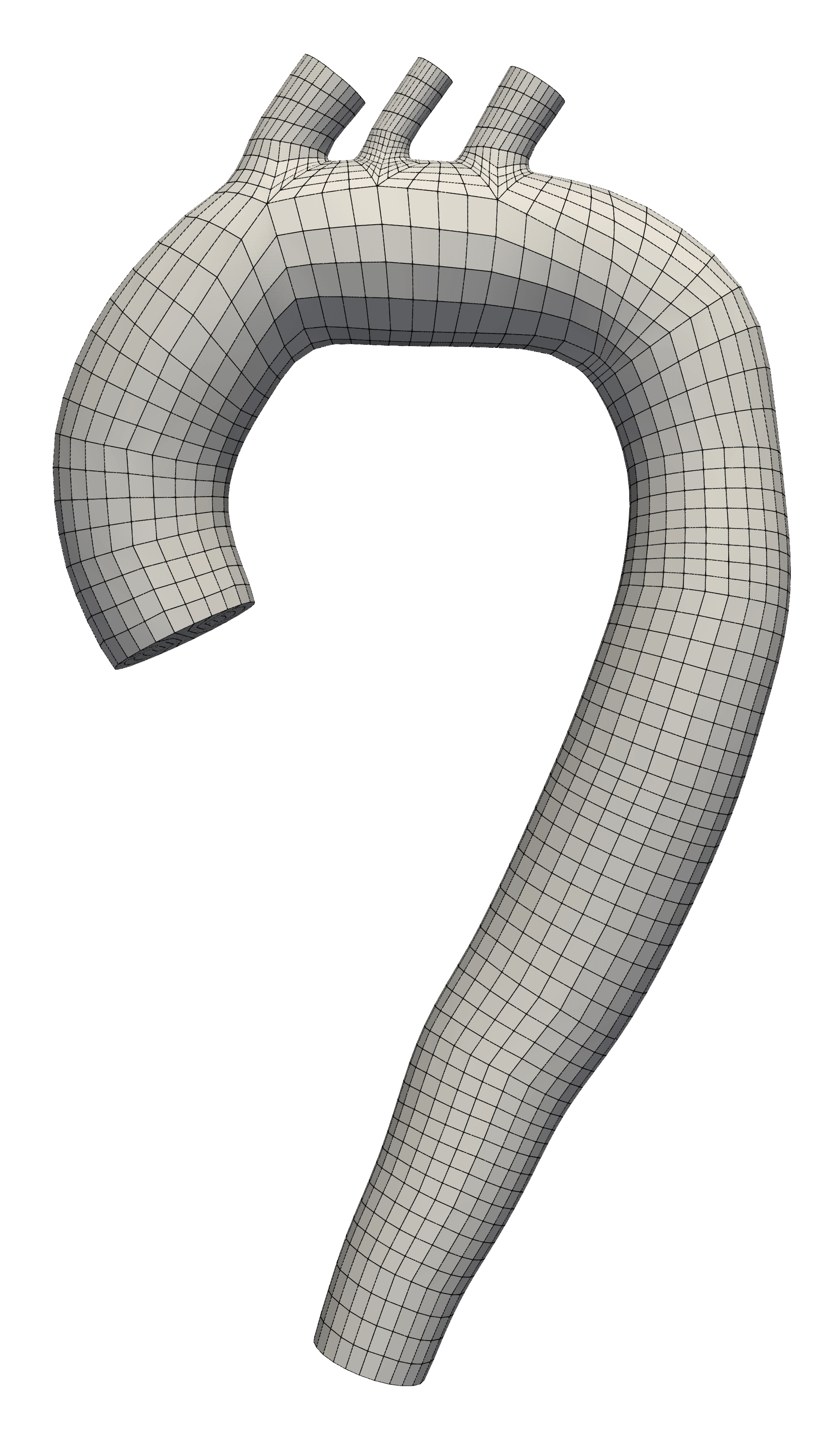}
        \hspace{0.00\linewidth}
        \includegraphics[width=0.10\linewidth, draft=\draftHeavy
        ]{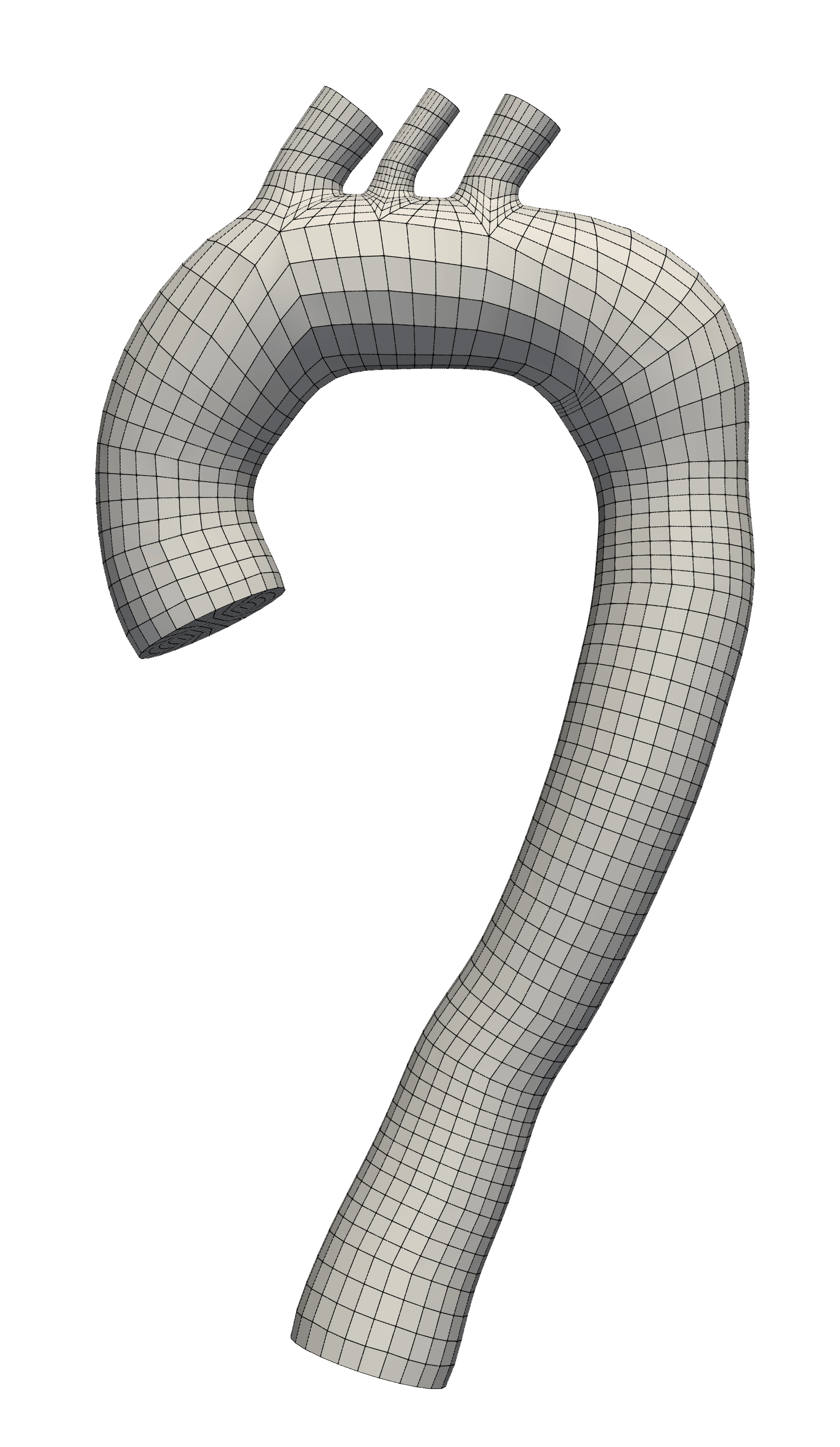}
        \hspace{0.00\linewidth}
        \includegraphics[width=0.10\linewidth, draft=\draftHeavy
        ]{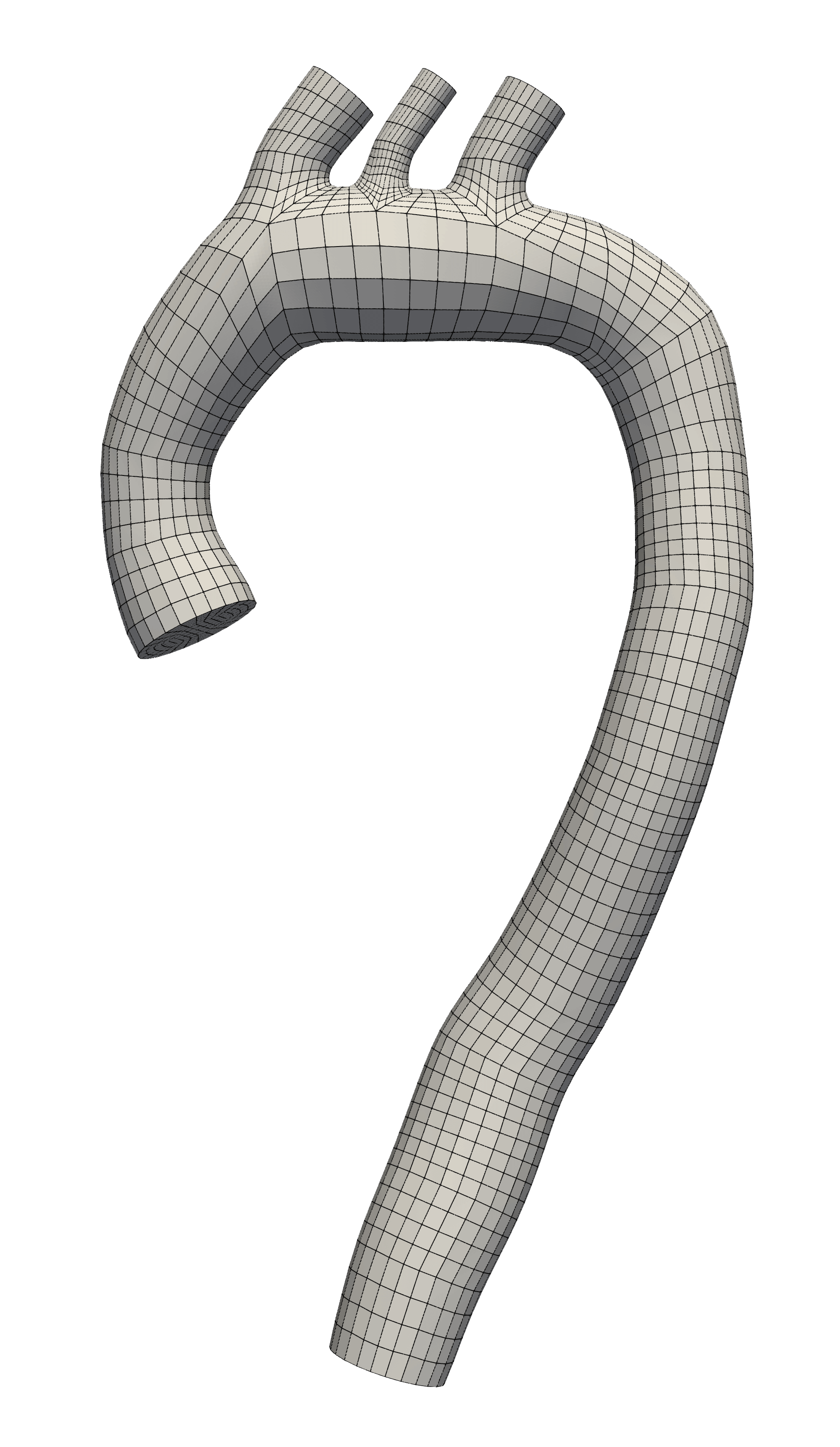}
        \\
        \includegraphics[width=0.10\linewidth, draft=\draftHeavy
        ]{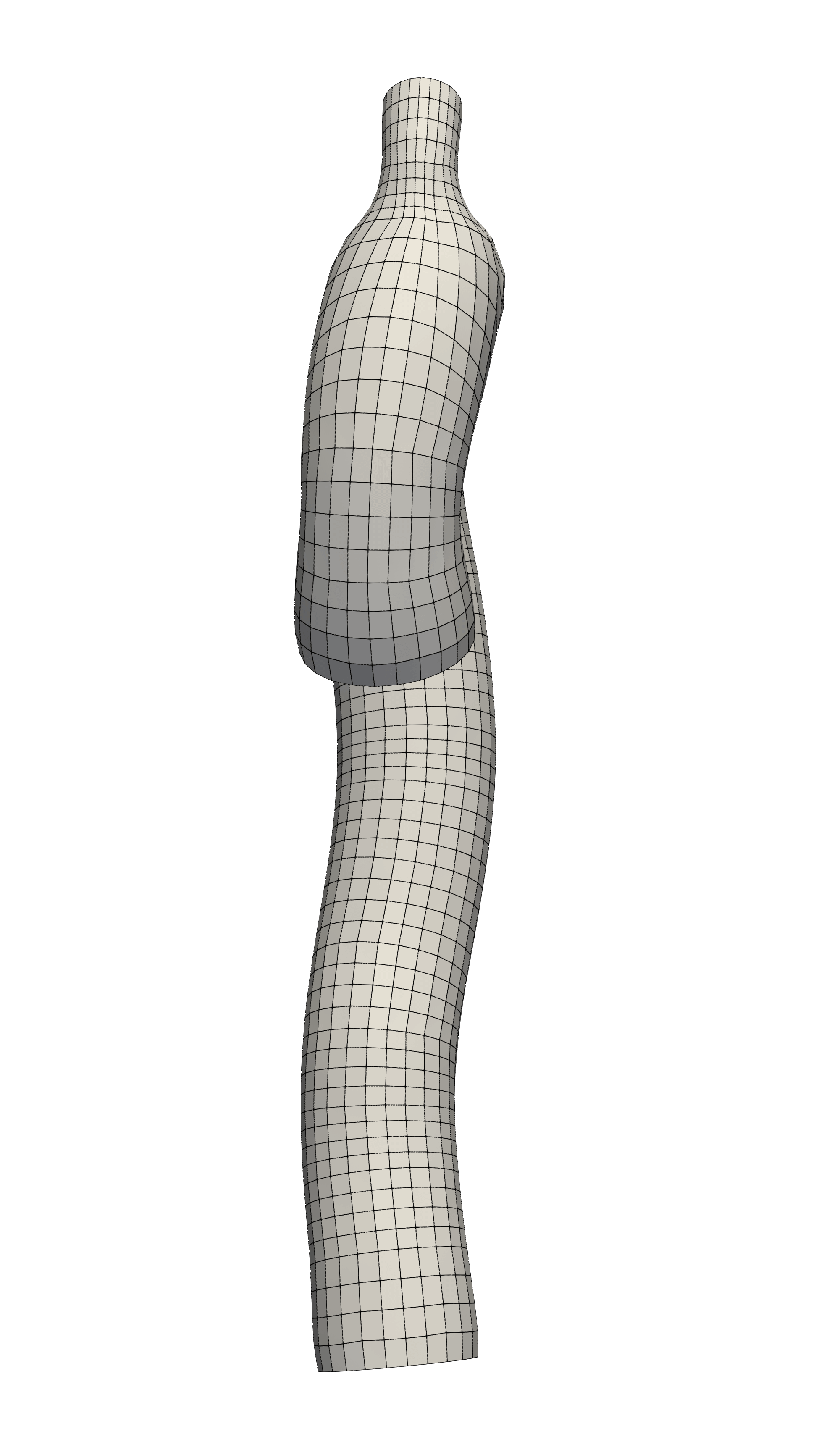}
        \hspace{0.00\linewidth}
        \includegraphics[width=0.10\linewidth, draft=\draftHeavy
        ]{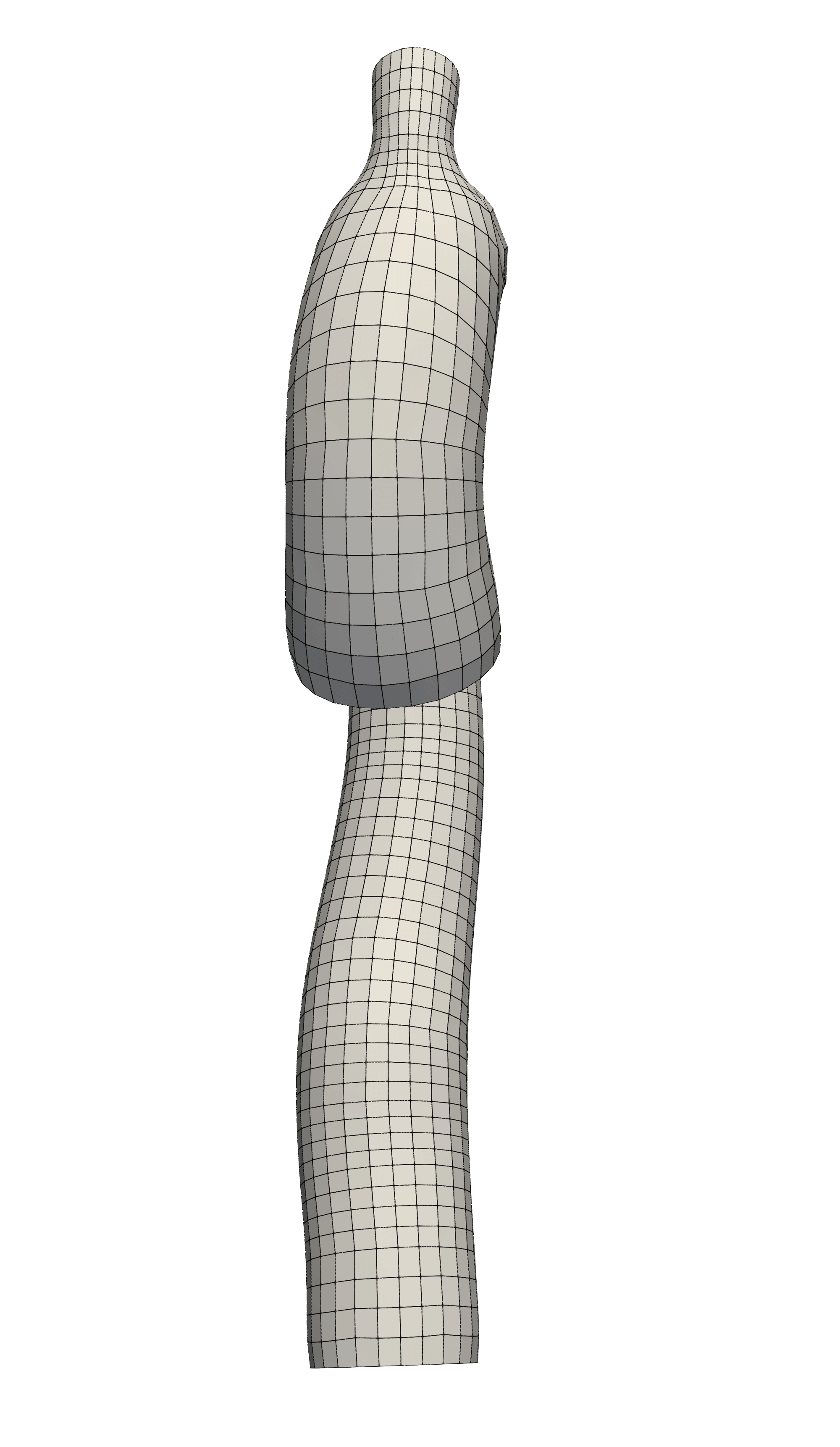}
        \hspace{0.00\linewidth}
        \includegraphics[width=0.10\linewidth, draft=\draftHeavy
        ]{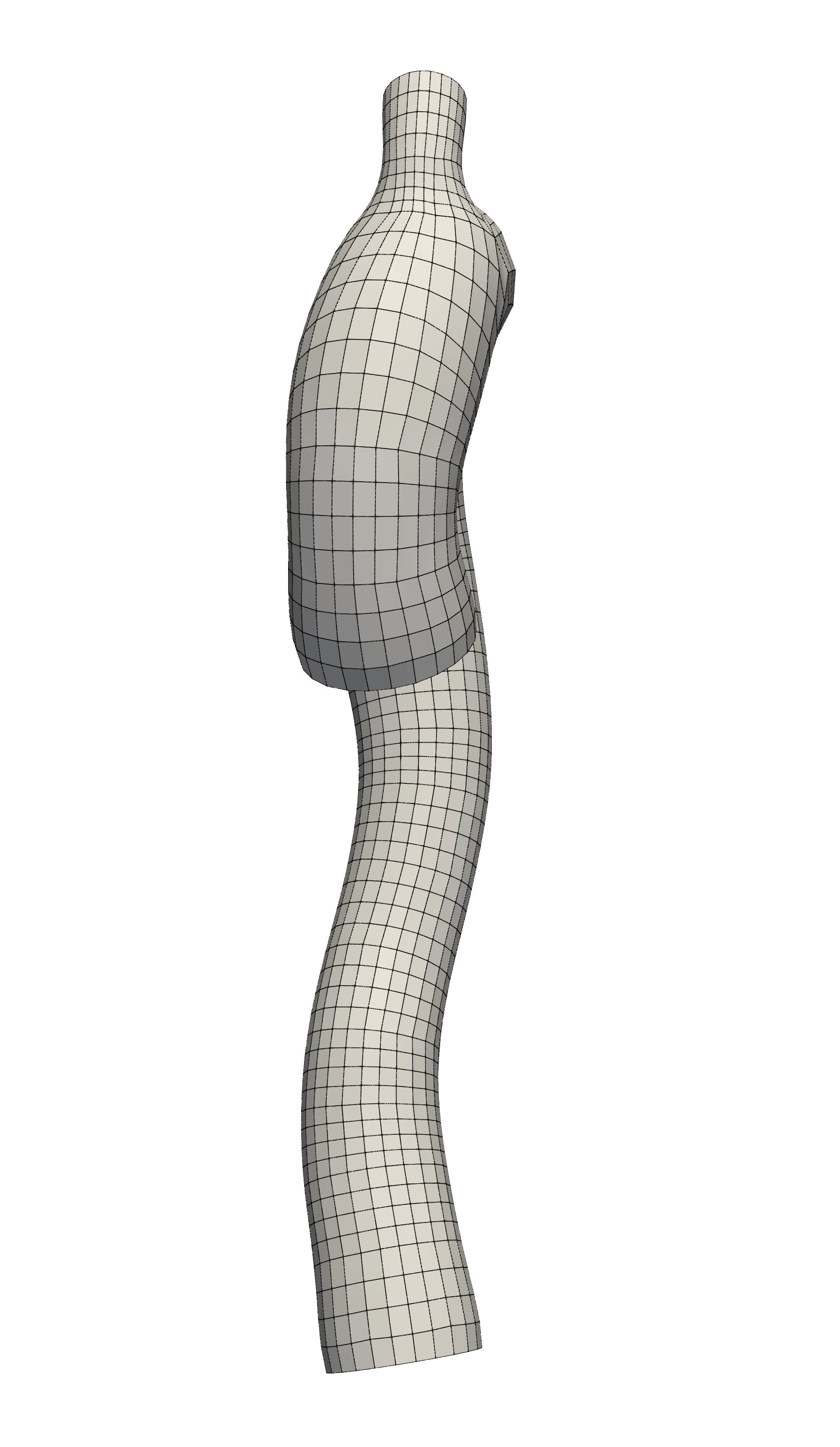}
        \hspace{0.00\linewidth}
        \includegraphics[width=0.10\linewidth, draft=\draftHeavy
        ]{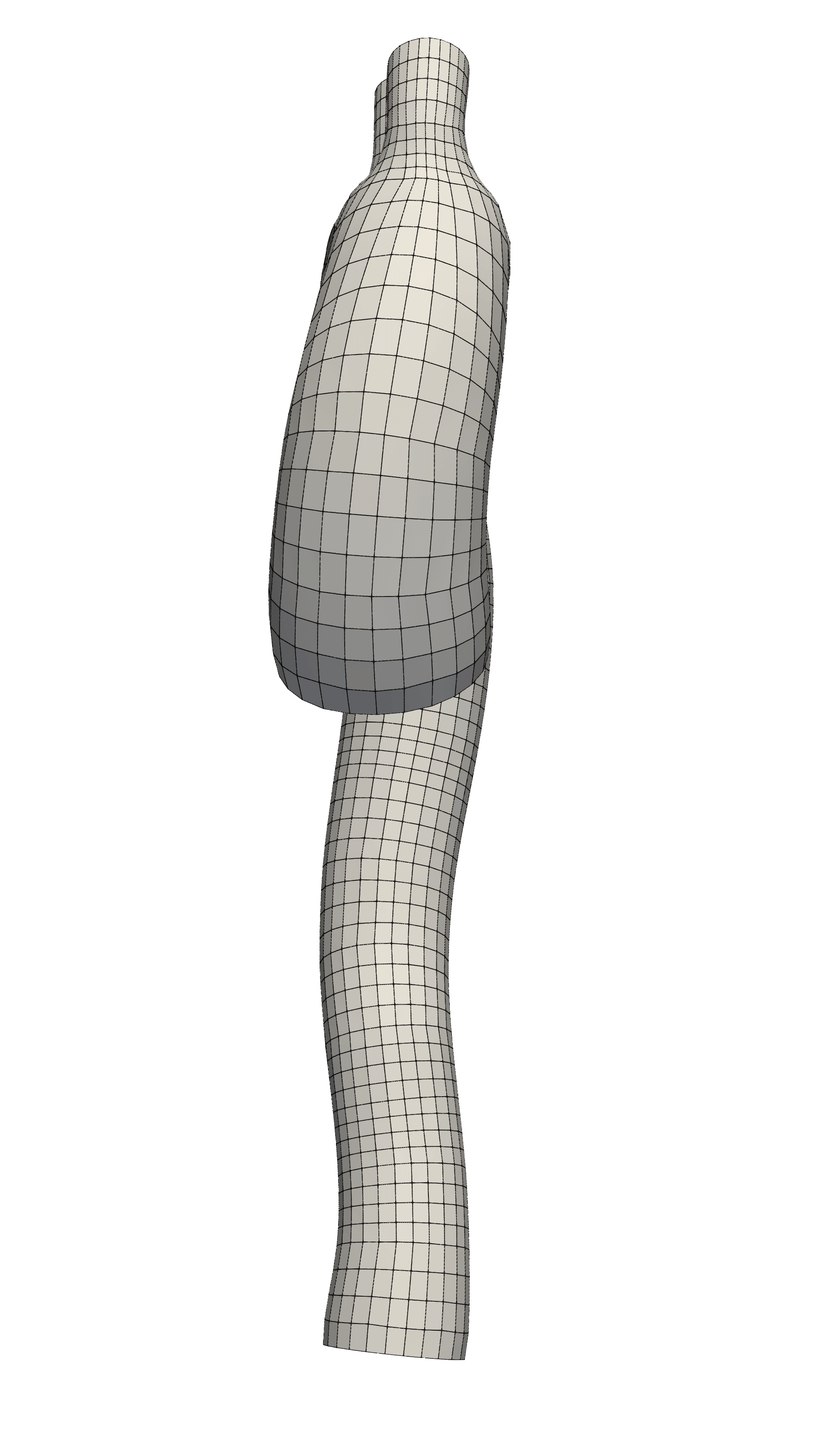}
        \hspace{0.00\linewidth}
        \includegraphics[width=0.10\linewidth, draft=\draftHeavy
        ]{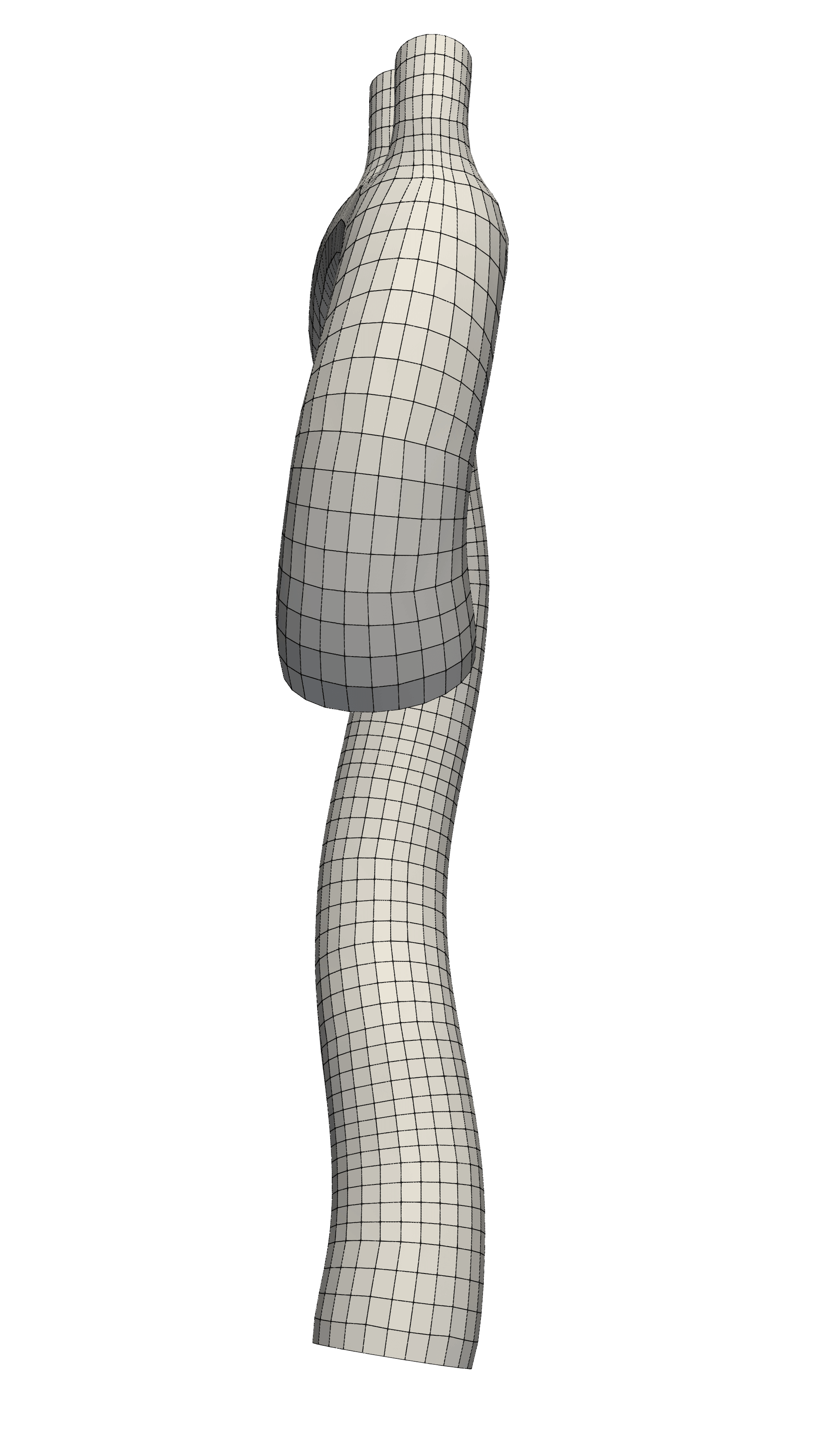}
        \hspace{0.00\linewidth}
        \includegraphics[width=0.10\linewidth, draft=\draftHeavy
        ]{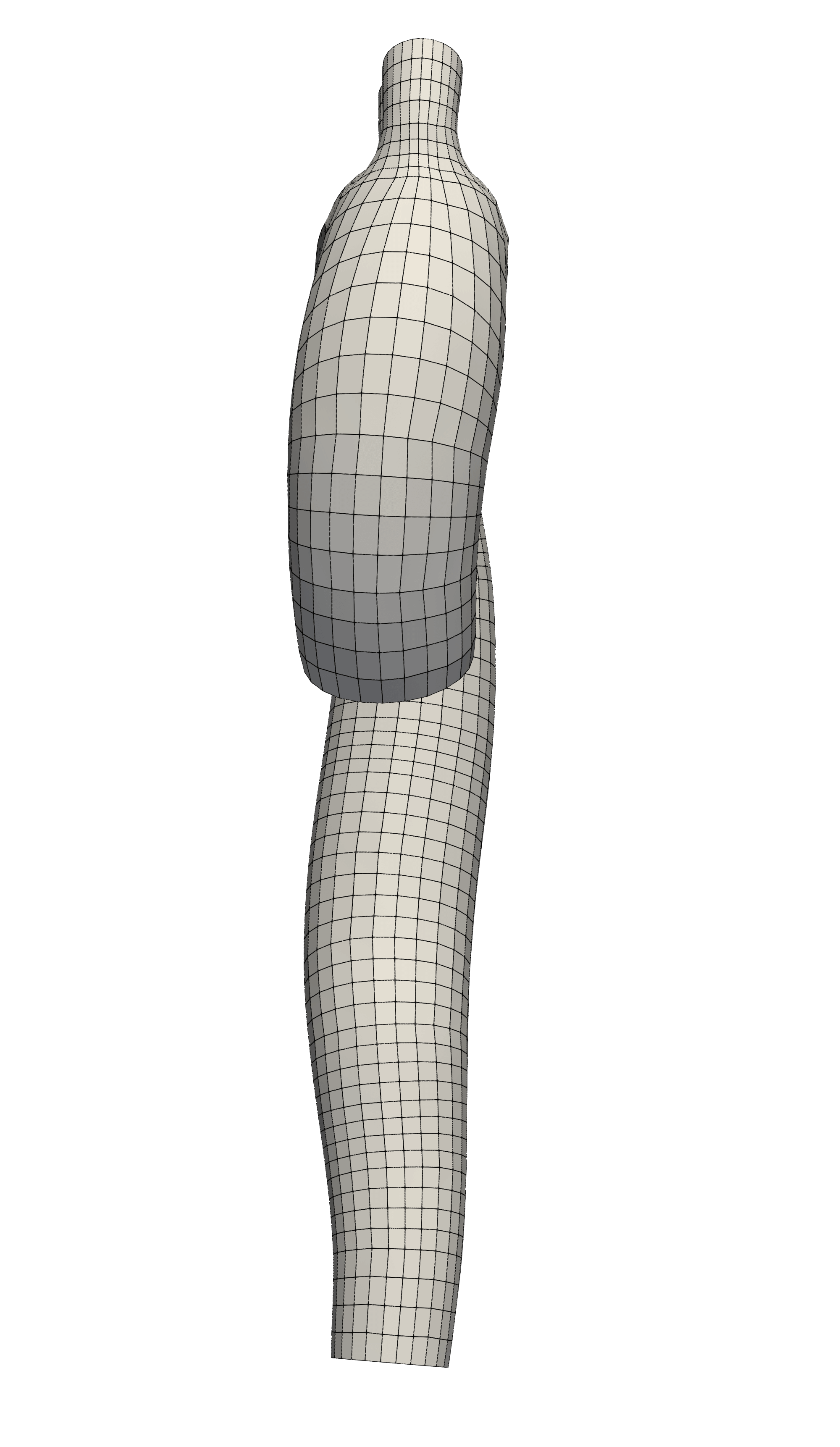}
        \hspace{0.00\linewidth}
        \includegraphics[width=0.10\linewidth, draft=\draftHeavy
        ]{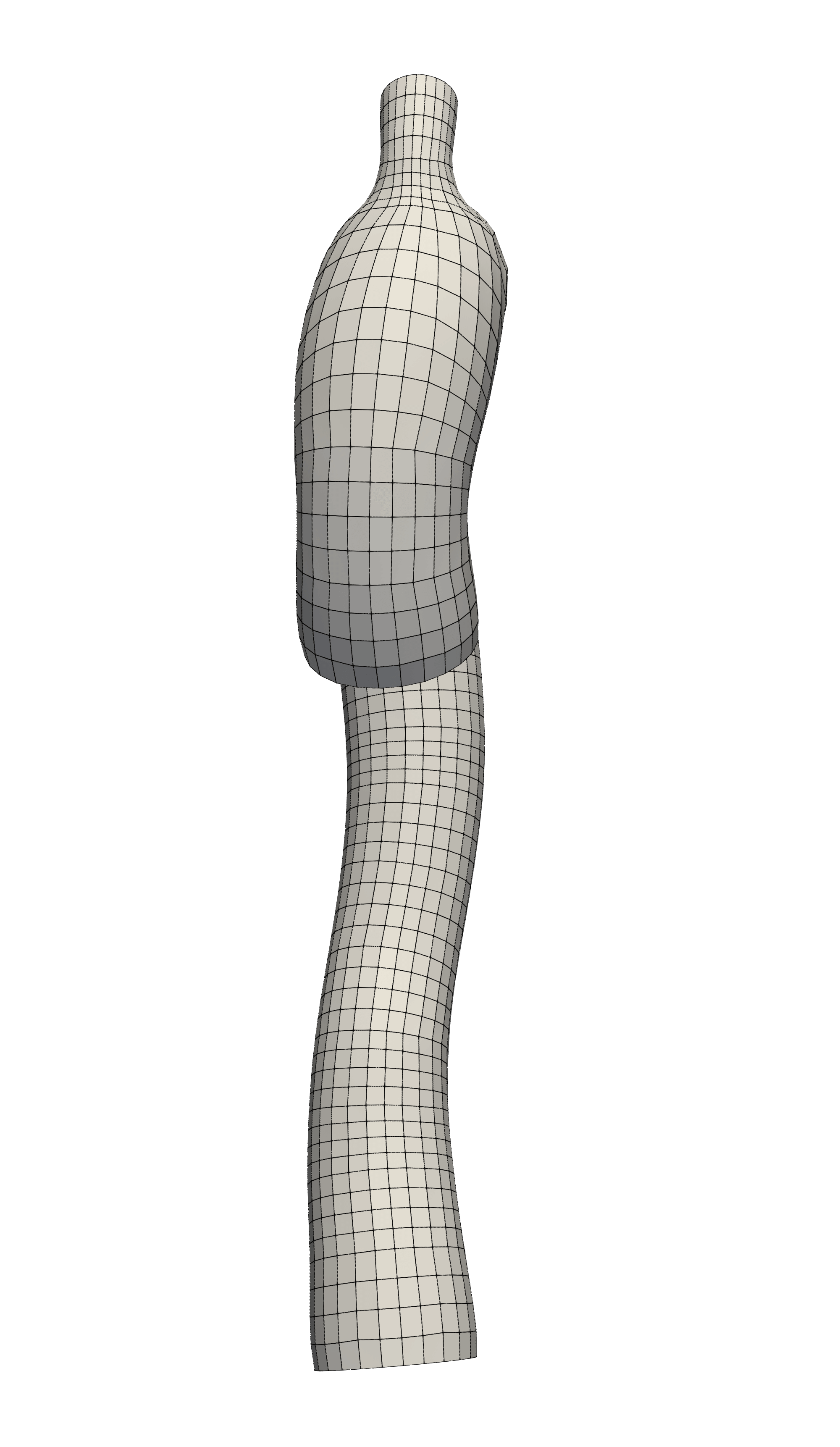}
        \hspace{0.00\linewidth}
        \includegraphics[width=0.10\linewidth, draft=\draftHeavy
        ]{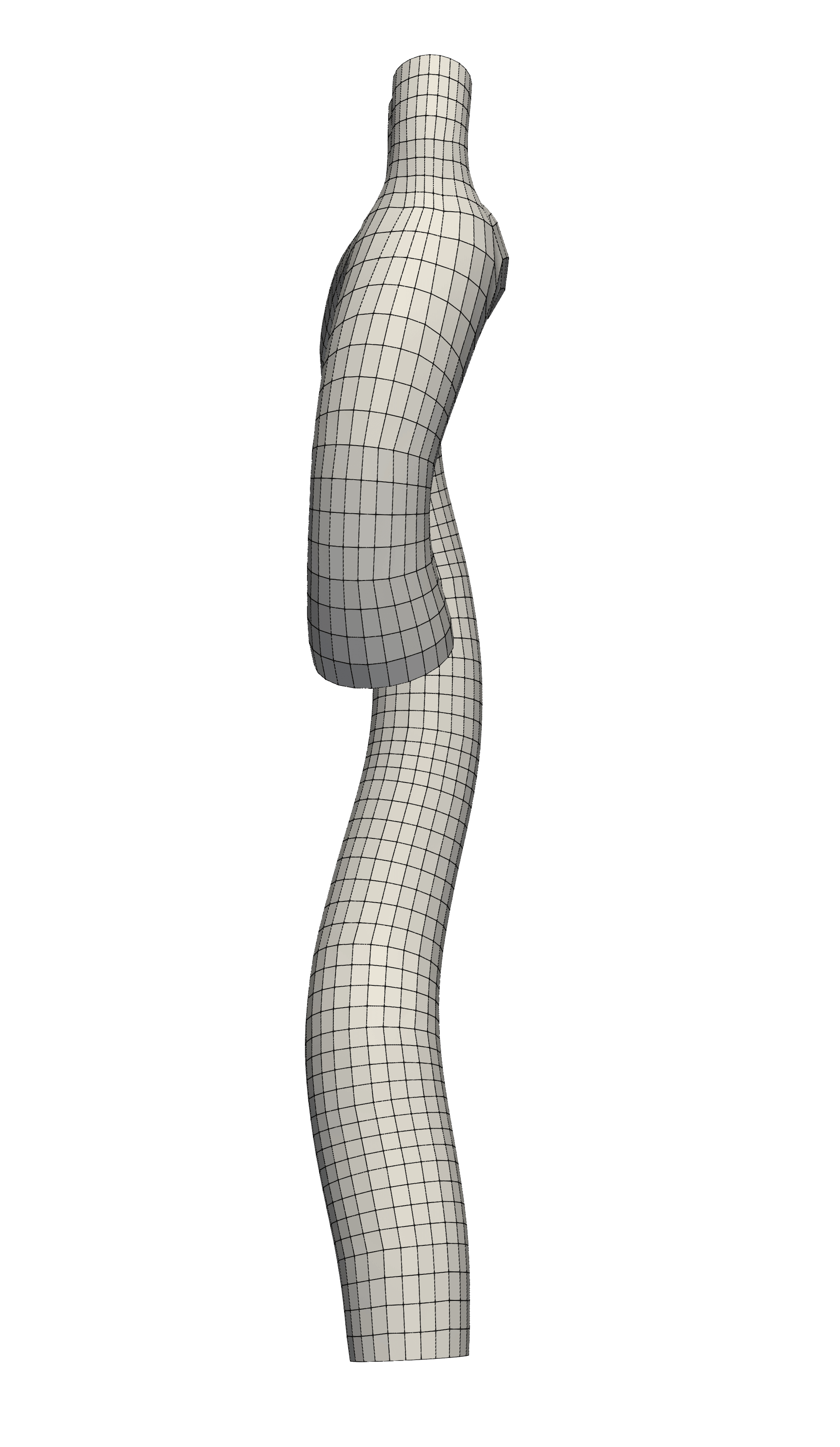}
        \caption{Sample cases of the SynthAorta dataset~\cite{Bosnjak2025SynthAorta}
            with Latin hypercube sampling of geometric hyperparameters
            showcasing the geometric variability within the training dataset:
            frontal view (top row) and lateral view (bottom row) of cases 1 to 8.}
        \label{fig:SynthAorta_sample_cases}
    \end{figure}

    For the purposes herein, we consider cases 1 to 1000
    of the SynthAorta dataset~\cite{Bosnjak2025SynthAorta}, wherein the
    hyperparameters are sampled with a Latin hypercube method, covering the entire
    parameter space. To showcase the geometric variability within the training and
    test data, Figure~\ref{fig:SynthAorta_sample_cases} shows eight
    samples.

    \subsection{Vascular flow modeling}
    \label{sec:vascular_flow}
    \subsubsection{Blood flow}
    To generate synthetic training data, we model the blood flow as an
    incompressible fluid. Namely, blood velocity and (kinematic)
    pressure are assumed to be described by the Navier--Stokes equations in a
    domain $\Omega \subset \mathbb{R}^3$, representing any aortic geometry in the
    dataset, and in a selected time interval $[0,T]$:
    \begin{align}
        \frac{\partial}{\partial t}\ve{u}
        +
        \ve{u} \cdot \nabla \ve{u}
        -
        2 \nabla \cdot \left(\nu(\ve{x}) \nabla^\mathrm{S}\ve{u} \right)
        +
        \nabla p
        &=
        \ve{0}
        & \text{in }\Omega \times (0,T]
        \,,
        \label{eqn:hemodynamics_momentum_balance}
        \\
        \nabla \cdot \ve{u}
        &=
        0
        & \text{in }\Omega \times [0,T]
        \label{eqn:hemodynamics_continuity}
        \,,
    \end{align}
    where $\nu$ denotes the kinematic viscosity, and the symmetric gradient is
    defined as
    $\grads (\cdot) \coloneqq \frac{1}{2}\left[\nabla (\cdot) + \nabla (\cdot)^\top\right]$.
    To account for the blood's rheological behavior, a generalized Newtonian
    model is considered, where the
    apparent viscosity depends on the local shear rate,
    $\dot{\gamma}(\ve{x}) \coloneqq 
    \sqrt{2 \nabla^\mathrm{S} \ve{u} : \nabla^\mathrm{S} \ve{u}}$, 
    via the Carreau model, i.e.,
    \begin{equation}\label{eq:carreau}
        \nu (\dot\gamma) = \eta_\infty + (\eta_0 - \eta_\infty)
        \left[ 1 + (\lambda \dot\gamma)^2 \right]^{\frac{n-1}{2}}
        \,,
    \end{equation}
    with $\eta_0$, $\eta_\infty$, $\lambda$, and $n$ being rheological parameters.
    The physiological fluid parameters for Equation~\eqref{eq:carreau} are taken from~\cite{Ranftl2021Bayesian}.
    In addition to the primary unknowns, i.e., the fluid velocity $\ve{u}$ 
    and pressure $p$, and their averages on the inlet and outlets,
    the WSS vector $\ve{\tau}_\mathrm{wall}\in\mathbb{R}^3$, defined as
    \begin{equation}
        \label{eqn:wall_shear_stress_vector}
        \ve{\tau}_\mathrm{wall}
        :=
        2 \rho\nu \, \ve{n} \cdot \grads \ve{u}
        -
        2 \rho\nu
        \left[ \ve{n} \cdot \left(\ve{n} \cdot \grads \ve{u} \right) \right]\ve{n}
        \,,
    \end{equation}
    where $\ve{n}$ is the unit outward normal vector and $\rho$ the fluid density,
    plays a critical role in various pathologies such as abdominal aortic aneurysms~\cite{Doyle2014Aneurysm} or
    aortic dissection~\cite{Rolf2025MechanismsOfAorticDissection}. Defined on $\Gamma_\mathrm{wall}$ only,
    it is obtained in a postprocessing step. For the sake of brevity, other biomarkers relevant in hemodynamics are
    not discussed, noting that such QoIs depend on the fluid velocity and pressure in some way and
    might hence be inferred from the predicted primary unknowns.

    \subsubsection{Boundary conditions}
\begin{figure}[!ht]
        \centering
        \begin{overpic}[height=4cm, draft=\draftHeavy
            ]{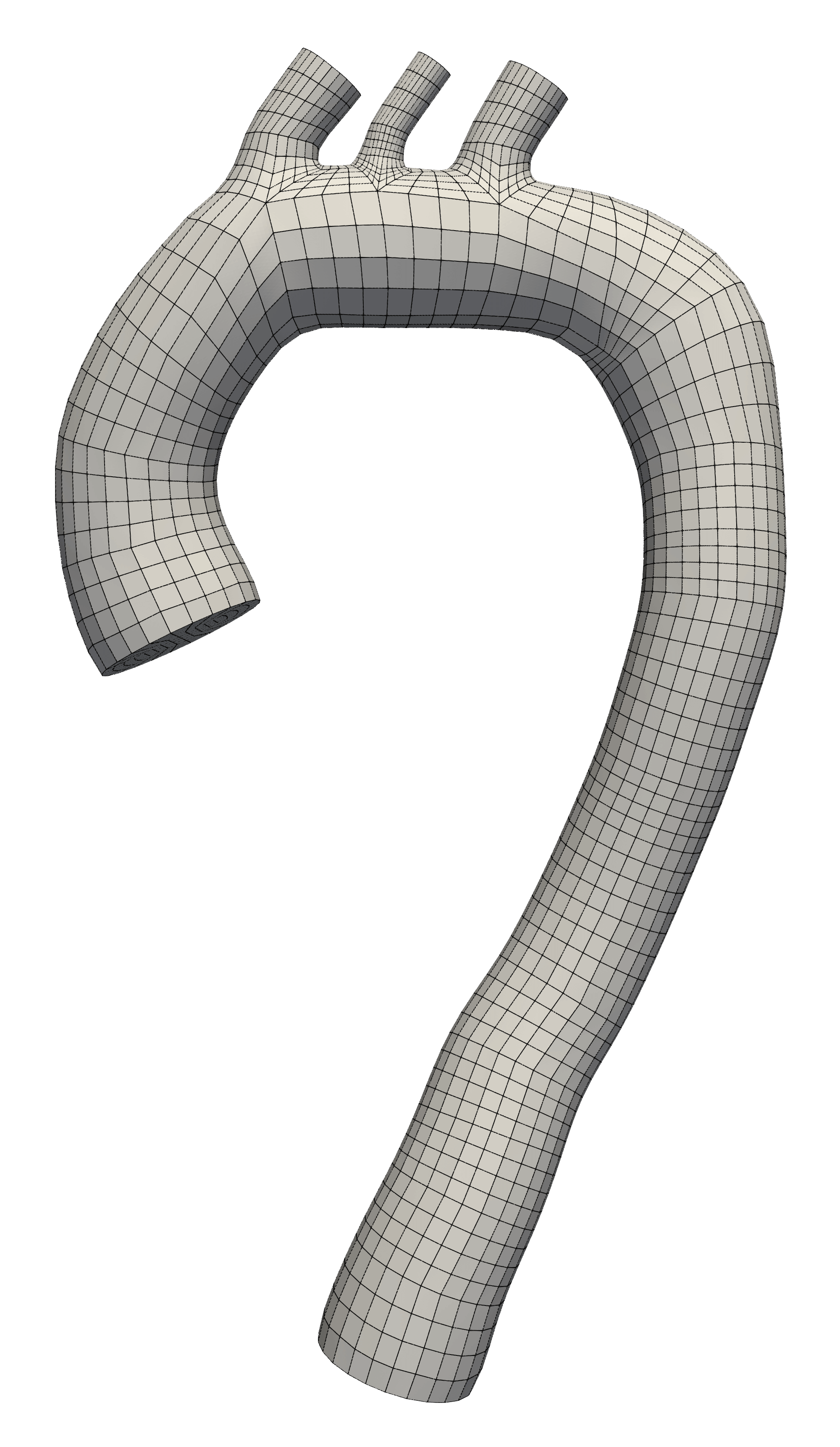}
            \put(120,460){$\Gamma_0$}
            \put(160,990){$\Gamma_1$}
            \put(260,990){$\Gamma_2$}
            \put(360,990){$\Gamma_3$}
            \put(120,60){$\Gamma_4$}
            \put(230,700){$\Gamma_\mathrm{wall}$}
        \end{overpic}
        \hspace{0.02\linewidth}
        \begin{overpic}[height=4cm, draft=\draftHeavy
            ]{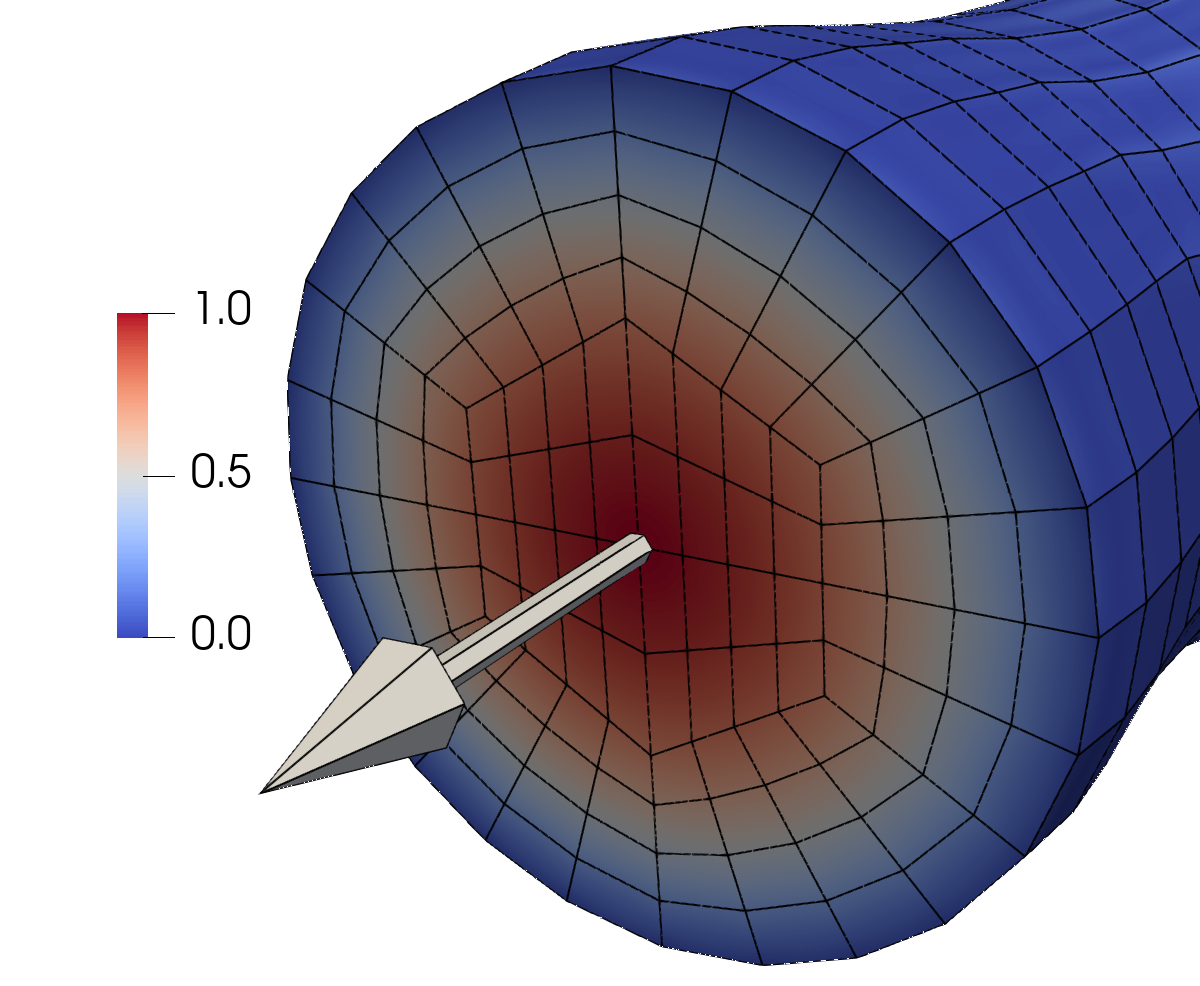}%
            \put(80,630){$\eta_\ve{x}$}
            \put(150,150){$\ve{n}$}
        \end{overpic}
        \hspace{0.06\linewidth}
        \begin{overpic}[height=4cm, draft=\draftHeavy
            ]{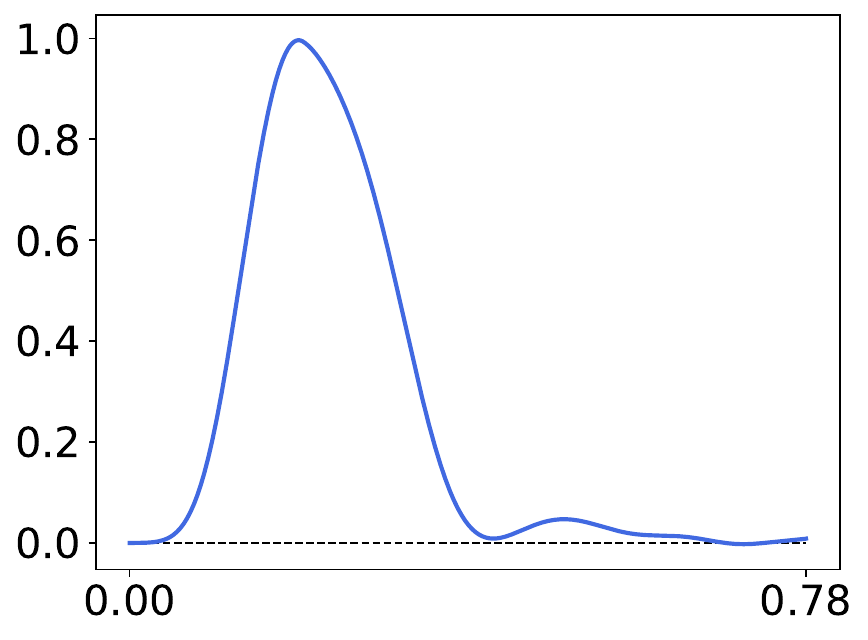}
            \put(375,0){Time $t$ in s}
            \put(-75,300){\rotatebox{90}{Scale $\eta_t$}}
        \end{overpic}
        \caption{Sample aortic geometry considered with inlet $\Gamma_0$ at the aortic root, supra-aortic outlets
            (brachiocephalic trunk $\Gamma_1$, left common carotid artery $\Gamma_2$, and left subclavian artery $\Gamma_3$),
            abdominal aorta outlet $\Gamma_4$, and aortic wall $\Gamma_\mathrm{wall}$ (left). Spatial (middle) and temporal (right) scaling of the inflow profile.}
        \label{fig:bcs_Gamma_i_definition_and_inlet_scale}
    \end{figure}

   The system of Equations~\eqref{eqn:hemodynamics_momentum_balance}--\eqref{eqn:hemodynamics_continuity} needs to be completed by suitable boundary conditions, which
   play a crucial role in recovering realistic and physiological flow fields.
   These boundary conditions depend on the considered patient geometry and on a set of parameters which, typically, need to be tuned according to the specific case.
   To handle this aspect, we consider an automatic parameter tuning procedure for each new geometry, with the goal of adapting the boundary conditions to maintain similar flow conditions across the dataset, corresponding to the physiological regime (see, e.g., \citet{Mills1970PressureFlowRelationships}).
   In what follows, we consider the domain boundary to be decomposed as
   \begin{equation}
   \partial \Omega = \Gamma_0 \cup \Gamma_{\rm wall} \cup \left( \bigcup_{i=1}^4 \Gamma_i \right) \,,
   \end{equation}
   i.e., inlet, vessel wall, and (four) outlet boundaries, respectively (see Figure~\ref{fig:bcs_Gamma_i_definition_and_inlet_scale}, left).
   The definition of boundary conditions is then based on the following assumptions:
    \begin{enumerate}
        \item We assume the heart rate and inflow profile's temporal scaling to be equal for all considered cases, defining the spatial scaling according to
        the inlet geometry.
        \item We neglect wall deformation,
        prescribing no-slip conditions at the vessel wall
        ($\Gamma_\mathrm{wall}$) for all cases.
        \item The flow split is dictated by the outlet surface area, similar
        to~\cite{Grinberg2008WindkesselTuning, Romor2025Assimilation}.
    \end{enumerate}
    Physiological flow conditions are enforced by controlling target
    flow rates $Q_i$ and mean outlet pressures $\overline{p}_i$ on $\Gamma_i$
    defined as
    \begin{align*}
        Q_i \coloneqq \int_{\Gamma_i} \ve{u} \cdot \ve{n}~\mathrm{ds}\ve{x}
        ,
        \quad
        \overline{p}_i \coloneqq |\Gamma_i|^{-1} \int_{\Gamma_i}
        p~\mathrm{ds}\ve{x}
        ,
        \quad
        \mathrm{where~}
        |\Gamma_i| \coloneqq \int_{\Gamma_i}
        1~\mathrm{ds}\ve{x}
        \quad
        \mathrm{for~}
        i=1,\dots,4\,.
    \end{align*}

    \paragraph{Inflow boundary condition}

    At the inlet surface $\Gamma_0$, i.e., the cross section at the aortic root
    (see Figure~\ref{fig:bcs_Gamma_i_definition_and_inlet_scale}), we prescribe a
    Dirichlet condition of the form 
    \begin{equation}\label{eq:u_in}
        \ve{u}(t, \ve{x})
        =
        -\ve{n} \,\,
        \eta_x(\ve{x}) \,\, \eta_t(t) \,\, \eta_Q
        \quad \text{on }
        \Gamma_0 \times (0,T]\,,
     \end{equation}
     where $\ve{n}$ stands for the unit outward normal vector at the inlet
     surface, $\eta_x(\ve{x})$ is a quasi-parabolic spatial profile, constructed
     by adopting the technique proposed in \cite{Takizawa2010SpaceTime}, and
     $\eta_t(t)$ is a time-dependent profile obtained from clinical measurements
     from a specific patient (see
     Figure~\ref{fig:bcs_Gamma_i_definition_and_inlet_scale}, right).
    Finally, the scaling factor $\eta_Q$ is introduced to account for volumetric
    flow variation across geometries. \citet{Westerhof1991} found stroke volume,
    \begin{align*}
      SV \coloneqq \int_{t=0}^{T_p} Q_\mathrm{in}(t)~\mathrm{d}t
      ,
      \qquad
      \text{with cardiac period $T_p$}\,,
    \end{align*}
    per unit body mass to be independent of organism size.
    Assuming geometric similarity, organism mass is proportional to a volume
    scale, $r_0^3$, which implies $SV \propto r_0^3$.
    Consequently, preserving $Q_0 / SV$, i.e., the heart
    rate $1/T_p$, requires $Q_0\propto r_0^3$,
    i.e., a constant $Q_0/r_0^3$,
    in accordance with Murray's cubic law \citep{Murray1926}.

\begin{figure}[!ht]
        \centering
        \includegraphics[width=0.48\linewidth, draft=\draftHeavy%
        ]{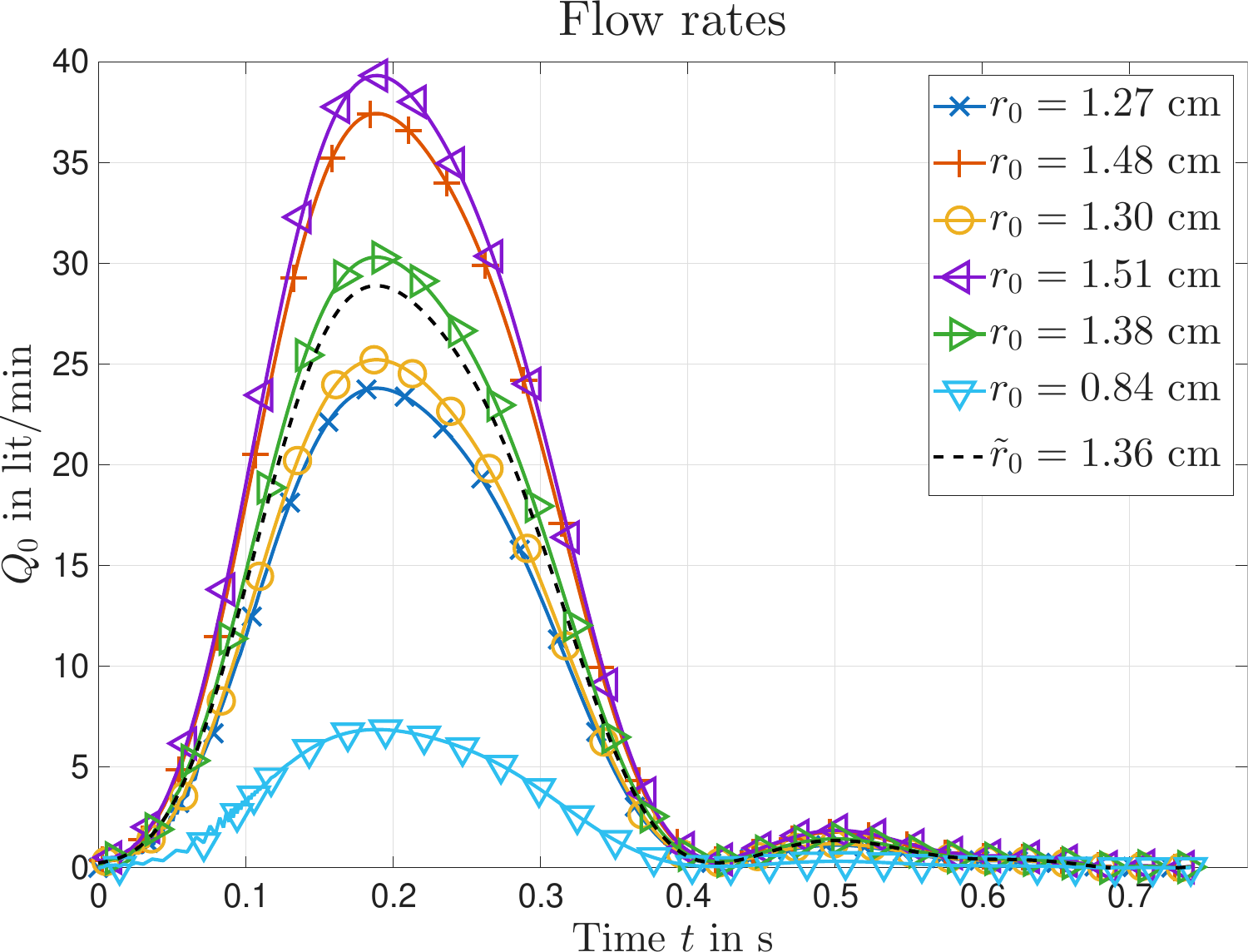}
        \hspace{0.02\linewidth}
        \includegraphics[width=0.48\linewidth, draft=\draftHeavy
        ]{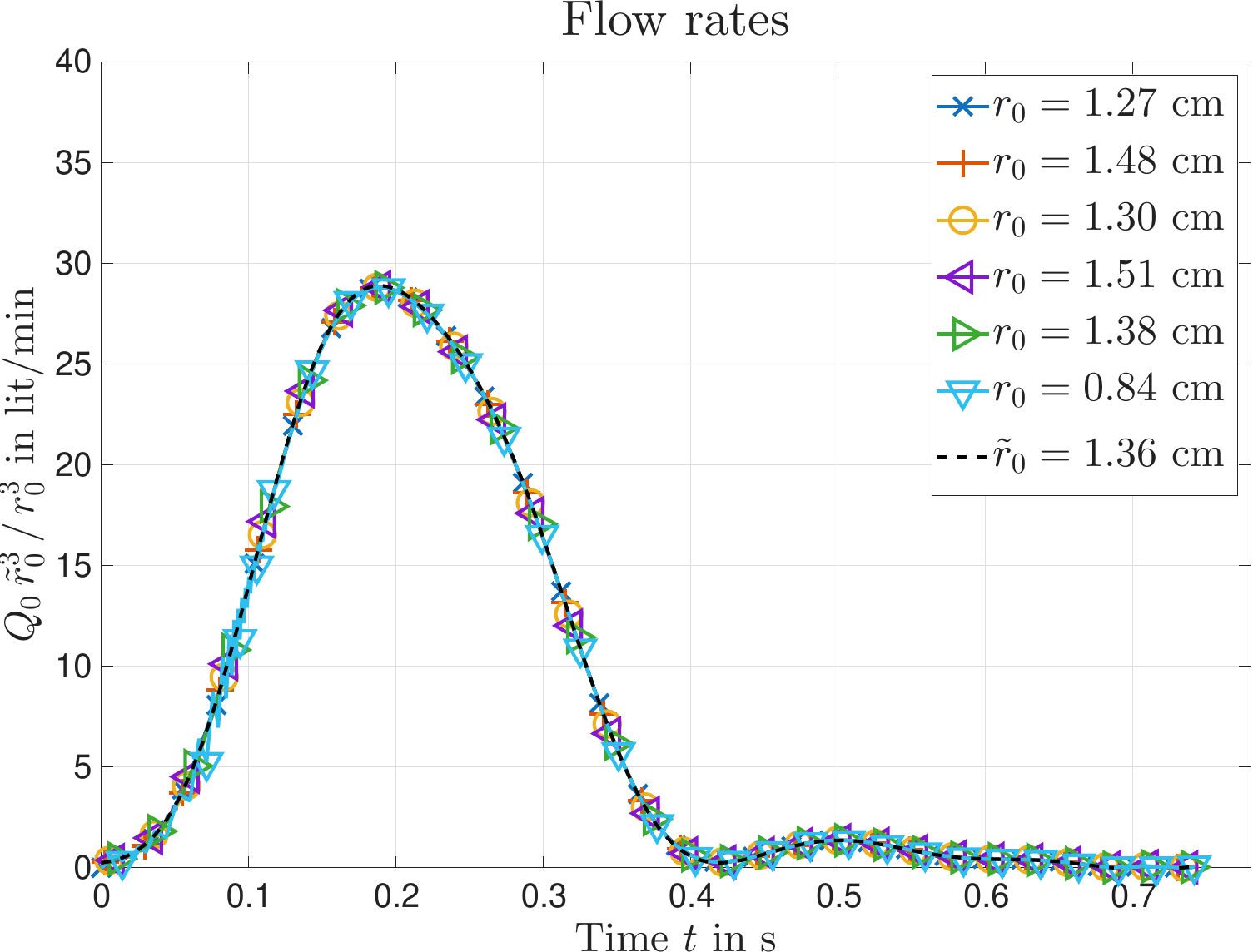}
        \caption{Left: Inlet flow rates $Q_0$ for a selection of cases of the 
        SynthAorta dataset with scaled inflow profiles according to Equation~\eqref{eq:u_in} 
        to preserve $Q_0/r_0^3$ according to Murray's law~\cite{Murray1926}.
        Right: Scaled flow rate recovering the reference inlet flow
        rate~\cite{Baumler2020FSIinAD}, illustrating the scaling.}
        \label{fig:bcs_inlet_velocity_scaling}
    \end{figure}

    Denoting quantities defined on the reference case analyzed in \cite{Baumler2020FSIinAD} with $\widetilde{(\cdot)}$,
    we compute the inlet mean radius of the reference case from $\pi \widetilde{r}_0^2 = \widetilde{A}_0$
    given the inlet area $\widetilde{A}_0$.
    That is, we scale the inlet velocity by a constant factor $\eta_Q$ per geometry according to
    \begin{equation}
        \label{eqn:bcs_inflow_scaling}
        \frac{\widetilde{Q}_0}{\widetilde{r}^{\,3}_0} = \frac{Q_0}{r^3_0}
        \quad\Leftrightarrow\quad
        \frac{\pi \widetilde{r}^{\,2}_0 \, \widetilde{\ve{u}}_\mathrm{mean}}{\widetilde{r}^{\,3}_0}
        = \frac{\pi {r}^2_0 \,\ve{u}_\mathrm{mean}}{{r}^3_0}
        \quad\Leftrightarrow\quad
        \widetilde{\ve{u}}_\mathrm{mean} \, \frac{r_0}{\widetilde{r}_0} = \ve{u}_\mathrm{mean}
        \quad\Rightarrow\quad
        \eta_Q = \frac{r_0}{\widetilde{r}_0}\,.
    \end{equation}
    Using this scaling, we obtain the inflow rates shown in
    Figure~\ref{fig:bcs_inlet_velocity_scaling} (left).
    For demonstration purposes, Figure~\ref{fig:bcs_inlet_velocity_scaling} (right)
    also contains a consistency check: inserting the numerically obtained flow
    rate into
    $\widetilde{Q}_0 = \nicefrac{\widetilde{r}^{\,3}_0}{r^3_0} \, Q$ yields the reference flow rate
    from~\citet{Baumler2020FSIinAD} itself.

    \paragraph{Windkessel parameter tuning}
    At the outlet, boundary conditions are imposed through lumped-parameter models, i.e., relating the mean outlet pressure to the outgoing flow. A popular choice consists of a 3-element Windkessel model on each outlet,
    in which the compliant pressure $p_{c,i}$ on the outlet $\Gamma_i$, to be
    enforced as the mean pressure $\overline{p}_i$ on that outlet, is computed as
    the solution of an ordinary differential equation of the form
    \begin{align}
        C_{i} \, \frac{\partial}{\partial t} p_{p,i}  + \frac{p_{p,i}-p_{d,i}}{R_{d,i}} &= Q_i
        \,,
        \label{eqn:bcs_wk_1}
        \\
        p_{c,i} - p_{p,i} &= R_{p,i} \, Q_i
        \,,
        \label{eqn:bcs_wk_2}
    \end{align}
    depending on the proximal and distal resistances, $R_{p,i}$ and $R_{d,i}$,
    accounting for the flow resistance of proximal and distal vasculature, respectively, on a capacitance $C_{i}$, accounting for vessel elasticity, and on the distal (also venous or basal) pressure $p_{d,i}$.
    In Equations~\eqref{eqn:bcs_wk_1}--\eqref{eqn:bcs_wk_2}, the \textit{proximal} pressure $p_{p,i}$ is an
    intermediate unknown, modeling the pressure at the compliant node between the proximal and distal resistances.

    In this work, we consider the scheme by~\citet{Grinberg2008WindkesselTuning}, neglecting the proximal resistances $R_{p,i}$ and yielding the 2-element (RC) model
    \begin{equation*}
      R_{d,i} C_i \, \frac{\partial}{\partial t} p_{c,i} + p_{c,i} =
      R_{d,i} \, Q_i + p_{d,i}
      \,.
    \end{equation*}

\begin{figure}[!ht]
        \centering
        \includegraphics[width=0.48\linewidth, draft=\draftHeavy
        ]{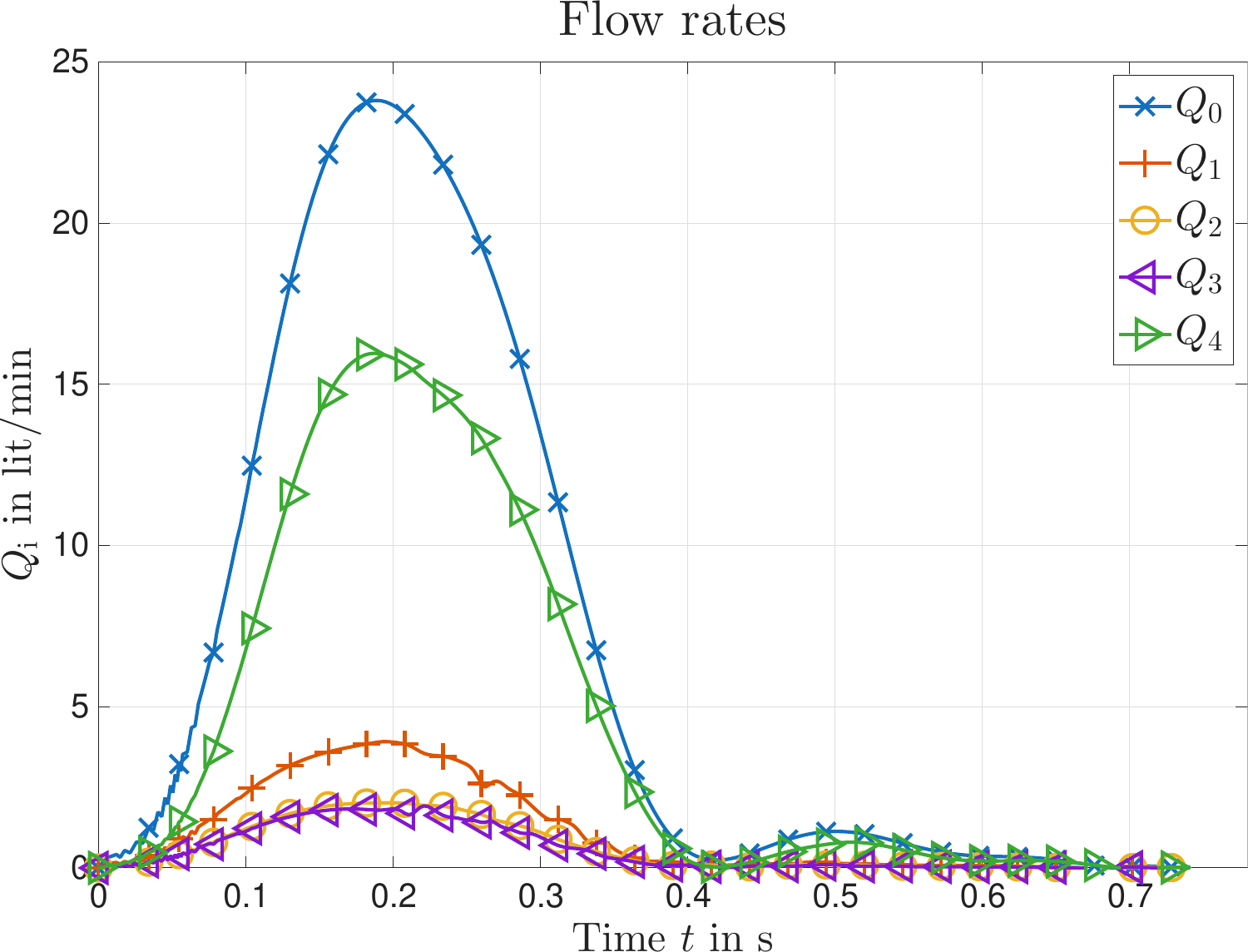}
        \hspace{0.02\linewidth}
        \includegraphics[width=0.48\linewidth, draft=\draftHeavy
        ]{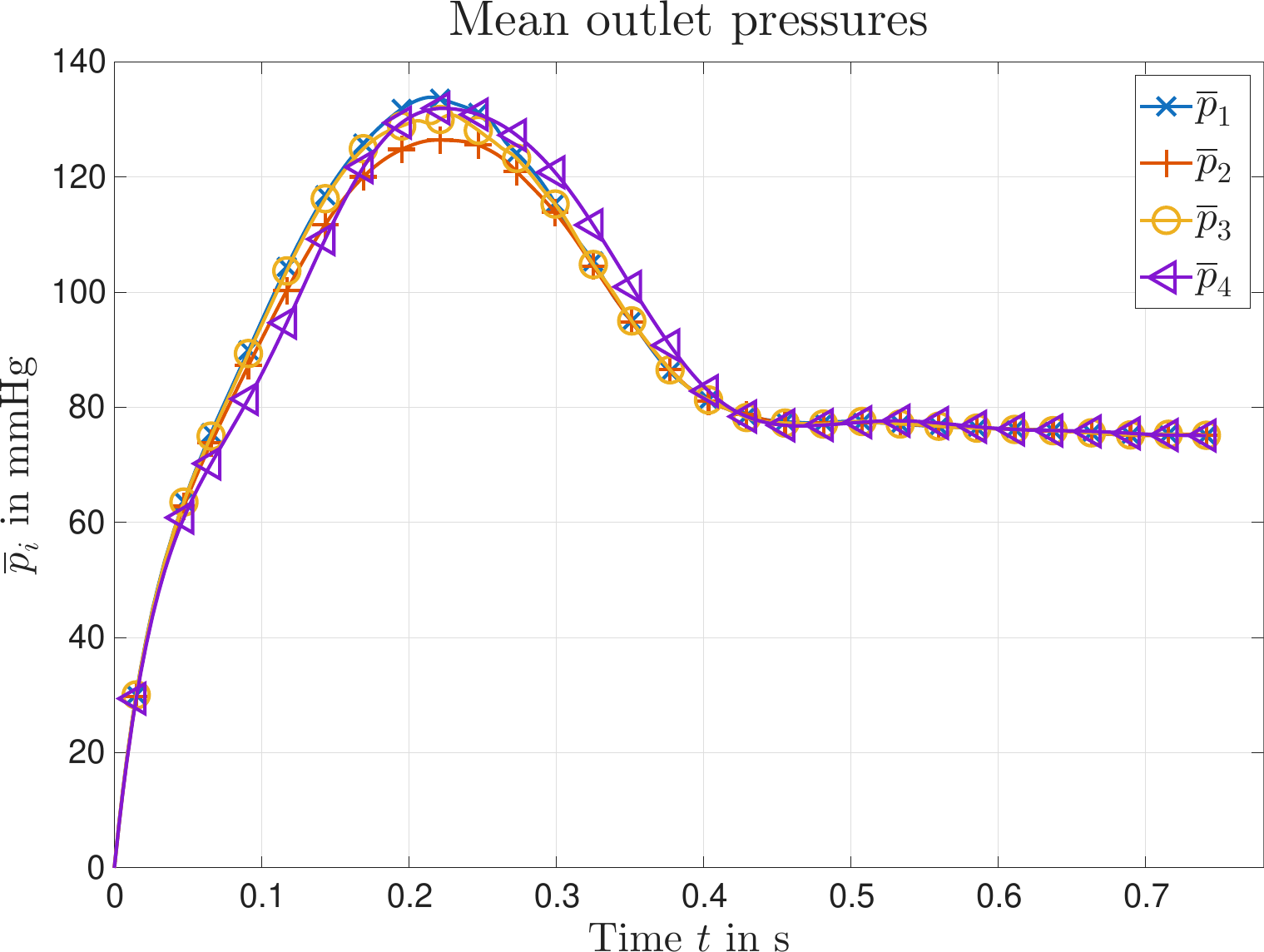}
        \caption{Flow rates $Q_i$, $i=0,\dots,4$ (left), and mean pressures $\overline{p}_i$, $i=1,\dots,4$ (right) at the different outlets for the reference case 1.}
        \label{fig:bcs_synthaorta_case_1_Qi_pi}
    \end{figure}

    The parameters of the Windkessel model are tuned in order to prescribe a systemic vascular pressure and a target flow split among the four outlets, according to the relation
    $\nicefrac{Q_i}{Q_j} \approx \nicefrac{R_{d,j}}{R_{d,i}}$ ($i,j = 1,\dots, 4$), which relies on a simplified zero-dimensional representation of the aortic domain~\cite{Grinberg2008WindkesselTuning}.
    We consider the flow split used in \citet{Katz2023ImpactTurbulence}, i.e., 16.81\% brachiocephalic trunk (BCT), 8.6\% left common carotid artery (LCC), 8.21\% left
    subclavian artery (LSA), and 66.38\% thoracic aorta (TA).
    In particular, the resistance parameters are tuned for the first patient in the SynthAorta dataset (case 1), then scaled to other geometries in a second step.
    The distal pressures are set to $p_{d,i} = 75~\text{mmHg}$. For case 1, we also fix the capacitance $C_4 = 12 \times 10^{-10}~\text{m}^3/\text{Pa}$ at the TA outlet.
    This procedure leads to the value $\hat{R}_{d,4} = 2.5 \times 10^7~\text{Pa\,s}/\text{m}^3$ for the TA outlet, achieving
    physiological target pressure levels under resting conditions of $\approx130~\text{mmHg}$ in systole and $\approx80~\text{mmHg}$ in diastole, see, e.g.,
    \citet{Mills1970PressureFlowRelationships}.  The resulting flow rates and pressures for case 1 are shown in Figure~\ref{fig:bcs_synthaorta_case_1_Qi_pi}.

    Since the inlet velocity is scaled across the database according to Equation~\eqref{eqn:bcs_inflow_scaling}, the proximal and distal resistances and the capacitance at each outlet
    are scaled accordingly to achieve similar pressure levels for the other cases.
    Moreover, to preserve the convergence properties,
    the capacitances $C_i$, $i=1,2,3,4$, are scaled to keep the product $R_{d,i}C_i$ constant on each outlet.

    Based on the inflow scaling in Equation~\eqref{eqn:bcs_inflow_scaling} and assuming that the flow split remains
    unchanged across the geometries,
    we can deduce scaling parameters by matching coefficients in
    \begin{align*}
        \widetilde{R}_{d,i} \widetilde{C}_i \, \frac{\partial}{\partial t} \widetilde{p}_{c,i} + \widetilde{p}_{c,i}
        &=
        \widetilde{R}_{d,i} \, \widetilde{Q}_i + \widetilde{p}_{d,i}
        \qquad
        \text{for the reference case 1}
        \,,
        \\
        R_{d,i} C_i \, \frac{\partial}{\partial t} p_{c,i}  + p_{c,i}
        &=
        R_{d,i} \, Q_i + p_{d,i}
        ,
        \qquad
        \text{for a particular case}
        \,,
    \end{align*}
    to obtain $p_{c,i} \equiv \widetilde{p}_{c,i}$,
    which for $Q= \nicefrac{r^3_0}{\widetilde{r}_0^{\,3}}\,\widetilde{Q}$ from Equation~\eqref{eqn:bcs_inflow_scaling} gives
    \begin{equation}
        R_{d,i} = \frac{\widetilde{r}_0^{\,3}}{r^3_0} \widetilde{R}_{d,i}
        ,
        \quad
        p_{d,i} = \widetilde{p}_{d,i}
        ,
        \quad
        C_i = \frac{r^3_0}{\widetilde{r}_0^{\,3}} \widetilde{C}_{i}
        \,.
        \label{eqn:Rd_C_scaling_under_geometry_variation}
    \end{equation}

    Note that the resistance scaling is consistent with \citet{Westerhof1991}: since volume and
    mass scale similarly with organism size (assuming some constant
    average density of the organism), the chosen volume ratio in
    Equation~\eqref{eqn:Rd_C_scaling_under_geometry_variation} leads to a linear relationship
    between the resistance and organism size.

\begin{figure}[!ht]
        \centering
        \includegraphics[width=\linewidth, draft=\draftHeavy
        ]{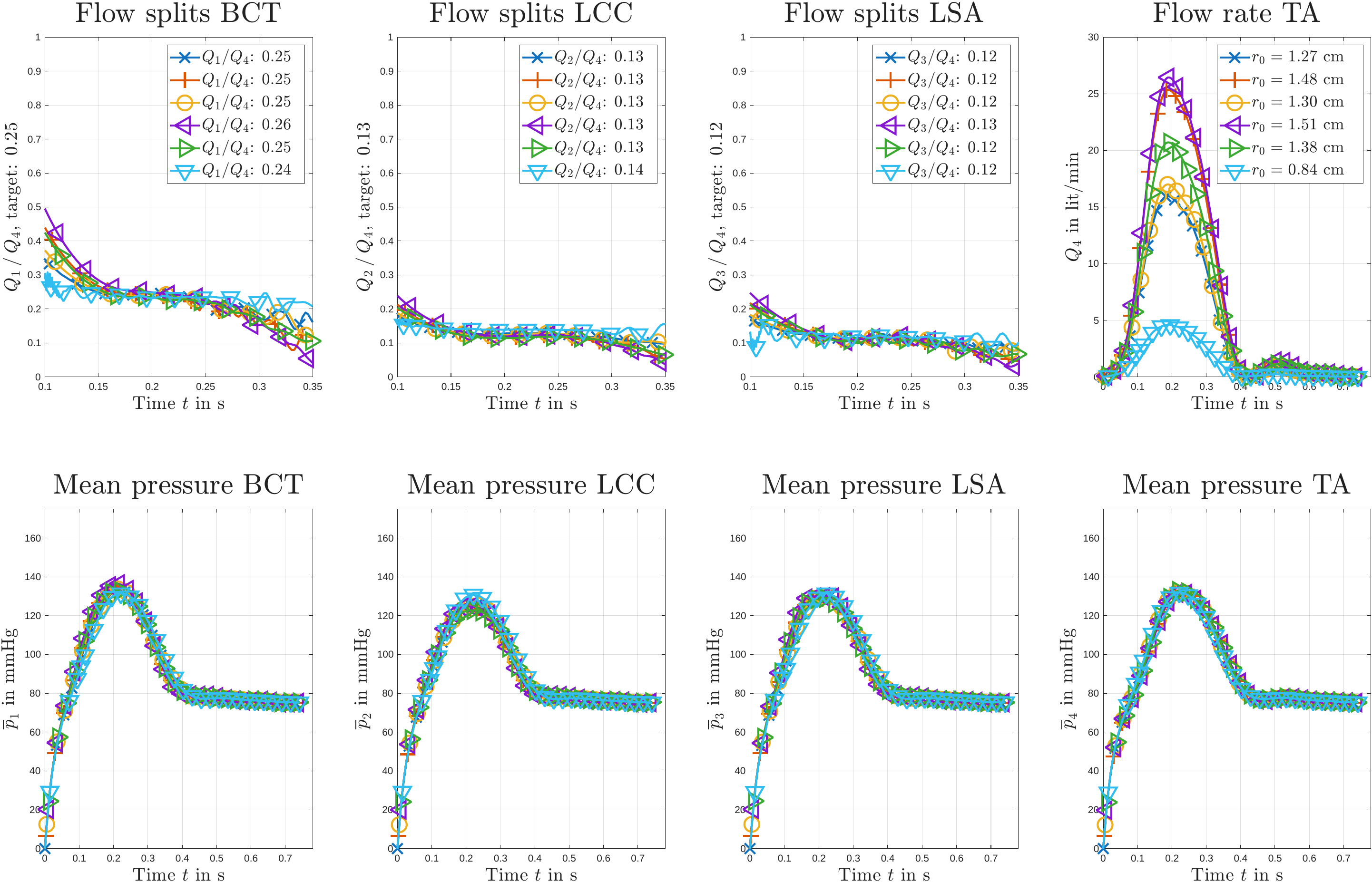}
        \caption{Top row: From left to right, flow split $\nicefrac{Q_i}{Q_4}$, 
              $i=1,2,3$ compared to the target values from 
              \citet{Katz2023ImpactTurbulence}, and
              flow rate $Q_4$ at the TA\@. Bottom row: Mean pressures at outlets
              $\overline{p}_i$, $i=1,\dots,4$ for BCT, LCC, LSA, and TA.}
        \label{fig:bcs_pressure_levels_flow_splits}
    \end{figure}

    The resulting flow split and mean outlet pressure curves are given in 
    Figure~\ref{fig:bcs_pressure_levels_flow_splits}
    for six selected cases from the SynthAorta dataset, showing that the chosen 
    scaling achieves the desired flow distribution among outlets and
    that the pressure level is kept roughly constant, despite geometric variations.
    Setting a constant distal pressure $p_{d,i} = 75~\text{mmHg}$
    for all cases is vital for fast convergence to the periodic state and for
    achieving the desired pressure levels, as discussed in Section~\ref{sec:grid_selection}.

    Note, however, that the capacitance is scaled with the inverse factor of
    $R_{d,i}$, which can trigger numerical instabilities
    when the distal resistance is increased and the capacitance is decreased
    according to the proposed tuning procedure. This might lead to some
    oscillations for the cases with the smallest radii.

    \subsection{Computational fluid dynamics}
    \label{sec:cfd_solver_and_results}

    The Navier--Stokes equations for incompressible flow of generalized Newtonian
    fluids, see Equations~\eqref{eqn:hemodynamics_momentum_balance}--\eqref{eqn:hemodynamics_continuity},
    are solved using an $L^2$-conforming discontinuous Galerkin approach based
    on a time-splitting scheme~\cite{Schussnig2025DGvariableViscosity}, in which
    the solution is split into a convective step with a fully explicit
    extrapolation of the nonlinear term, a pressure Poisson step, and a
    step projecting the velocity onto a divergence-free space.
    The formulation enforces weak mass conservation and energy stability by
    additional stabilization terms, which are critical for robust and accurate
    discretizations of turbulent flows on non-resolved scales.
    This strategy avoids solving linearized or nonlinear coupled velocity--pressure
    systems~\cite{Schussnig2021Robust}, typically leading to higher
    throughput in vascular flow. We formally achieve
    accuracy of order 2 in time, which is lower
    than that of consistent splitting schemes~\cite{Pacheco2021Efficient, Still2025SplittingDG},
    but sufficient given the spatial discretization accuracy within the
    SynthAorta dataset and practical performance considerations
    rooted in the problem parameters.

    The solver is implemented in ExaDG~\cite{Arndt2020ExaDG}, using the
    matrix-free framework provided by deal.II~\cite{Arndt2025dealii9c7}.
    The preconditioning technique for the solution of the linear systems for the
    pressure Poisson equation and the momentum equation is based on a
    mixed-precision, hybrid multigrid method
    with matrix-free Chebyshev-accelerated Jacobi
    smoothers~\cite{Fehn2020hybridMultigrid, KronbichlerWall2018performanceCGvsDGmatrixFree}.
    Moreover, backflow stabilization via a velocity penalization
    method based on~\cite{Bruneau1994} is used on the Neumann
    boundaries to prevent instabilities for re-entrant flow (see
    also~\cite{Gravemeier2012BackflowStabilization, Schussnig2022PhD}).

    \subsubsection{Convergence studies and grid selection}
    \label{sec:grid_selection}

    The numerical approach has to be verified in terms of spatiotemporal accuracy,
    while ensuring that the periodic state is reached up to the desired tolerance.
    The grids employed for CFD analysis, which feature trilinear mappings provided
    in the SynthAorta dataset, are depicted in Figure~\ref{fig:SynthAorta_sample_cases}.
    We employ an adaptive backward differentiation formula of order 2
    for time stepping based on the
    Courant--Friedrichs--Lewy
    (CFL) number, such that smaller grids
    automatically translate to smaller time steps.
    Second-order discontinuous polynomials for the velocity approximation
    and discontinuous linear
    polynomials for the pressure approximation are used.
    A grid sensitivity study on the whole set of geometries is not practically feasible. Hence, to validate the solver, we perform convergence studies for selected cases ($i=1, 1000, 2000, 3000, 4000, 5000$)
    of the SynthAorta dataset, which are considered to be representative of the main features of the dataset.
    The geometric representation is trilinear, and the points inserted during refinement
    conform to the underlying geometry, see~\cite{Bosnjak2025SynthAorta}, such that the geometric approximation improves
    under grid refinement.

\begin{figure}[!ht]
        \centering
        \includegraphics[height=3.8cm, draft=\draftHeavy
        ]{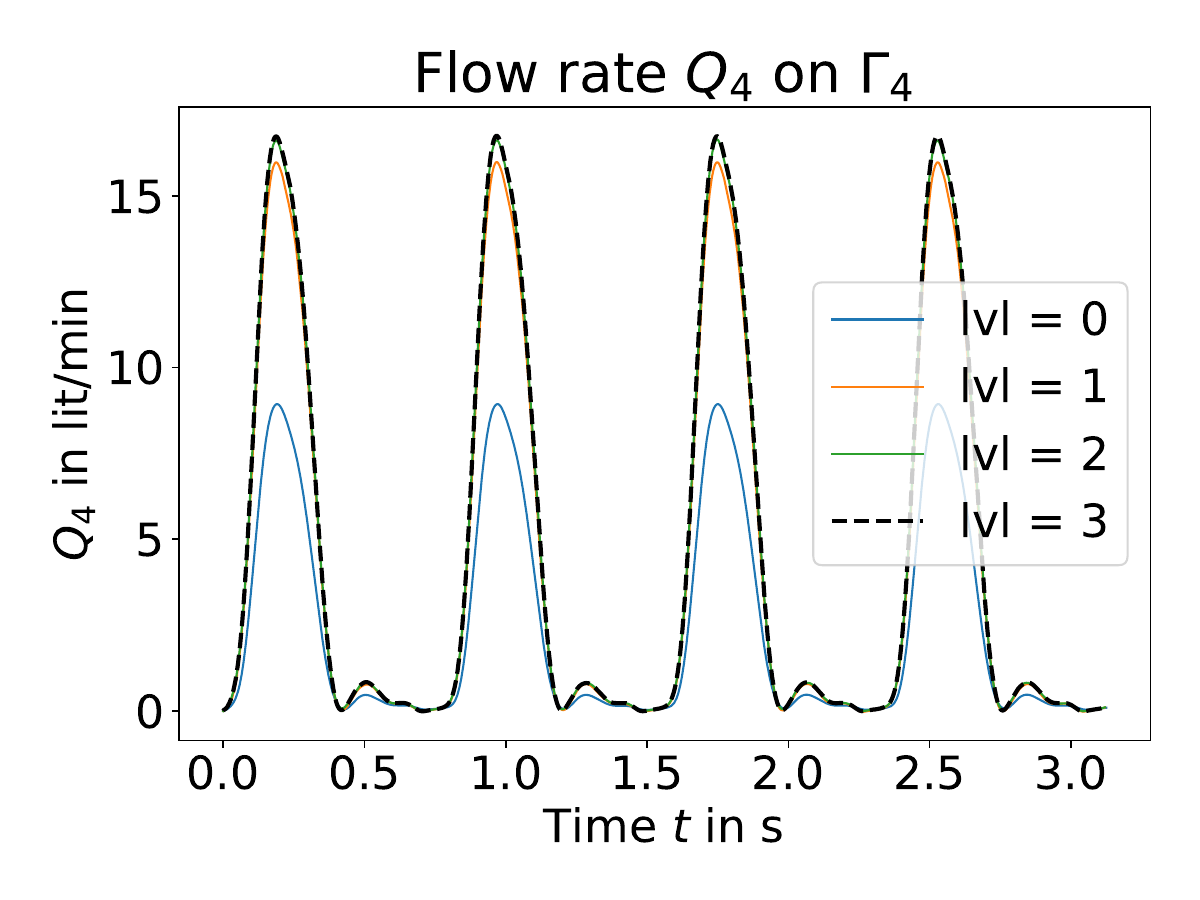}
        %
        %
        \includegraphics[height=3.8cm, draft=\draftHeavy
        ]{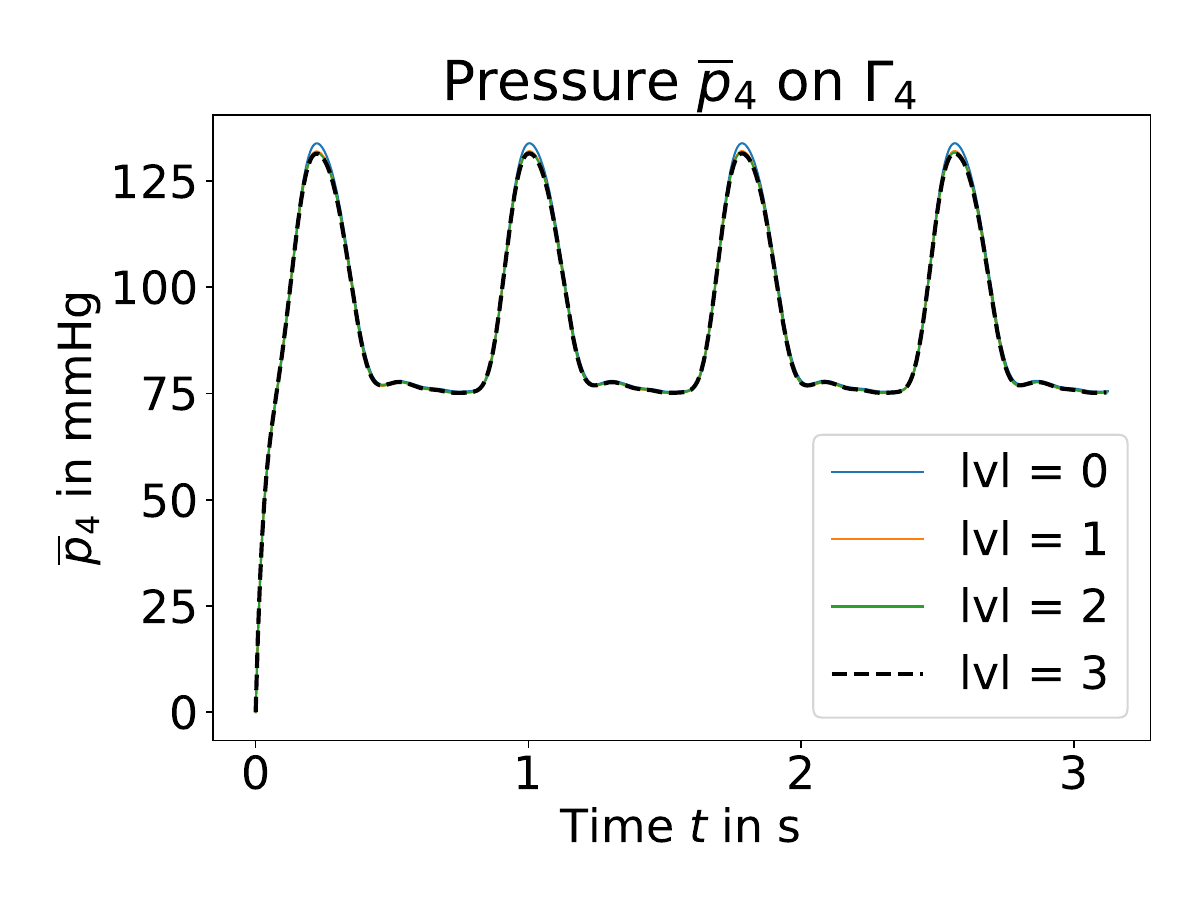}
        %
        %
        %
        \includegraphics[height=3.8cm, draft=\draftHeavy
        ]{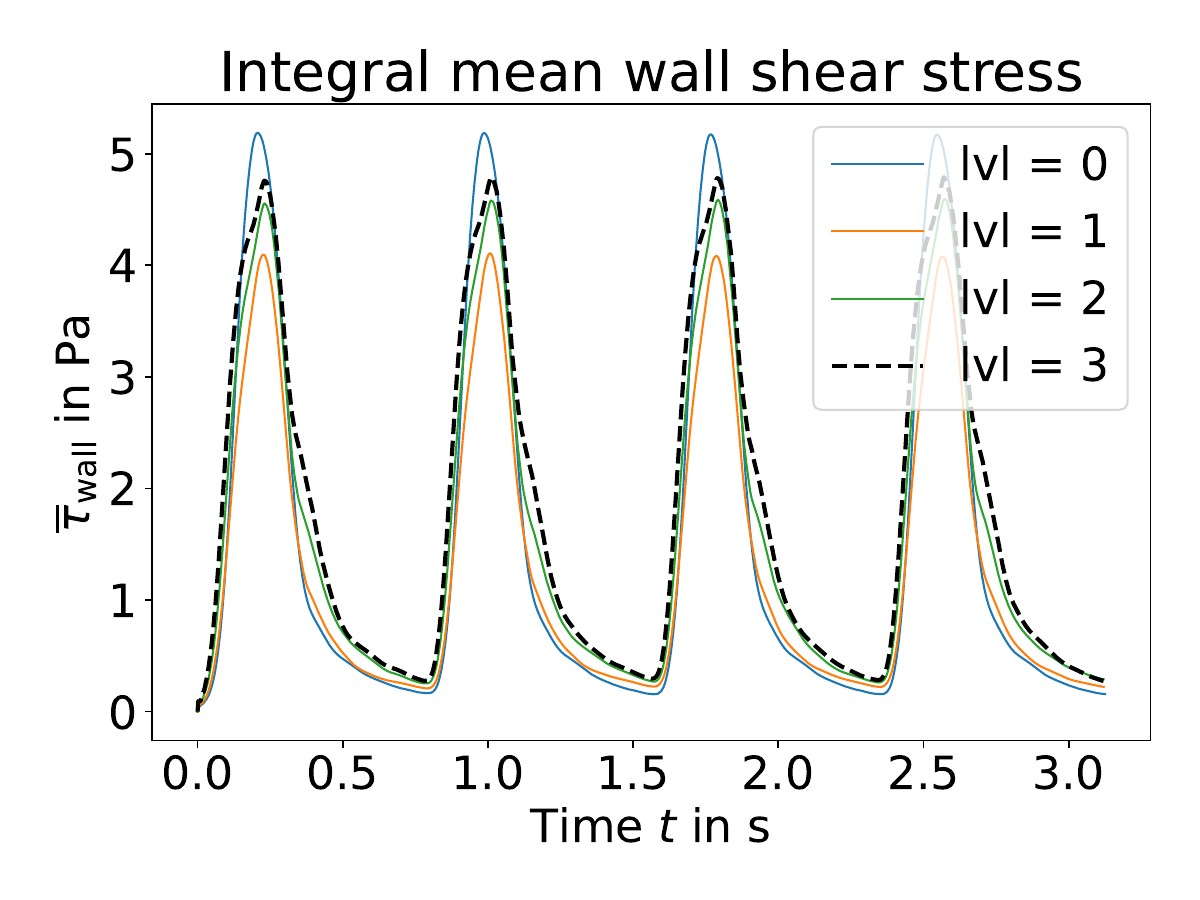}
        \caption{Flow rate $Q_4$ (left), outlet mean pressure $\overline{p}_4$ (middle), and
            integral mean WSS $\overline{\tau}_\mathrm{wall}$ (right) of case 1 of the
            SynthAorta dataset over time, considering four grid refinement levels and 
            an adaptive time-stepping technique (keeping the CFL number constant).
            The pressure approximation is accurate on all considered levels, while the 
            flow rates require level $l=1$ or higher. The WSS approximation requires 
            $l=2$ to achieve relative errors typically lying around $10\%$.
            }
        \label{fig:convergence_pi_Qi_tau}
    \end{figure}

    The observed quantities are the mean pressures and flow rates on the outlets,
    $\overline{p}_i$ and $Q_i$, $i=1,2,3,4$, respectively, and the averaged WSS,
    \[
    \overline{\tau}_\mathrm{wall} \coloneqq |\Gamma_\mathrm{wall}|^{-1}
        \int_{\Gamma_\mathrm{wall}} \|\ve{\tau}_\mathrm{wall}\|~\mathrm{ds}\ve{x}
        \,,
    \]
    where $\ve{\tau}_\mathrm{wall}$ is the WSS defined in Equation~\eqref{eqn:wall_shear_stress_vector}. For these quantities, we
    define the time-averaged error
    \begin{equation}
    \epsilon_{\phi_i}
        \coloneqq
        \frac{%
            \int_{t=0}^T
            \left| \phi_i(t) - \phi_i^\mathrm{ref}(t)\right|
            \mathrm{d}t
        }{%
            \int_{t=0}^T
            \left| \phi_i^\mathrm{ref}(t)\right|
            \mathrm{d}t
        }, \; i=1,\hdots,4\,,
    \end{equation}
    for $\phi=Q, \overline{p}, \overline{\tau}_\mathrm{wall}$, with respect to a reference solution computed on a fine grid containing $\approx 10.21\times10^6$ degrees of freedom (DoFs)
    and approximately $8\times10^4$ time steps (uniform refinement level $l=3$).
    We show the results of the convergence study for case $1$, which yielded the largest error in the
    WSS, the most critical quantity.
    In particular, relative errors under uniform grid refinement are reported in Table~\ref{tab:convergence_study}.
    The mean outlet pressure shows relatively low errors already for grid level $l=0$,
    while the flow rates are more prone to approximation errors. The relative difference in
    the integral mean WSS is the most sensitive QoI,
    with errors in the range of 10\% even for a grid uniformly refined twice, and errors up to 60\% in most geometries on levels 0 and 1.
    Based on the velocity and pressure errors, we select level $l=1$, noting, however, that
    $l=2$ would be required for reasonably accurate WSS\@.
    The corresponding transient profiles of the monitored quantities are depicted in Figure~\ref{fig:convergence_pi_Qi_tau}.
    \begin{table}[ht!]
        \centering
        \caption{Relative errors for case 1 of the SynthAorta dataset using
          refinement levels $l=0,1,2$, and refinement level $l=3$ as reference
          solution.}
        \vspace{2mm}
        \label{tab:convergence_study}
        \begin{tabular}{||c|c|c|c|c||c|c|c|c||c||}
            \hline
            &&&&&&&&&\\[-1.6ex]
            $l$ &
            $\epsilon_{\overline{p}_1}$&$\epsilon_{\overline{p}_2}$&$\epsilon_{\overline{p}_3}$&$\epsilon_{\overline{p}_4}$ &
            $\epsilon_{Q_1}$&$\epsilon_{Q_2}$&$\epsilon_{Q_3}$&$\epsilon_{Q_4}$&$\epsilon_{\overline{\tau}_\mathrm{wall}}$
            \\[1.5ex]
            \hline\hline
            &&&&&&&&&\\[-1ex]
            0
            &0.0112
            &0.0056
            &0.0089
            &0.0082
            &0.4595
            &0.4823
            &0.4659
            &0.4682
            &0.2235
            \\
            1
            &0.0019
            &0.0022
            &0.0019
            &0.0017
            &0.0572
            &0.0657
            &0.0641
            &0.0479
            &0.2716
            \\
            2
            &0.0012
            &0.0014
            &0.0013
            &0.0007
            &0.0214
            &0.0266
            &0.0301
            &0.0086
            &0.1076
            \\[1.4ex]
            \hline
        \end{tabular}
    \end{table}

    \paragraph{Periodicity}
    In addition to the overall accuracy, we investigate the relative difference between the solutions over different cardiac cycles, i.e.,
    within time intervals $[0.1, 0.78] + 0.78 \, n$, $n=0,1,2,3$.
    Results are presented only for the reference case 1 of the dataset.
    Namely, to investigate the convergence to the periodic state, we consider, for each refinement level $l$, as the reference solution the one computed on the same level $l$ but for the fourth cardiac cycle.
    Table~\ref{tab:periodicity_error} shows negligible errors with respect to
    periodicity compared to the spatiotemporal errors shown in Table~\ref{tab:convergence_study}, already for the first cycle (for $t\geq0.1~\text{s}$).
    Overall, we conclude that the first cardiac cycle is acceptable for all grid levels, since the solutions have mostly
    reached the periodic state, which is attributable to the specific Windkessel parameters chosen.

    \begin{table}[ht!]
        \scriptsize
        \centering
        \caption{Relative errors in \% for case 1 of the SynthAorta dataset comparing the first two
            and the final fourth cardiac cycle. Compared with Table~\ref{tab:convergence_study}, we note that the spatiotemporal error
            dominates for $t\geq0.1~\text{s}$, while the first cardiac cycle can already be considered.
            The first cardiac cycle already yields solutions sufficiently close to a periodic state,
            considering engineering accuracy requirements and neglecting the first $0.1$~s.
            }
        \vspace{2mm}
        \label{tab:periodicity_error}
        \begin{tabular}{||c||c|c|c|c||c|c|c|c||c|c|c|c||}
            \hline
            &
            \multicolumn{4}{|c||}{\,}&
            \multicolumn{4}{|c||}{\,}&
            \multicolumn{4}{|c||}{\,}\\[-1.6ex]
            \multirow{2}{*}{\vspace*{-0mm}Cardiac} &
            \multicolumn{4}{|c||}{$\epsilon_{\overline{p}_4}$ in \%}&
            \multicolumn{4}{|c||}{$\epsilon_{Q_4}$ in \%}&
            \multicolumn{4}{|c||}{$\epsilon_{\overline{\tau}_\mathrm{wall}}$ in \%}
            \\[-1.5ex]
            \multirow{2}{*}{\vspace*{-3mm}cycle}
            &
            \multicolumn{4}{|c||}{\,}&
            \multicolumn{4}{|c||}{\,}&
            \multicolumn{4}{|c||}{\,}\\
            \cline{2-13}
            &&&&&&&&&&&&
            \\[-1.5ex]
            &$l=0$&$l=1$&$l=2$&$l=3$&$l=0$&$l=1$&$l=2$&$l=3$&$l=0$&$l=1$&$l=2$&$l=3$
            \\[.5ex]
            \hline\hline
            &&&&&&&&&&&&\\[-1ex]
            1
            & 0.153
            & 0.136
            & 0.135
            & 0.129
            & 0.231
            & 0.236
            & 0.151
            & 0.179
            & 0.720
            & 1.129
            & 1.381
            & 1.087
            \\
            2
            & 0.032
            & 0.026
            & 0.028
            & 0.018
            & 0.193
            & 0.266
            & 0.221
            & 0.182
            & 0.385
            & 0.703
            & 0.933
            & 0.961
            \\[1.4ex]
            \hline
        \end{tabular}
    \end{table}

    \subsubsection{Uncertainty of QoIs}
\begin{figure}[!ht]
        \centering
        \includegraphics[height=3.8cm, draft=\draftHeavy
        ]{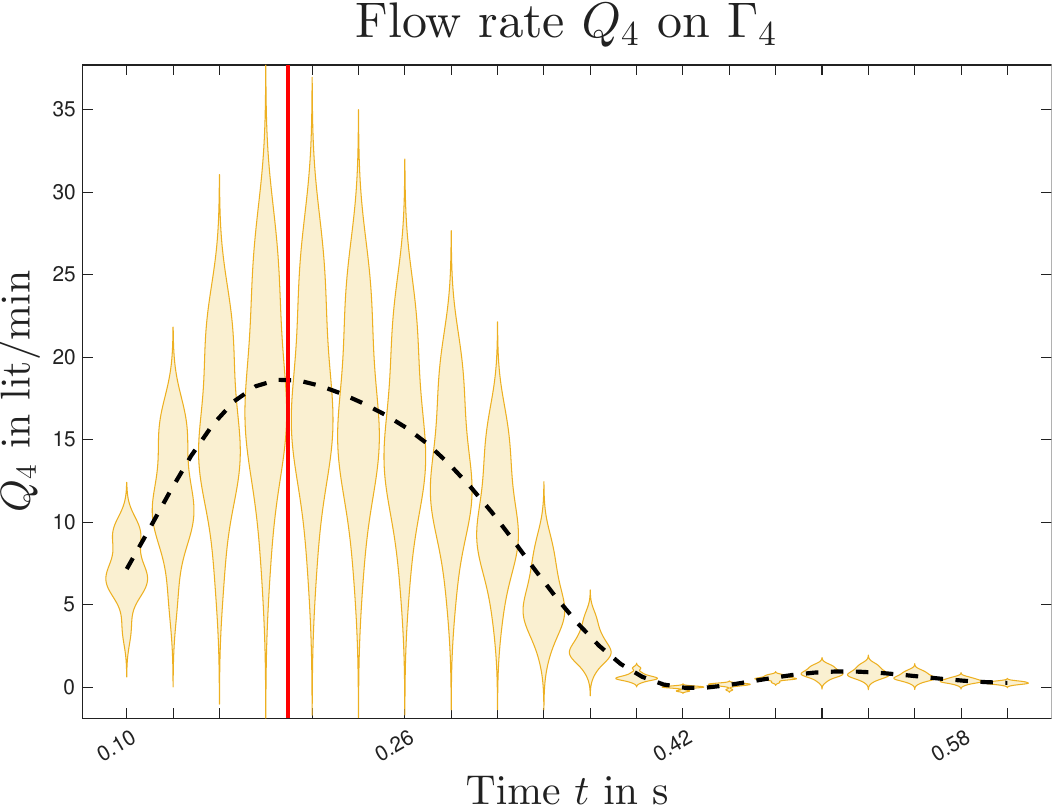}
        %
        %
        \includegraphics[height=3.8cm, draft=\draftHeavy
        ]{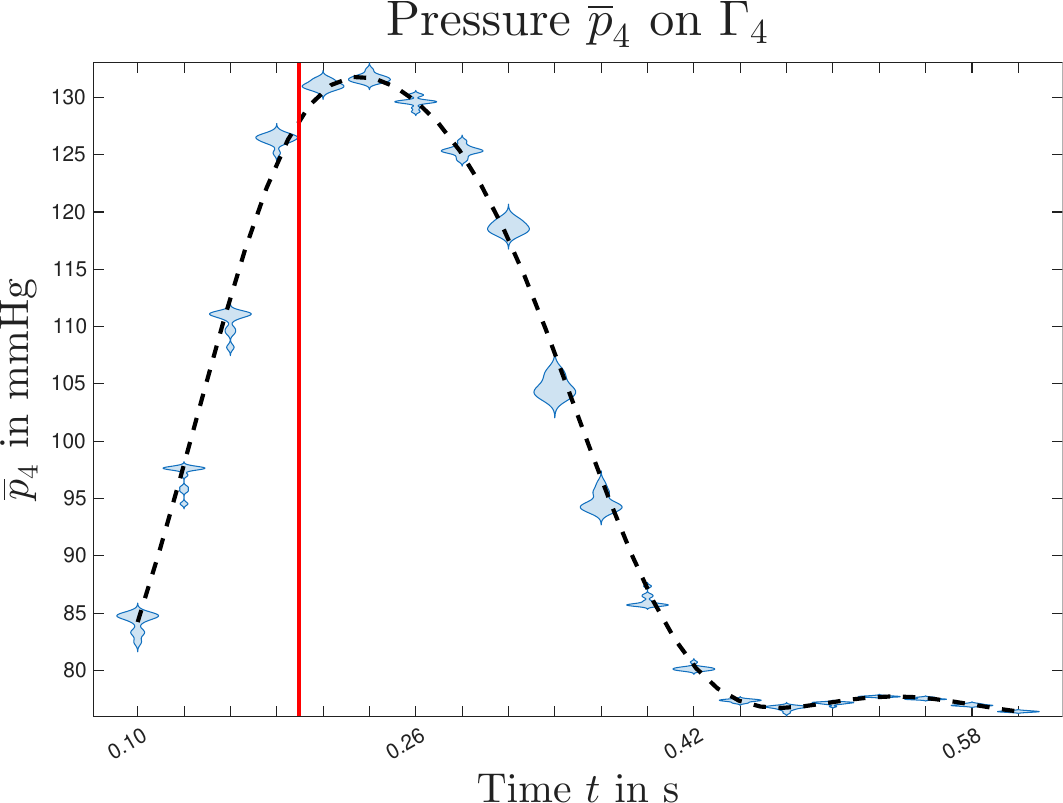}
        %
        %
        %
        \includegraphics[height=3.8cm, draft=\draftHeavy
        ]{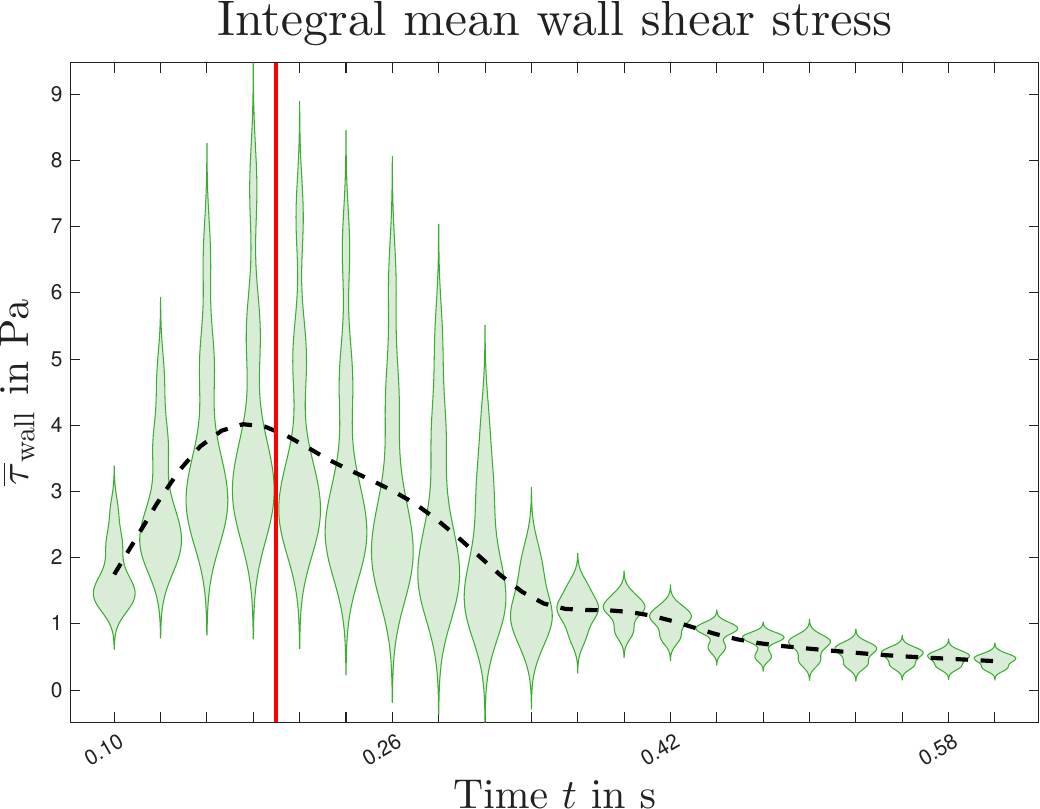}
        \caption{Violin plots of the temporal trace of the flow rate $Q_4$ (left),
            outlet mean pressure $\overline{p}_4$ (middle), and integral mean
            WSS $\overline{\tau}_\mathrm{wall}$ (right) for cases 1 to 1000 of the
            SynthAorta dataset.
            The red line shows the design point at $t=0.19~\text{s}$ for snapshot creation.
            }
        \label{fig:synthaorta_case_1_1000_uq}
    \end{figure}

    The inherent complexity of the vascular flow data generated from the SynthAorta 
    dataset can be illustrated through the resulting uncertainty in the considered QoIs.
    Figure~\ref{fig:synthaorta_case_1_1000_uq} depicts violin plots for
    cases 1 to 1000 over the cardiac cycle,
    demonstrating (i) the effectiveness of the algorithm used to tune the
    Windkessel parameters across geometries, leading to very little
    fluctuation from the target pressure levels despite vast geometric variation, and (ii) the
    impact of the geometric variation, explicitly through changes of the cross section and
    implicitly through the scaling of volumetric flow rate $Q_0$ via
    $Q_0 = \nicefrac{r^{\,3}_0}{\widetilde{r}^3_0} \, \widetilde{Q}$, see
    Equation~\eqref{eqn:bcs_inflow_scaling}.

    The variations in the WSS are directly
    related to the variations in the velocity: the violin plots of the flow rate show
    large variations in systole, while during the diastolic phase, variations are limited.
    The WSS shows similar variations over the cardiac cycle, but the widths of the violins
    are distinctly different. The flow rates show top-heavy, more uniform distributions,
    while the WSS features bottom-heavy distributions with thin but non-vanishing top regions.
    In diastole, the variation in the WSS seems more pronounced, although the mean absolute value
    is also smaller. In summary, peak systole shows the largest variations,
    while the correlation under geometric variation between
    velocities and flow rates, on the one hand, and WSS, on the other hand, is non-trivial.

    Since the observed absolute values and their variations
    are largest at peak systole, we choose the point of maximum inflow at $t=0.19~\text{s}$
    as the design point for our surrogates.

    \subsection{Dataset of full-order solutions}\label{sec:filtering}
    Maximizing the number and/or the fidelity of the CFD solutions considered
    for training and testing the ML models, while keeping the
    computational resources required to generate them tractable,
    requires handling the increasing amount of error within the data.
    In the considered cases, we fix the spatiotemporal resolution, as described in
    Section~\ref{sec:grid_selection}, and we assume that the solution reaches a
    periodic state from $0.1~\text{s}$ on. In this setting, the solver tolerances
    and related numerical settings
    are chosen to obtain results as fast as possible while preserving overall
    acceptable
    accuracy and stability. This drive for efficiency leads to decreased robustness
    of the linear solvers. Hence, given a certain computational budget,
    a trade-off between accuracy/stability and number of samples has to be found.

\begin{figure}[!ht]
        \centering
        \includegraphics[width=0.24\linewidth, draft=\draftHeavy
        ]{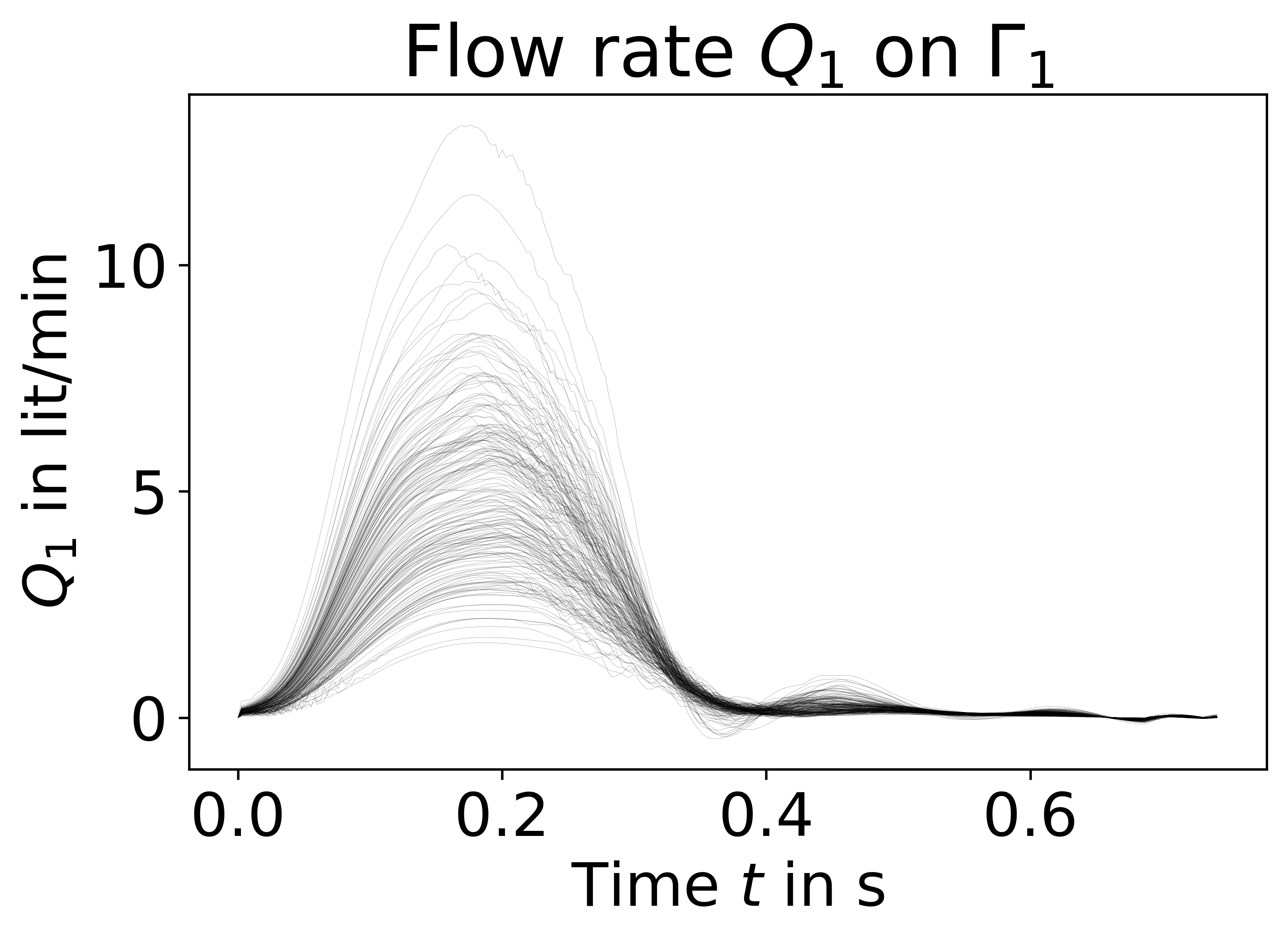}
        \hspace{0.0\linewidth}
        \includegraphics[width=0.24\linewidth, draft=\draftHeavy
        ]{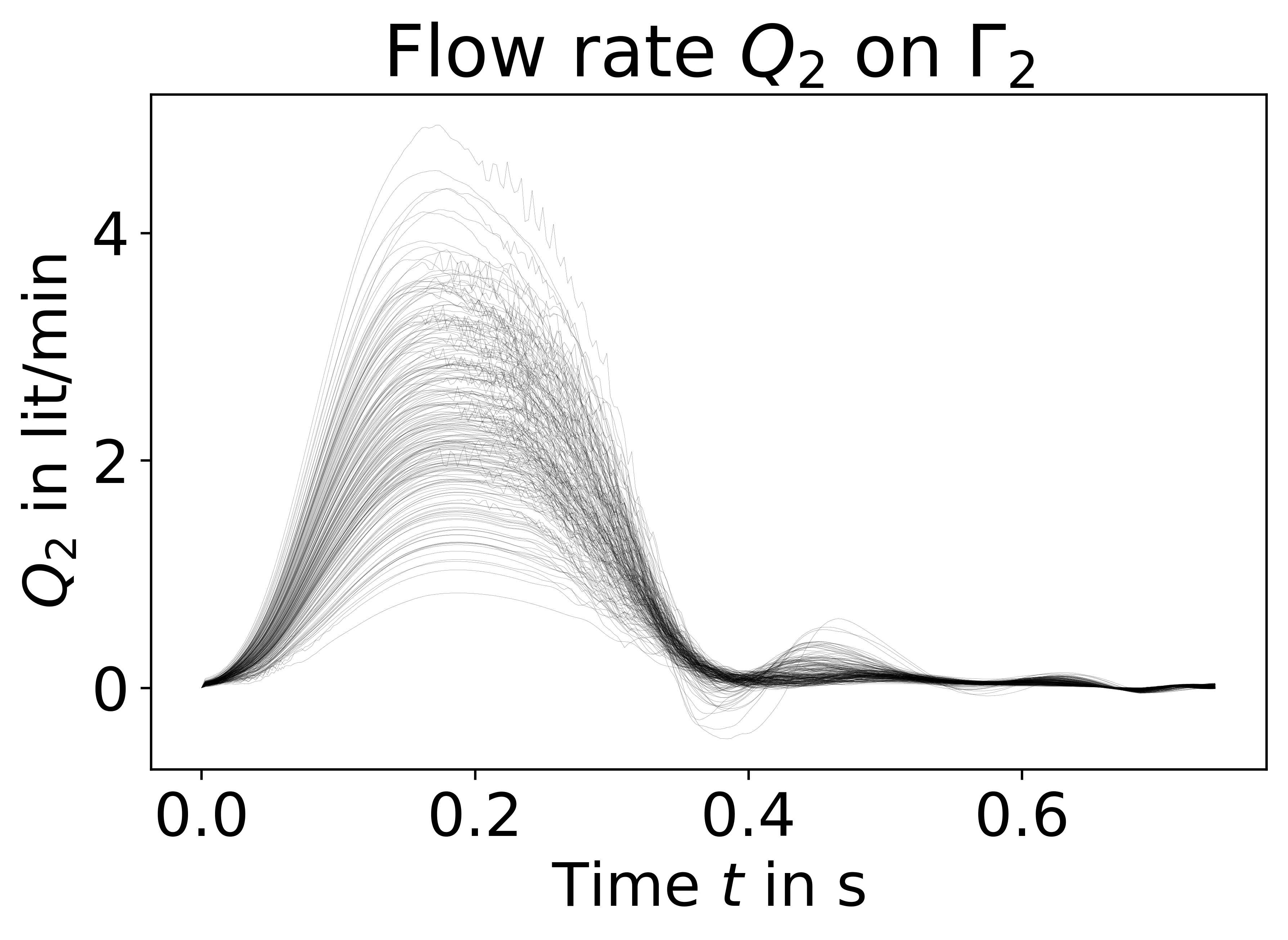}
        \hspace{0.0\linewidth}
        \includegraphics[width=0.24\linewidth, draft=\draftHeavy
        ]{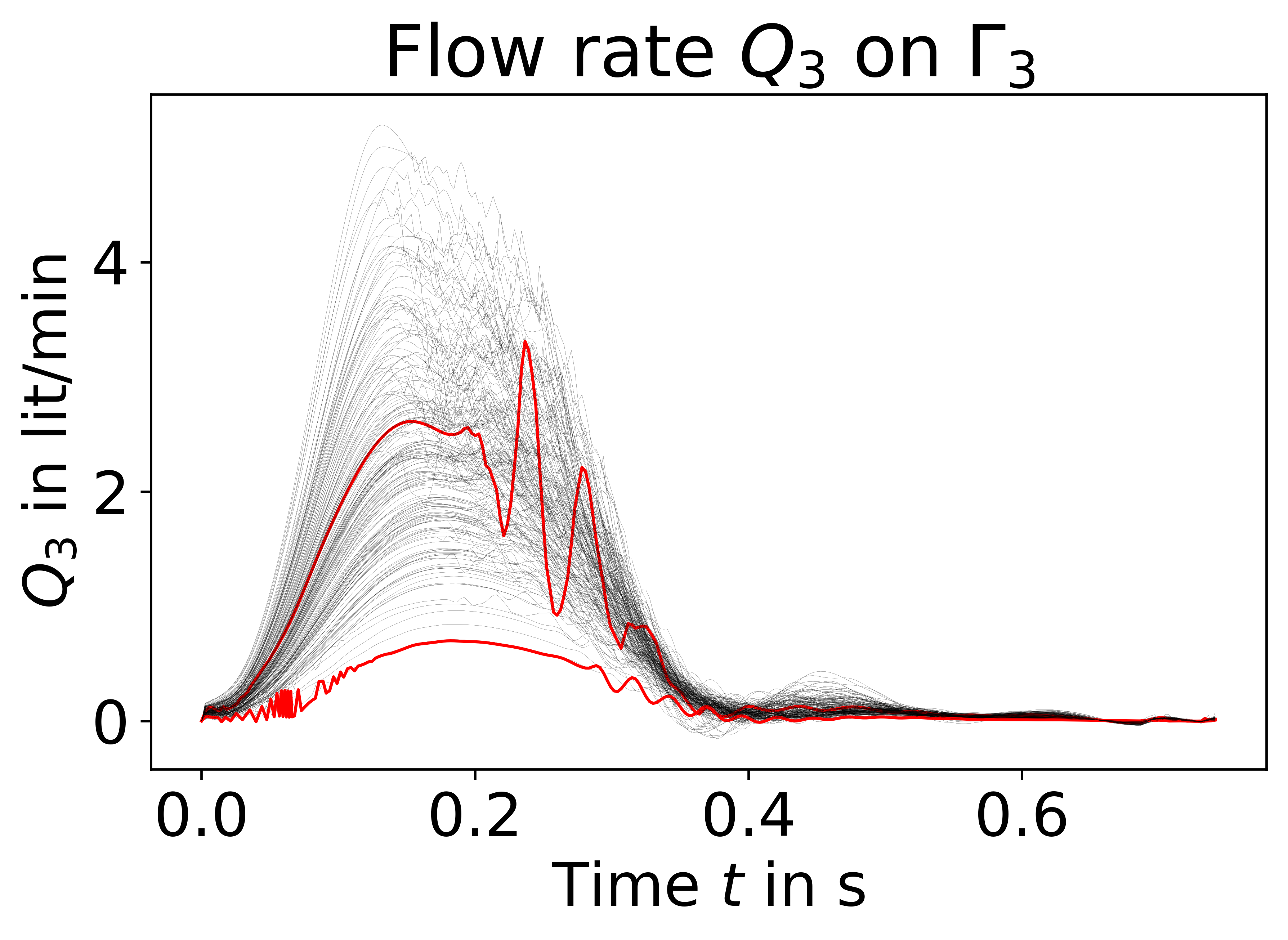}
        \hspace{0.0\linewidth}
        \includegraphics[width=0.24\linewidth, draft=\draftHeavy
        ]{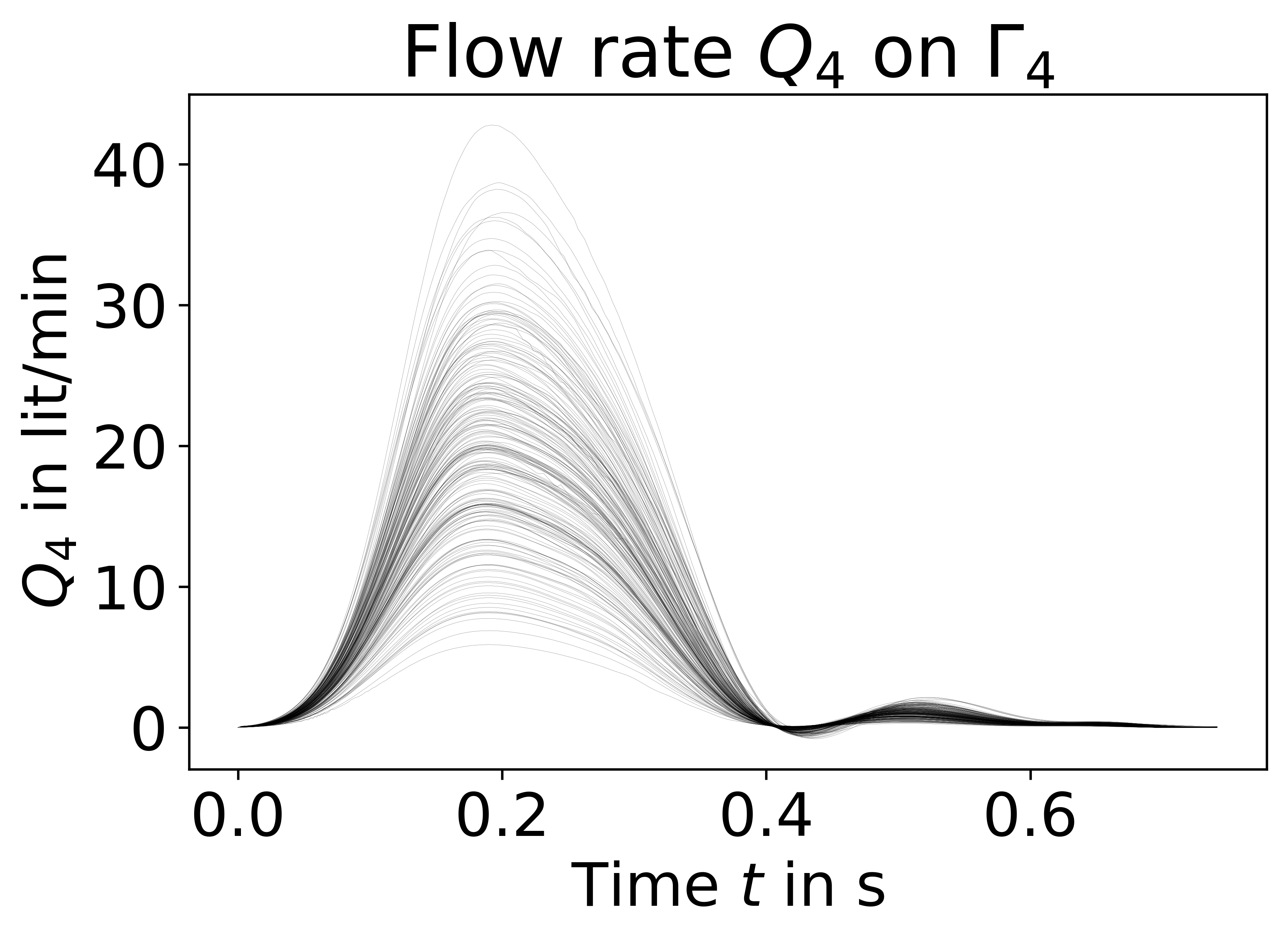}
        \\
        \includegraphics[width=0.24\linewidth, draft=\draftHeavy
        ]{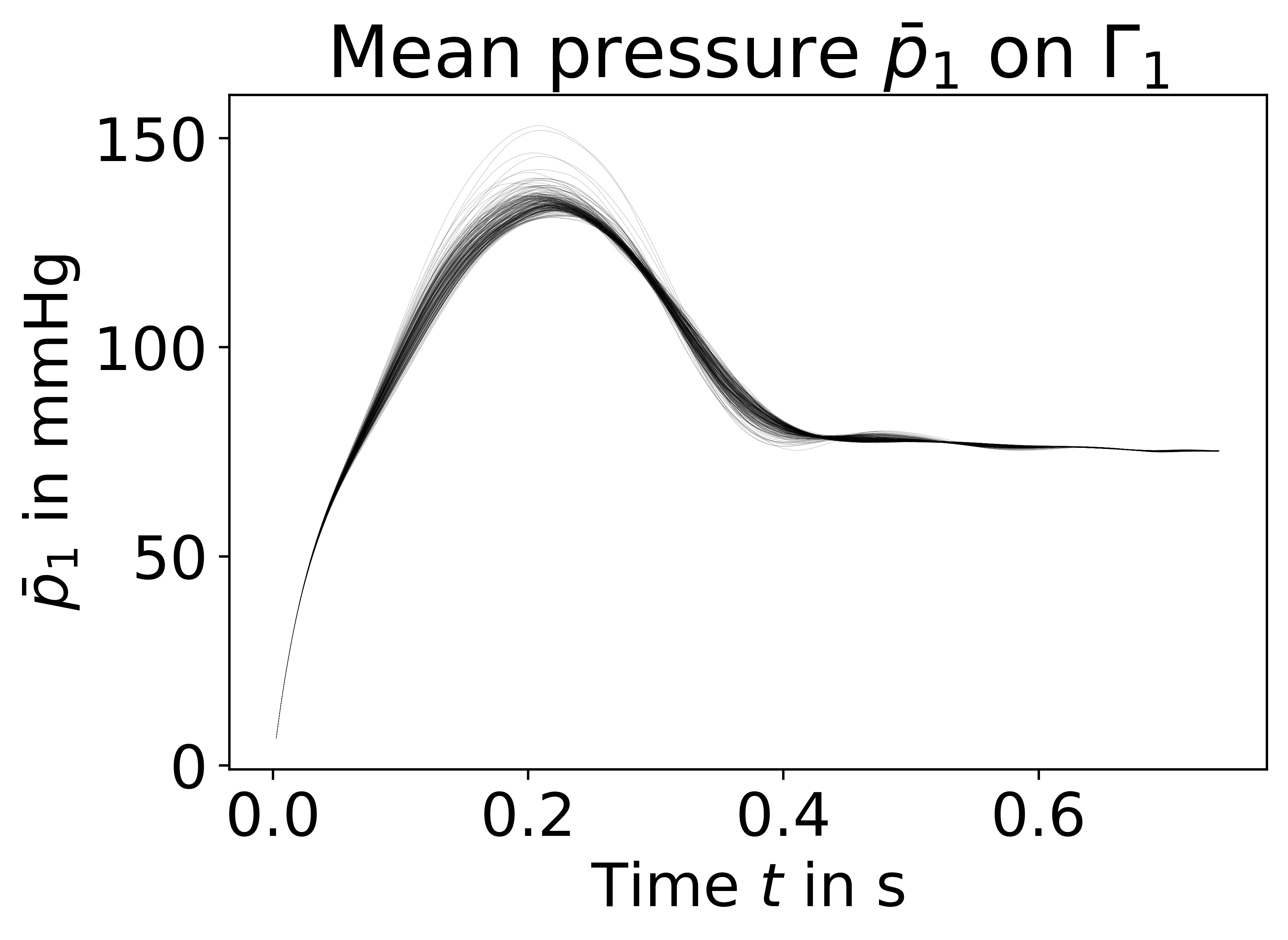}
        \hspace{0.0\linewidth}
        \includegraphics[width=0.24\linewidth, draft=\draftHeavy
        ]{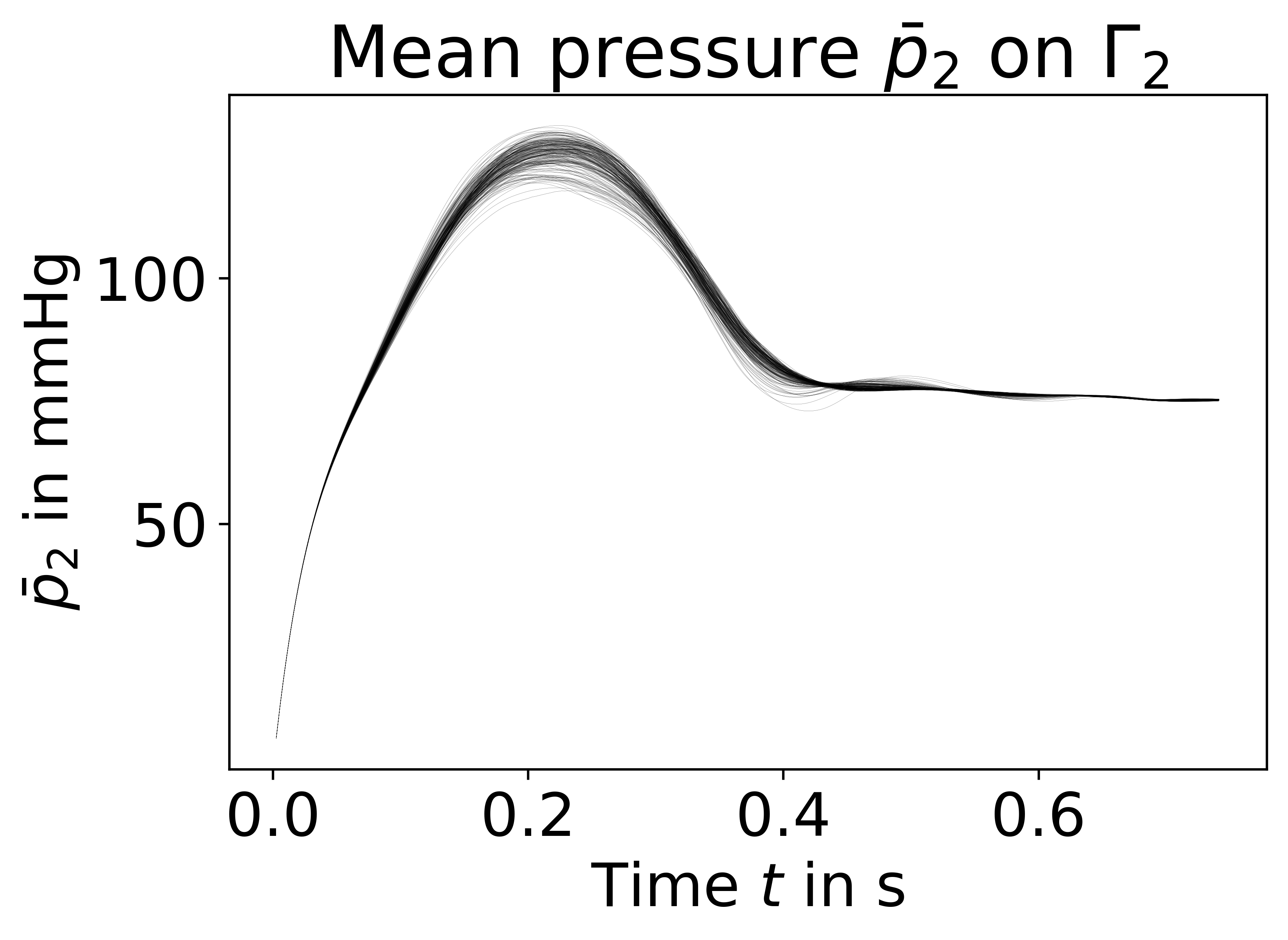}
        \hspace{0.0\linewidth}
        \includegraphics[width=0.24\linewidth, draft=\draftHeavy
        ]{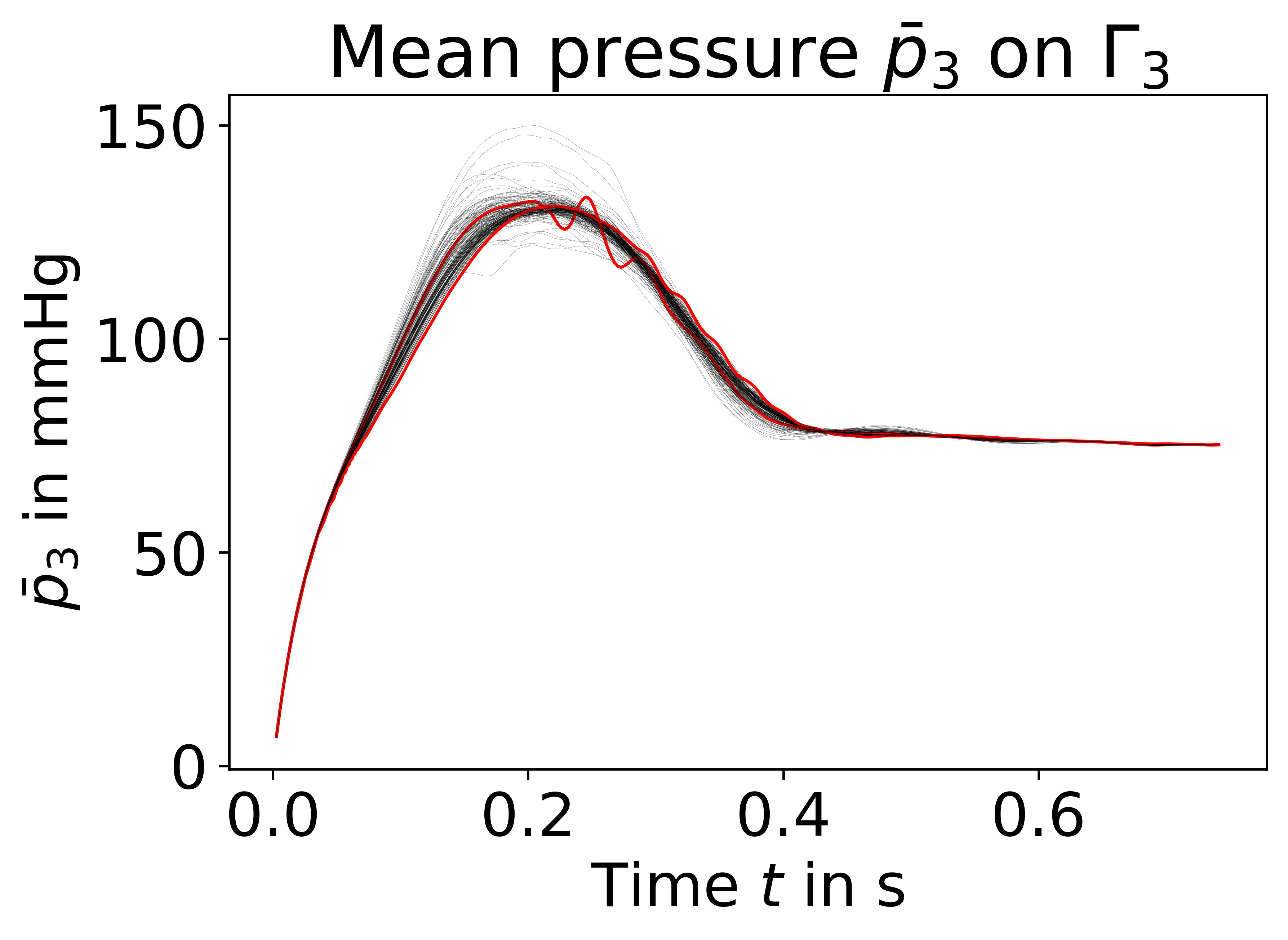}
        \hspace{0.0\linewidth}
        \includegraphics[width=0.24\linewidth, draft=\draftHeavy
        ]{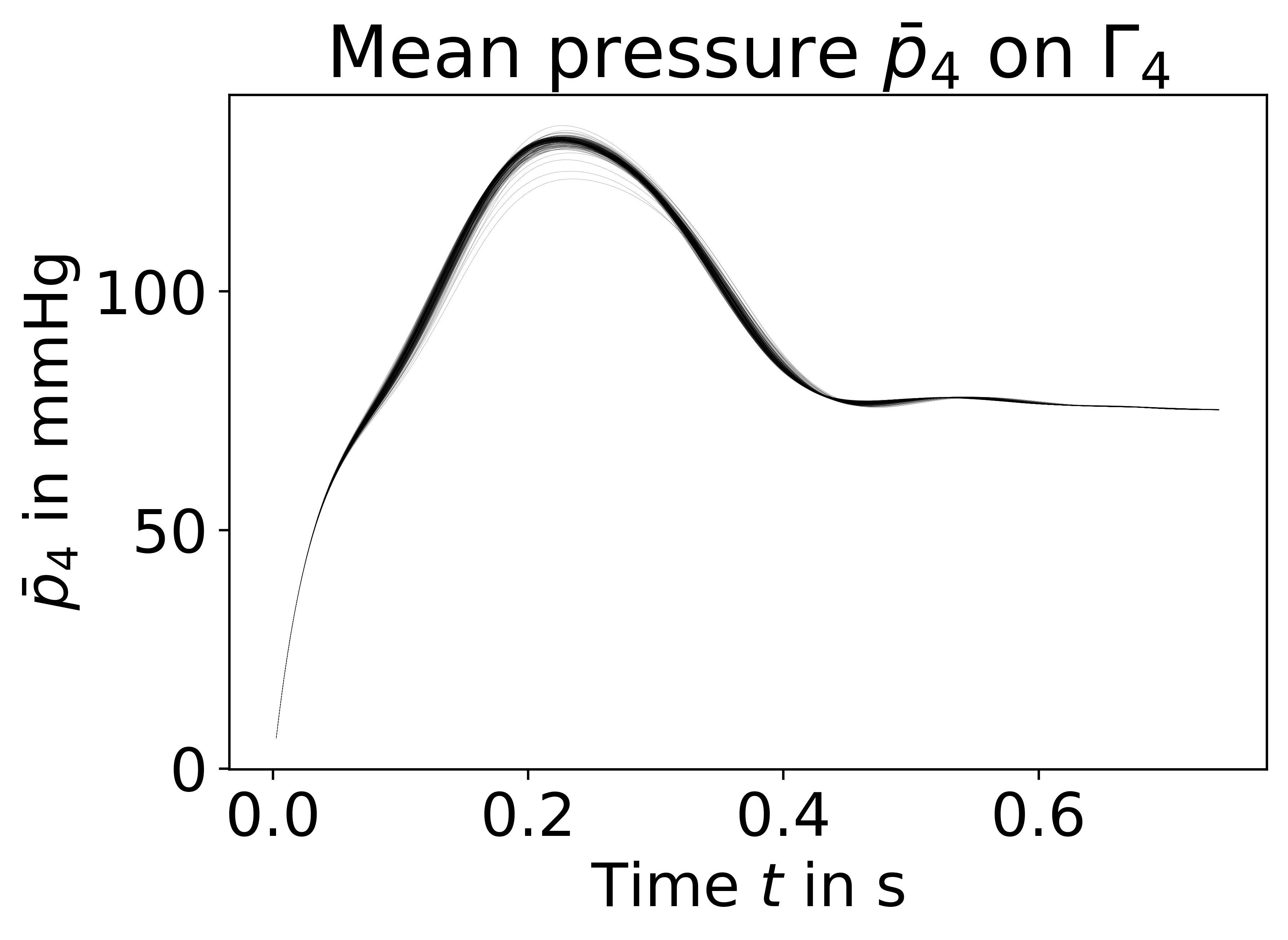}
        \caption{Flow rates $Q_i$ (top row) and outlet mean pressures
            $\overline{p}_i$ (bottom row), $i=1,\dots,4$,
            of the first 1000 cases of the SynthAorta dataset. The two filtered
            cases out of 1000 are highlighted in red.}
        \label{fig:filtering_pi_Qi}
    \end{figure}

    Herein, we choose the route of postprocessing the simulation dataset to detect
    parameter configurations that deliver unphysical or diverging solutions,
    by (i) evaluating whether the final simulation time of $0.78~\text{s}$ was
    reached, and (ii) introducing a tolerance on the second derivative of the
    flow rate $Q_i$ and on the mean outlet pressure $\overline{p}_i$.
    A total of 15 of the first 1000 cases fail within the first time step due to bad
    element quality, while 2 cases are flagged
    as unphysical due to the variation in the pressure and flow rate, 
    see Figure~\ref{fig:filtering_pi_Qi}.
    These cases can be successfully simulated using lower grid resolution and/or
    different CFD solver settings (results omitted), indicating that the
    geometric parameters and the Windkessel tuning themselves are not problematic,
    but the CFD solver settings combined with
    the element quality are troublesome. In order not to introduce bias into the
    data, we restrict ourselves to these filtering criteria, although they are rather
    loose, so that more than 98\% of the first 1000 cases remain in the dataset.

    Additional filtering is performed during the processing of training and test
    data subsequently used in ML models, to remove cases which
    might not represent relevant physiological regimes.
    Specifically, we remove geometries in which the corresponding pressure solution
    yielded a maximum pressure drop at peak systole higher than a prescribed
    threshold of $5~\text{mmHg}$.
    This threshold value is motivated by the distributions of filtered
    average and maximum values of the velocity, pressure drop, displacement, and
    WSS fields for the first 5000 cases of the SynthAorta dataset.

\begin{figure}[htp!]
      \centering
      \includegraphics[width=1\linewidth
      ]{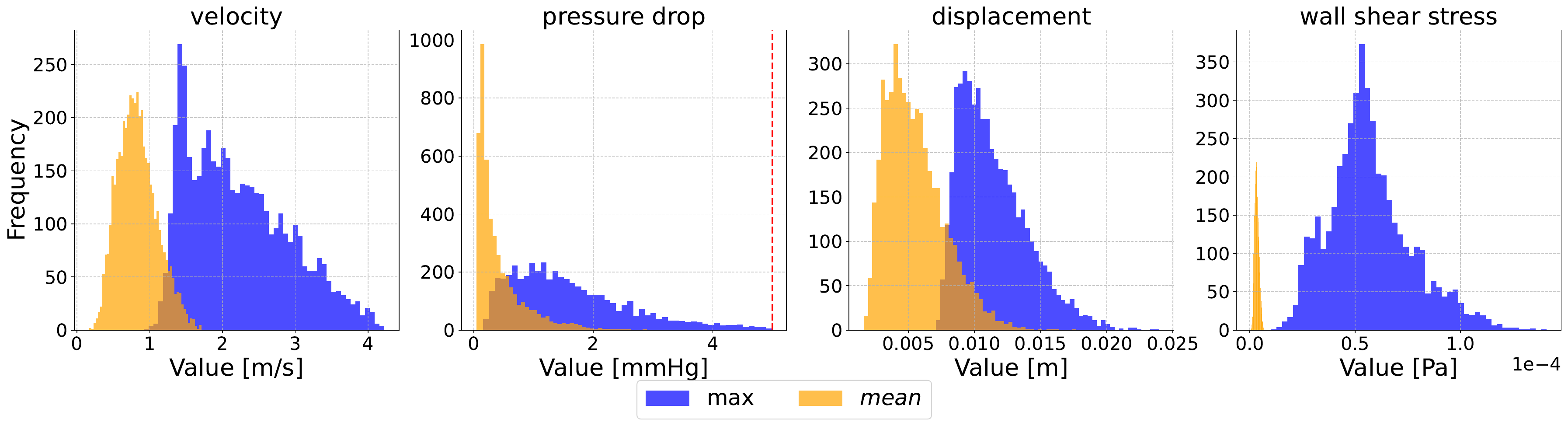
      }
      \caption{From left to right: Distribution of the maximum and mean values of
        the velocity magnitude, pressure drop, displacement magnitude from the
        common reference geometry, and WSS
        magnitude fields at peak systole for the considered
        $4660$ accepted geometries ($93.2\%$ of the first $5000$ cases) of the
        SynthAorta dataset. The red dashed line indicates the threshold of
        $5~$mmHg on the pressure drop used for filtering out non-physiological
        cases.}
      \label{fig:max_mean_values_snapshots_distribution_all}
    \end{figure}

    Adopting the filtering techniques described above, the
    resulting total number of accepted geometries is $4660$ out of $5000$
    ($93.2\%$). Out of these $4660$ cases, $n_{\text{train}}=4560$ are used for
    training and $n_{\text{test}}=100$ for testing.
    The resulting distributions of the average and maximum values of the velocity,
    pressure drop, displacement, and WSS fields of the final training and test
    datasets are shown in
    Figure~\ref{fig:max_mean_values_snapshots_distribution_all}.

    To reduce the dimension of the reduced space, the discontinuous second-order
    velocity $\ve{u}$, the discontinuous first-order pressure $p$, and the
    continuous second-order nodal displacement field $\ve{d}$ are interpolated
    onto \emph{continuous} finite element spaces on uniformly refined grids. This
    step accounts for the higher-order contributions, since standard Graph U-Nets
    have no notion of higher-order elements.
    This leads to a scalar finite element space with $16125$ nodes, so
    $N_u=N_d=48375$ DoFs for vector fields (velocity and displacement),
    and $N_p=16125$ DoFs for the pressure field. The WSS
    $\ve{\tau}_\mathrm{wall}$ is obtained in postprocessing.

    \subsection{Reconstruction error on reduced-dimensional spaces}
    An effective way to explore the complexity of the snapshot dataset is to compute reduced-dimensional spaces via randomized singular value decomposition (rSVD) and
    investigate how the relative $\ell^2$-reconstruction error of the velocity and pressure drop decays with respect to its dimension.
    Knowing how the reconstruction error decays with respect to the reduced dimensions is crucial to select the most effective surrogate modeling approach. If the decay is fast or the threshold on the accuracy is reached
    with only a few rSVD modes, a linear subspace surrogate model might be sufficient (e.g., the reduced basis method or proper orthogonal decomposition combined with neural networks, POD-NN). Conversely, if the decay is slow, nonlinear surrogates and manifold learning approaches might be necessary and more effective.
    Given a set of training velocity and pressure snapshots $\{\ve{u}_i\}_{i=1}^{n_{\text{train}}}$ and $\{p_i\}_{i=1}^{n_{\text{train}}}$, with $\ve{u}_i\in\mathbb{R}^{N_u}$ and $p_i\in\mathbb{R}^{N_p}$, respectively,
    we define the snapshot matrices gathering the mean-centered snapshots as the column vectors:
    \begin{align*}
        \ve{U} 
        &=
        \big[\,\ve{u}_1-\ve{u}_\mathrm{mean}\ \big|\
        \cdots\
        \big|\ \ve{u}_{n_{\text{train}}}-\ve{u}_\mathrm{mean}\,\big]
        &
        \hspace{-3cm}
        \in\mathbb{R}^{N_u\times n_{\text{train}}}
        \,,
        \\
        \ve{P} 
        &=
        \big[\,p_1-p_\mathrm{mean}\ \big|\
        \cdots\
        \big|\ p_{n_{\text{train}}}-p_\mathrm{mean}\,\big]
        &
        \hspace{-3cm}
        \in\mathbb{R}^{N_p\times n_{\text{train}}}
        \,,
    \end{align*}
    where $\ve{u}_\mathrm{mean}$ and $p_\mathrm{mean}$ denote the means computed over the training snapshots.
    We then compute their rank-$r$ rSVD, i.e.,
    \begin{equation*}
        \ve{U}_r = \ve{\Phi}_r\,\ve{\Sigma}_r\,\ve{V}_r^\top = \sum_{j=1}^{r} \sigma_j\, \ve{\phi}_j\, \ve{v}_j^\top, \quad
        \ve{P}_r  = \ve{\Psi}_r\,\ve{\Lambda}_r\,\ve{Q}_r^\top = \sum_{j=1}^{r} \lambda_j\, \psi_j\, q_j^\top\,,
    \end{equation*}
    where $\ve{\Sigma}_r=\operatorname{diag}(\sigma_1,\dots,\sigma_r)$,
    $\ve{\Lambda}_r=\operatorname{diag}(\lambda_1,\dots,\lambda_r)$ contain the
    $r$ largest singular values in decreasing order,
    the $r$ columns of $\ve{\Phi}_r=[\ve{\phi}_1|\cdots|\ve{\phi}_r]$ and
    $\ve{\Psi}_r=[\psi_1|\cdots|\psi_r]$ are the leading left singular vectors
    (spatial modes),
    and the columns of $\ve{V}_r=[\ve{v}_1|\cdots|\ve{v}_r]$ and
    $\ve{Q}_r=[q_1|\cdots|q_r]$ are the corresponding right singular vectors.
    The columns of $\ve{\Phi}_r$ and $\ve{\Psi}_r$ define the bases of the
    (reduced) $r$-dimensional spaces that approximate the original set of
    snapshots.
    For both training and test snapshots, the reconstruction is defined by the
    orthogonal projection onto these spaces, using the same training means and
    bases, i.e.,
    \begin{equation*}
        \ve{u}_{r, i} = \ve{u}_\mathrm{mean} + \ve{\Phi}_r \ve{\Phi}_r^\top \big(\ve{u}_i - \ve{u}_\mathrm{mean}\big), \quad
        p_{r, i} = p_\mathrm{mean} + \ve{\Psi}_r \ve{\Psi}_r^\top \big(p_i - p_\mathrm{mean}\big)\,.
    \end{equation*}
    For the training snapshots, $i=1,\hdots,n_{\text{train}}$, this coincides with
    the columns of $\ve{U}_r$ and $\ve{P}_r$ shifted by the means, i.e.,
    $\ve{u}_{r, i} = \ve{u}_\mathrm{mean} + \sum_{j=1}^r \sigma_j (\ve{v}_j)_i\, \ve{\phi}_j$
    and
    $p_{r, i} = p_\mathrm{mean} + \sum_{j=1}^r \lambda_j (q_j)_i\, \psi_j$.
     The relative reconstruction error, for an arbitrary training or test index $i$,
     is then computed as the $\ell^2$-norm of the difference between the original
     and reconstructed snapshot divided by the $\ell^2$-norm of the original
     centered snapshot, i.e.,
    \begin{equation}
        \label{eq:err}
        \epsilon_{\ve{u}, i}(r)
        =
        \frac{||\ve{u}_i - \ve{u}_{r, i}||_{\ell^2}}{||\ve{u}_i-\ve{u}_\mathrm{mean}||_{\ell^2}},\quad
        \epsilon_{p, i}(r)
        =
        \frac{||p_i - p_{r, i}||_{\ell^2}}{||p_i-p_\mathrm{mean}||_{\ell^2}}\,.
    \end{equation}
    In the following study, we will consider the mean and maximum
    $\ell^2$-reconstruction errors on both the training and the test snapshot
    sets.

\begin{figure}[htp!]
        \centering
        \includegraphics[width=1\linewidth]{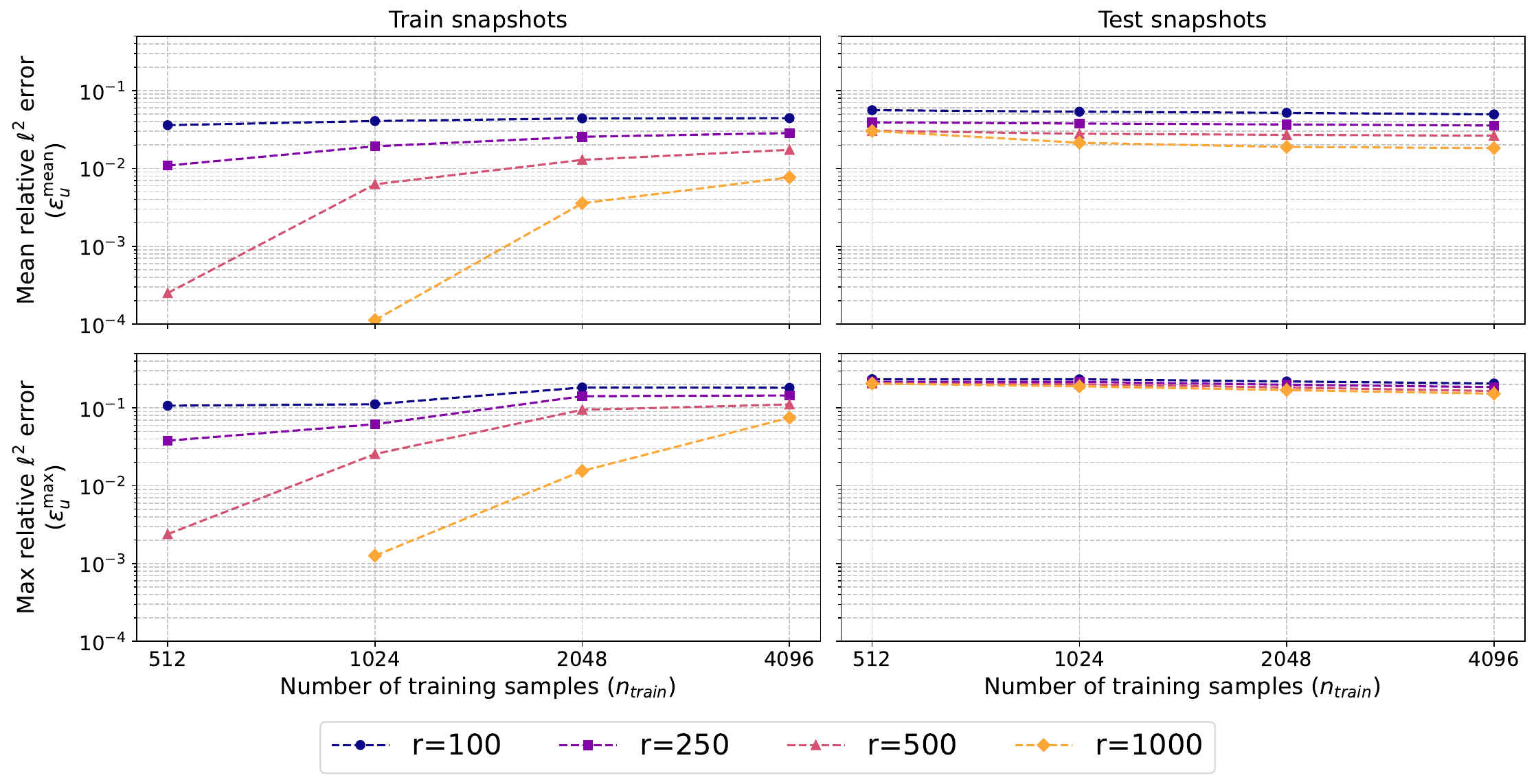}
        \caption{Relative $\ell^2$-reconstruction errors for the velocity field versus the number of training samples $n_{\text{train}}$, for different numbers of retained rSVD modes $r$. The left and right columns report the errors on the training and test snapshots, and the top and bottom rows the mean and maximum errors, respectively. Missing values correspond to reconstruction errors close to machine precision.}
        \label{fig:recu}
    \end{figure}

\begin{figure}[htp!]
        \centering
        \includegraphics[width=1\linewidth]{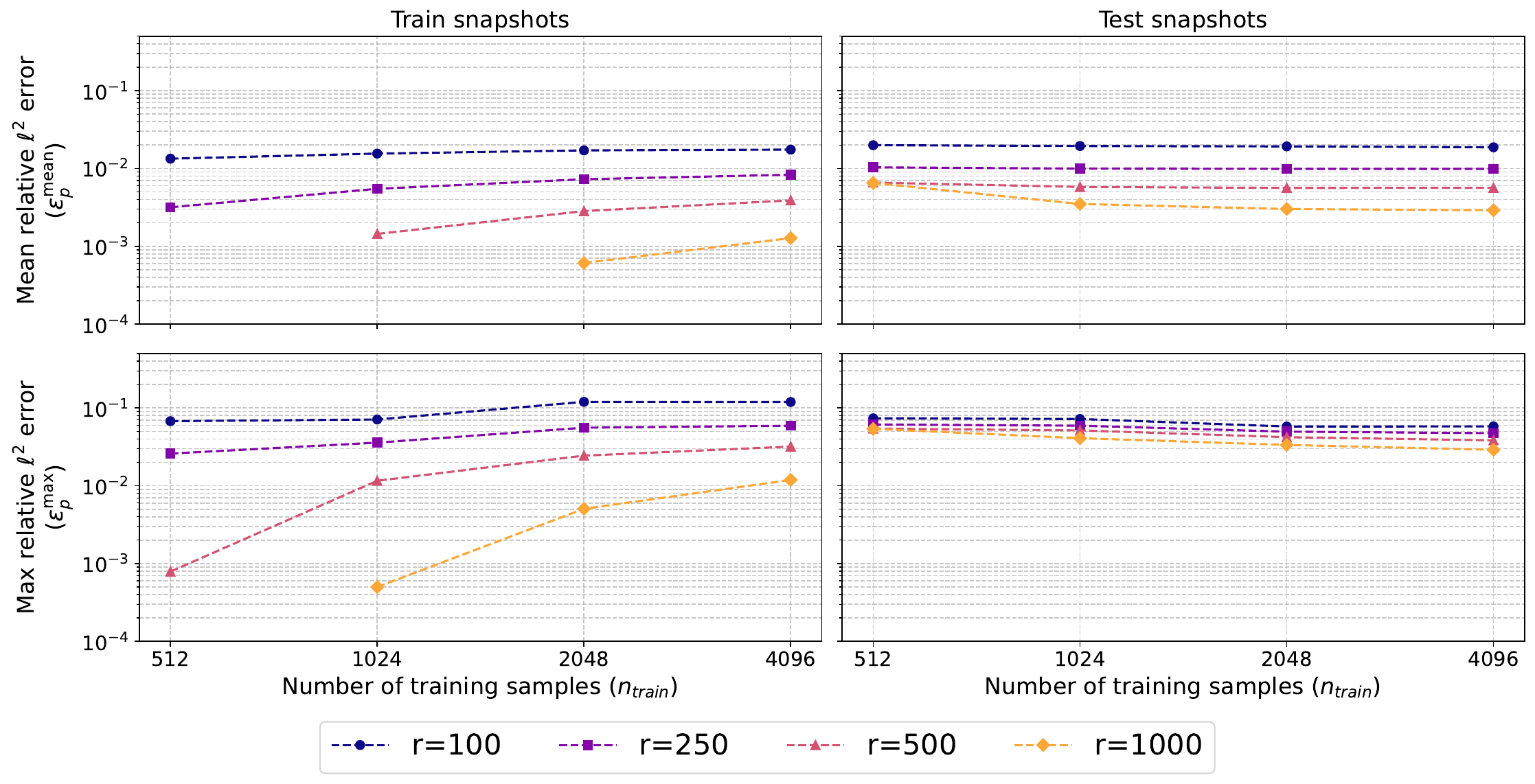}
        \caption{Relative $\ell^2$-reconstruction errors for the pressure field versus the number of training samples $n_{\text{train}}$, for different numbers of retained rSVD modes $r$. The left and right columns report the errors on the training and test snapshots, and the top and bottom rows the mean and maximum errors, respectively. Missing values correspond to reconstruction errors close to machine precision.}
        \label{fig:recp}
    \end{figure}

    Figures~\ref{fig:recu} and \ref{fig:recp} show the resulting relative
    reconstruction errors for increasing dimension $r$ of the reduced space for
    velocity and pressure, respectively.
    As expected, increasing $r$ yields a decrease in the reconstruction errors.
    Moreover, increasing $r$ from $500$ to $1000$ still yields a significant
    improvement, indicating the high complexity of the snapshot dataset.
    As the number of training samples increases, the test
    reconstruction error 
    is observed to practically stabilize during the transition
    from $2048$ to $4096$
    samples, albeit without reaching the training reconstruction error values. The training
    reconstruction error grows as the number of training
    samples increases to $4096$. This behavior also indicates that the
    considered dataset is highly complex and cannot be easily represented by a
    low-dimensional space.
    Comparing the velocity and pressure reconstruction errors reveals that
    the velocity error is in general higher than the pressure error, indicating
    that pressure fields are more easily represented using a reduced basis.

    \begin{remark}
        On the training set, the decay of the reconstruction errors with
        increasing $r$ is closely related to the decay of the singular values
        $\sigma_j$ and $\lambda_j$: faster singular-value decay indicates that
        fewer modes are needed to capture a given fraction of the centered
        snapshot energy. However, the singular values alone do not determine
        the mean and maximum of the per-snapshot relative errors considered here.

        Evaluating reconstruction errors on a held-out test set additionally
        assesses how well the reduced space computed from the training snapshots
        represents unseen snapshots drawn from the same underlying distribution.
        Applying this procedure to other datasets, e.g., for aortas affected by
        coarctation~\cite{Romor2025Assimilation} or for simplified aortic
        arches~\cite{pajaziti2023shape}, permits an empirical comparison of their
        linear approximation complexity, provided that sample sizes, sampling
        procedures, field representations, centering, and error norms and
        normalizations are comparable.
    \end{remark}

\begin{figure}[htp!]
        \centering
        \includegraphics[width=1\linewidth]{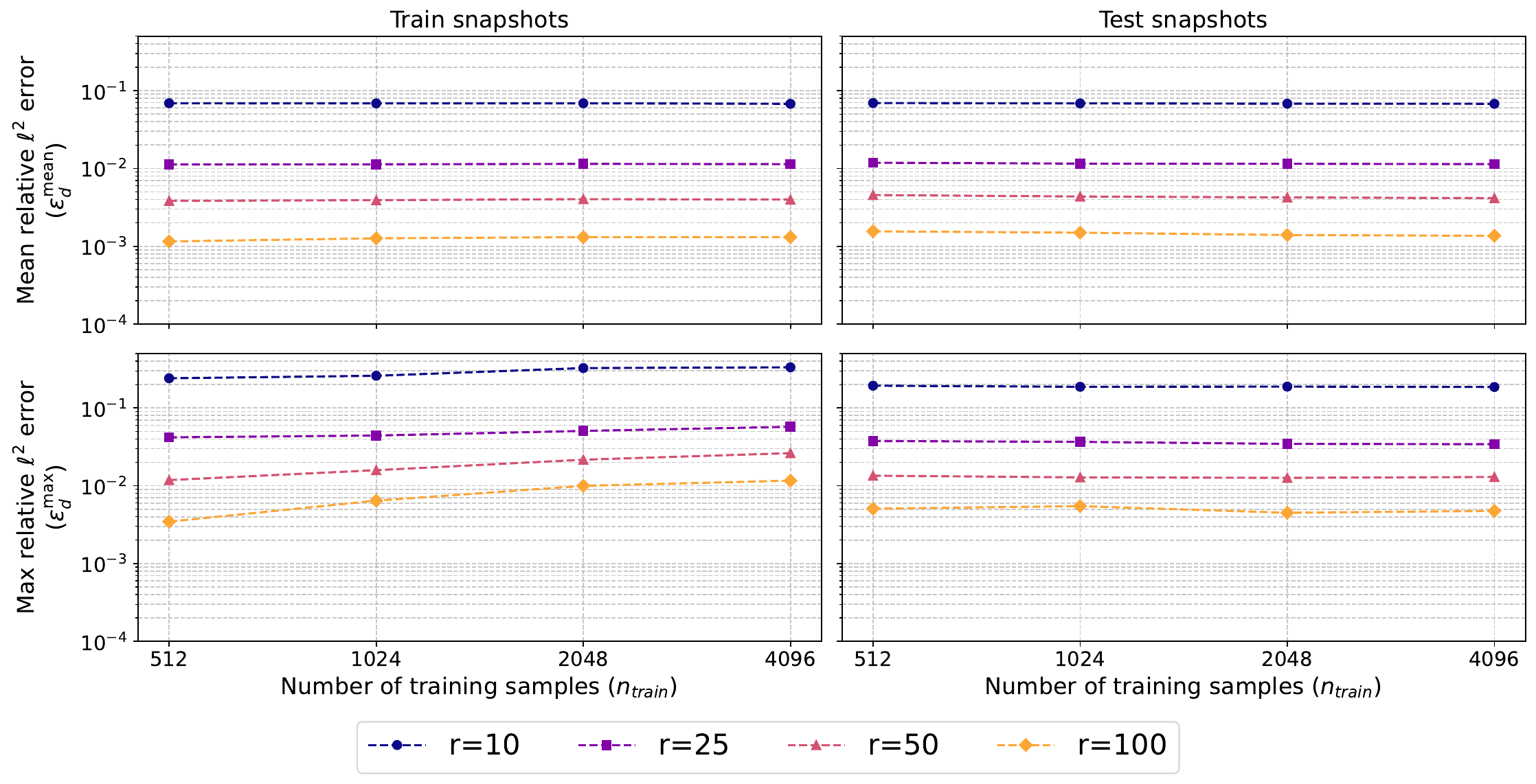}
        \caption{Relative $\ell^2$-reconstruction errors for the displacement field 
        versus the number of training samples $n_{\text{train}}$, for different
        numbers of retained rSVD modes $r$. The left and right columns report the
        training and test errors, and the top and bottom rows the
        mean and maximum errors, respectively.}
        \label{fig:recd}
    \end{figure}

    The relative $\ell^2$-reconstruction error can also be used to quantify the complexity of the parameter space for geometry generation.
    To this end, we consider the set of displacement fields $\{d_i\}_{i=1}^{n_{\text{train}}}$ and $\{d_i\}_{i=1}^{n_{\text{test}}}$, which map the reference geometry onto the other shapes, for the training and test sets, respectively,
    computing the rSVD and the corresponding relative reconstruction errors as described above. The resulting mean and maximum errors are shown in Figure~\ref{fig:recd}.
    The mean reconstruction error for the displacement field stays almost constant when increasing the number of samples from $512$ to $4096$, which indicates that the geometric parameter space is reasonably well represented using $512$ cases.
    However, the sensitivity of the Navier--Stokes equations to the domain variability causes the velocity snapshots to have a worse decay of the reconstruction error (Figure~\ref{fig:recu}).
    Concerning the displacement, increasing $r$ reduces the reconstruction error, achieving a mean relative error of $\approx 1 \times 10^{-3}$. This error is lower than the best mean reconstruction error obtained for the velocity and comparable to the mean reconstruction error for pressure using the same number of modes. The spatial complexity of the velocity snapshot data is thus---as expected---much higher.

\begin{figure}[ht!]
        \centering
        \includegraphics[width=1\textwidth]{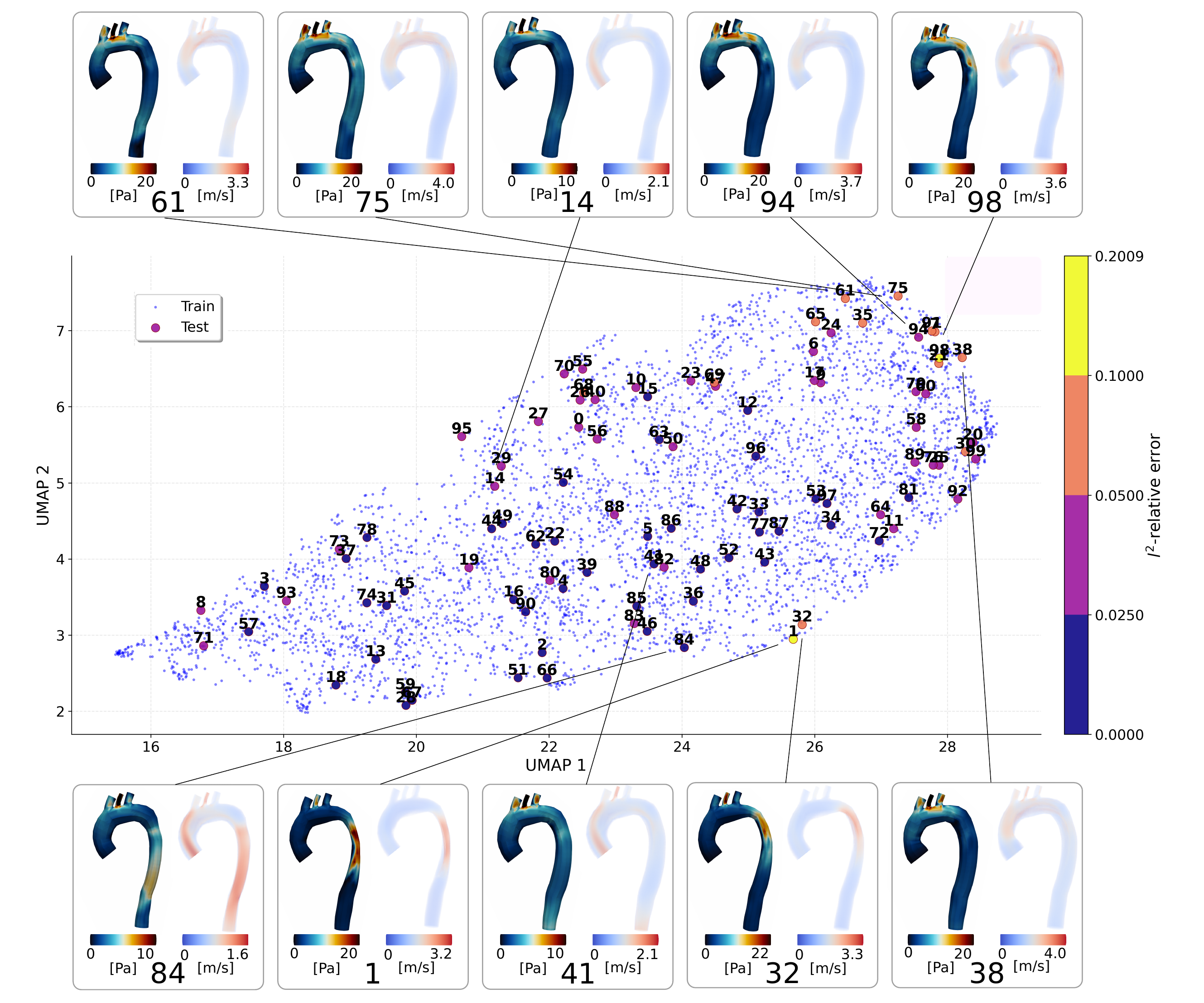}
        \caption{UMAP clustering of the velocity snapshots at peak systole using the truncated rSVD coefficients with $r=500$. The color indicates the test rSVD relative $\ell^2$ reconstruction error.}
        \label{fig:umap_clustering}
    \end{figure}

    To further characterize the snapshot data, we employ the Uniform Manifold
    Approximation and Projection (UMAP) technique~\cite{McInnes2018}.
    UMAP is a nonlinear dimensionality-reduction technique that maps 
    high-dimensional data to a low-dimensional embedding while approximately
    preserving the neighborhood relationships between data points, revealing
    structures such as clusters, transitions, and outliers that are difficult
    to identify in the original space. Figure~\ref{fig:umap_clustering} shows
    the UMAP of the velocity snapshot data in $\mathbb{R}^2$, for the
    case $r=500$, identifying which cases yield the highest reconstruction
    errors. It can be observed that cases with high test $\ell^2$-reconstruction
    errors lie close to the boundaries of the UMAP embedding, while snapshots
    yielding errors in the $5\%$--$20\%$ range are associated with outliers in
    the flow pattern distribution, including mild stenoses (cases $1$, $32$) and
    mild coarctation (cases $38$, $61$, $75$, $98$, $100$).

    \section{Inter-patient surrogate modeling}
    \label{sec:surrogate_modeling}

    This section focuses on surrogate models built
    on the dataset of CFD solutions across the shape database introduced in
    Section~\ref{sec:dataset_generation}.
    We consider two architectures for inter-patient surrogate modeling:
    (i) Graph U-Nets, relying on message passing on the hierarchy of reference
    graphs/meshes (level $1$, level $0$, and the reference geometry's centerline),
    and (ii) Transformer U-Nets, relying on convolutional and cross-attention
    layers on structured grids.
    These methods will be introduced in Section~\ref{sec:inference}. One of the
    most critical aspects is the high dimensionality of the space of possible
    aortic shapes, which can only be marginally approximated by the parametrization
    used to generate the training
    set~\cite{Bosnjak2025SynthAorta}.

    For this purpose, we benchmark several AL strategies used to
    select the most informative samples during training
    (Section~\ref{sec:active_learning}).
    Thereafter, we discuss possible approaches to improve the physical consistency of
    the model relying on the automatic differentiation of the associated PDE residual
    during training or within a fine-tuning step
    (Section~\ref{sec:physics_consistency}).
    Finally, we assess the out-of-distribution generalization of the surrogates
    on different databases, considering a dataset of synthetic aortic
    coarctation patients~\cite{Yevtushenko2022, Katz2026} (generated via
    statistical shape models) and the Vascular Model Repository
    (VMR)~\cite{Wilson2013} dataset in Section~\ref{sec:results_ood}. Every computational study
    in this section is performed on a single NVIDIA A100-SXM4-40GB.

    \subsection{Inference with Graph U-Nets and Transformer U-Nets}
    \label{sec:inference}

    \subsubsection{Dataset preprocessing}
    \label{subsec:dataset_preprocessing}

All geometries in the SynthAorta dataset are characterized by the same mesh
topology by definition,
i.e., retaining a one-to-one correspondence between nodes and elements.
Consequently, the nodal displacement field for a geometry $i$ is defined as
\begin{equation}\label{eq:disp_mesh}
\ve{d}^i_j = \ve{x}^i_j - \ve{x}^\mathrm{ref}_{j}\,,
\end{equation}
where $\ve{x}^\mathrm{ref}_{j}$ and $\ve{x}_j^i$ denote the positions of node
$j$ in the reference and in the considered geometry, respectively. The nodal
coordinates of the case-specific geometry can therefore be expressed as
$\ve{x}_j=\ve{x}^\mathrm{ref}_{j}+\ve{d}_j$, where no registration is necessary.

Based on the deformed geometry, we subsequently compute additional geometric
features for the network. In particular, we construct a five-dimensional
centerline encoding for each node by projecting it onto the case-specific,
deformed centerlines of the aortic arch and the three supra-aortic branches.
The encoding comprises the distance to the centerline, the arc-length
coordinate of the closest centerline point, and the three components of the
unit vector pointing from the node toward that point.

Each preprocessed training and test sample stores, at each node, the target
fields, namely, the velocity $\ve{u}$, pressure $p$, and WSS, along with the
quantities used as network inputs and conditioning:
\begin{enumerate}
  \item the case-specific nodal coordinate vector
  $\ve{x}=\ve{x}^{\mathrm{ref}}+\ve{d}$,
  \item the velocity field, provided either on the full domain or only on its
  boundary, depending on the field to be predicted, and
  \item the five-dimensional geometrical centerline encoding described above.
\end{enumerate}
All input and target fields are normalized component-wise to $[0,1]$ using
minima and maxima computed exclusively from the training set.

Concerning the input velocity field, for the network predicting the velocity,
we use only the boundary data, i.e., the velocity restricted to the inlet,
outlet, and wall nodes (zero-valued on all interior nodes), so that the model
reconstructs the interior flow from the boundary values and the geometry.
For the pressure and WSS networks, the full velocity field is provided---either
the ground-truth field or the one predicted by the velocity network---so
that $p$ and $\text{WSS}$ are inferred from the complete velocity field
and the geometric encoding.
This choice is motivated by recent results suggesting that exploiting the
correlation between velocity and pressure fields, rather than the correlation
between geometry and pressure, can yield a better outcome in the context of
surrogate modeling~\cite{Romor2025Assimilation}.

    \subsubsection{Architecture specifics}
\begin{figure}[htp!]
        \centering
        \includegraphics[width=1\linewidth]{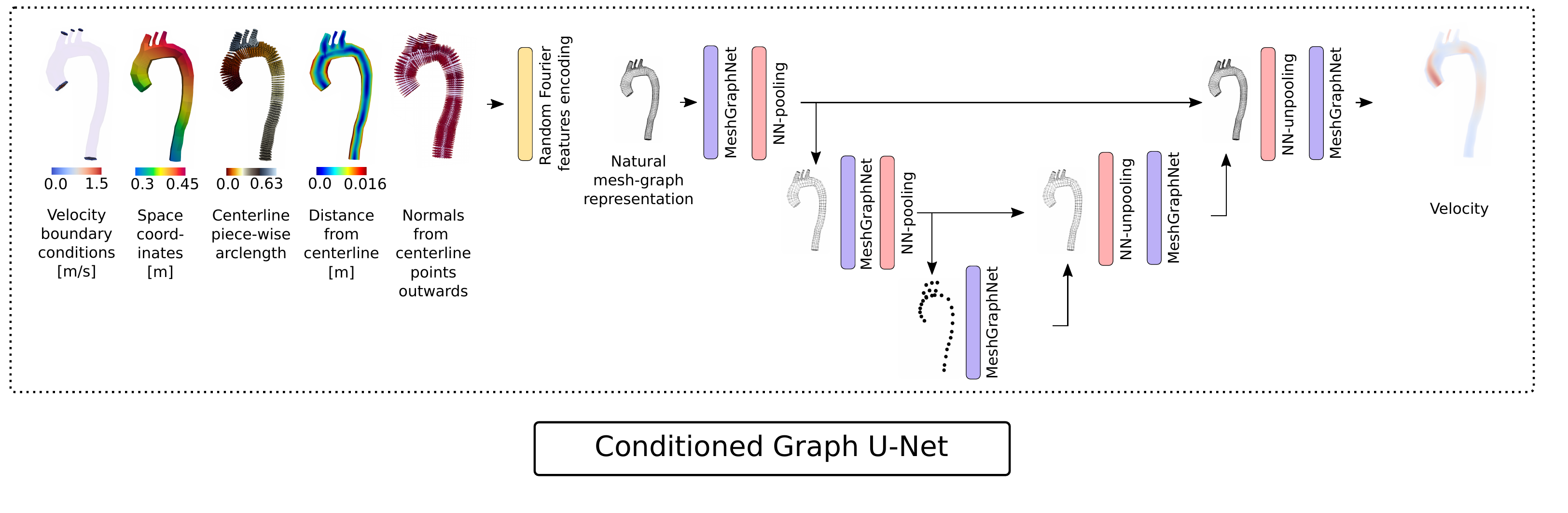}
        \caption{
          Schematic representation of the conditioned Graph U-Net. The nodal
          input features, i.e., velocity boundary conditions, spatial
          coordinates, and the centerline encoding (piecewise arc length,
          distance from the centerline, and normals pointing outward from the
          centerline), are lifted by a random Fourier feature encoding and
          processed by MeshGraphNet layers on the natural mesh graph.
          Nearest-neighbor pooling and unpooling layers connect the levels of
          the U-Net.}
        \label{fig:graph_unet_architecture}
    \end{figure}

\begin{figure}[htp!]
        \centering
        \includegraphics[width=1\linewidth]{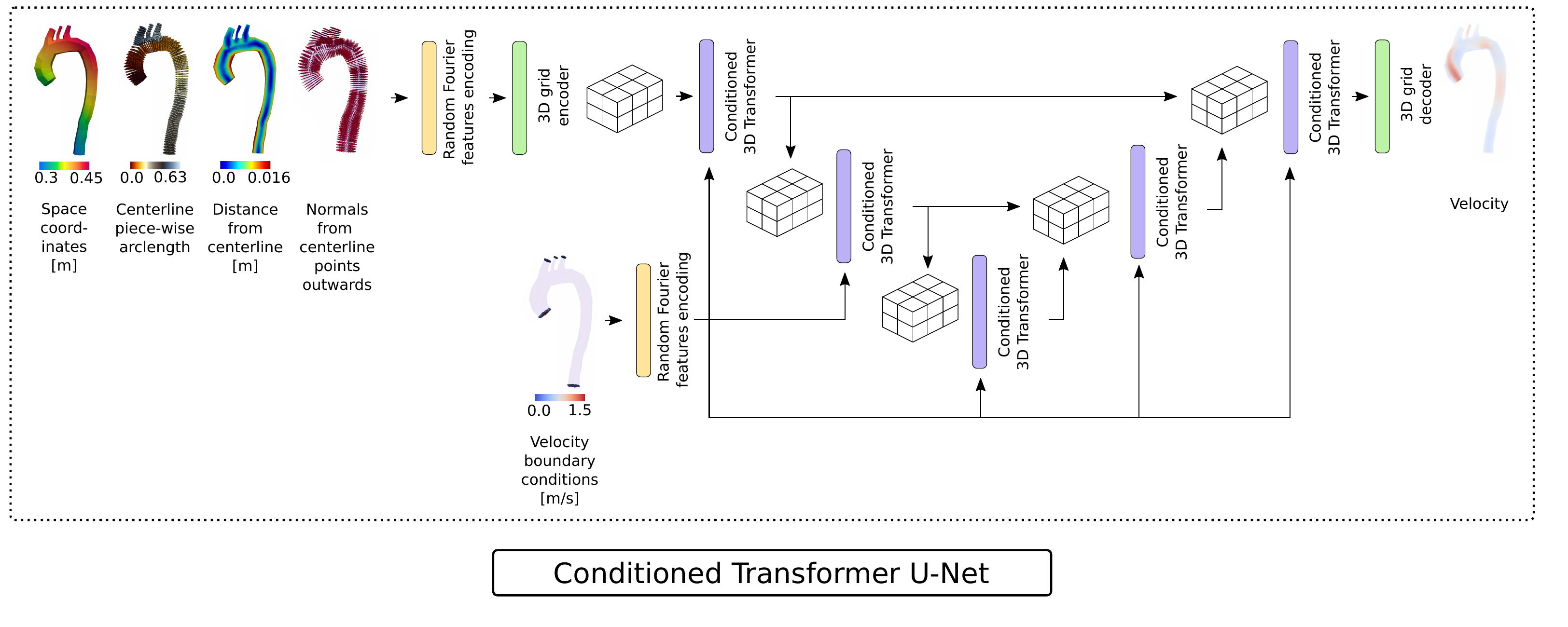}
        \caption{Schematic representation of the conditioned Transformer U-Net.
        The geometric inputs, i.e., spatial coordinates and centerline encoding,
        are lifted by a random Fourier feature encoding, mapped onto a 3D grid,
        processed by conditioned 3D transformer blocks at decreasing and
        increasing grid resolutions, and decoded back to the mesh nodes.
        The velocity boundary conditions are encoded separately and injected as
        a conditioning signal into each transformer block via cross-attention.}
        \label{fig:transformer_unet_architecture}
    \end{figure}

    The Graph U-Net is composed of MeshGraphNet
    layers~\cite{Pfaff2020GraphUNets, Hongyang2022GraphUNets} implemented in
    \texttt{PhysicsNeMo}~\cite{pnm2026shardtensor}, see
    Figure~\ref{fig:graph_unet_architecture}.
    The Transformer U-Net~\cite{bodnar2024foundationmodelearth} substitutes
    graph convolutional layers with attention layers applied to a 3D grid
    obtained by padding the input vector with zeros. Convolutions and transpose
    convolutions are applied in a U-Net fashion to the 3D grid, first decreasing
    the size of the 3D encoded grid and then increasing it back to the original
    size, see Figure~\ref{fig:transformer_unet_architecture}.

    The variables considered in the network differ between the two architectures,
    motivated by the significantly higher computational cost for training GNN
    layers.
    \begin{enumerate}
      \item Graph U-Net: the boundary velocity ($3$ components), the centerline
      encoding ($5$ components), and the deformed position ($3$ components) are
      concatenated into an $11$-dimensional nodal feature vector, mapped
      to a higher-dimensional embedding through a random Fourier feature (RFF)
      encoding followed by a feed-forward network (FFN), and fed as node
      features to the graph layers.
      \item Transformer U-Net: the boundary velocity ($3$ components), the
      centerline encoding ($5$ components), and the deformed position
      ($3$ components) are split into two separate streams that are encoded
      independently: a geometric stream gathering the centerline encoding and
      the deformed position ($8$ components), which provides the input tokens
      processed by the network, and a boundary stream gathering the boundary
      velocity ($3$ components), which is injected as a conditioning signal
      through the cross-attention layers. Each stream is given its own RFF
      encoding and FFN projection.
    \end{enumerate}

    In both cases, the per-node input features (see
    Section~\ref{subsec:dataset_preprocessing}) are first lifted to a
    $64$-dimensional embedding through an RFF map (with $2$ features over
    $16$ logarithmically spaced scales) followed by a two-layer FFN; the same
    embedding dimension is used throughout the network and as conditioning width.

    The Graph U-Net operates directly on the unstructured mesh, organized as a
    three-level hierarchy: the volumetric reference mesh at levels $1$ and $0$,
    and its centerline. The pooling and unpooling operators are parameter-free
    $k$-nearest-neighbor ($k=8$) linear interpolators that transfer node and
    edge features between consecutive levels. Each level applies a residual
    block composed of three MeshGraphNet sub-networks, each performing two
    message-passing steps with two-layer multilayer perceptrons, sigmoid linear
    unit activation functions, and mean aggregation; the node feature width
    grows along the encoder following the sequence $\{32,64,128\}$ and is
    mirrored along the decoder as $\{64,32,16\}$.

    The Transformer U-Net instead embeds the node features into a zero-padded
    $32^3$ voxel grid with a base width of $64$ channels. A depth-three encoder
    halves the grid resolution at each stage while doubling the number of channels
    ($64\!\to\!128\!\to\!256$). This is followed by a $512$-channel bottleneck
    composed of three transformer blocks, and a symmetric decoder restores the
    original resolution with skip connections. Each transformer block combines
    a cross-attention layer ($4$ heads), whose queries come from a grouped
    $3\times3\times3$ convolution on the volume and whose keys/values come from
    the boundary conditioning token, with a $1\times1\times1$ convolutional FFN
    expansion ($\times4$) and group normalization.

    We train three different networks, one for each of the three target fields
    (velocity $\ve{u}$, pressure $p$, and WSS), with output dimension $3$, $1$,
    and $3$, respectively.
    The main hyperparameters and the resulting number of trainable parameters
    are summarized in Table~\ref{tab:arch_params}, while Table~\ref{tab:arch_timings}
    details the required training time. The Transformer U-Net
    is faster per iteration despite its larger parameter count, owing to its
    dense voxel-grid representation compared to message passing on the
    unstructured mesh of the Graph U-Net.

    \begin{table}[!htb]
        \centering
        \renewcommand{\arraystretch}{1.2}
        \rowcolors{2}{RoyalBlue!7}{white}
        \begin{tabular}{lcc}
            \toprule
            \rowcolor{RoyalBlue!25}
            \textbf{Component / hyperparameter} & \textbf{Graph U-Net} & \textbf{Transformer U-Net} \\
            \midrule
            Spatial representation       & 3-level mesh hierarchy      & padded $32^3$ voxel grid \\
            Conditioning mechanism       & fused node features         & boundary cross-attention \\
            Input embedding dimension    & $64$                        & $64$ \\
            U-Net depth / levels         & $3$                         & $3$ \\
            Feature width (enc.\ stages) & $32,\,64,\,128$             & $64,\,128,\,256$ \\
            Bottleneck width             & $128$                       & $512$ \\
            Core block                   & MeshGraphNet ($\times3$)    & cross-attention $+$ conv-FFN \\
            Message passing / blocks per stage & $2$                   & $2$ \\
            Attention heads              & -       & $4$ (every block) \\
            \midrule
            \rowcolor{ForestGreen!12}
            Parameters, $\ve{u}$ net     & $4.36\times10^{6}$          & $22.17\times10^{6}$ \\
            \rowcolor{ForestGreen!12}
            Parameters, $p$ net          & $4.36\times10^{6}$          & $22.17\times10^{6}$ \\
            \rowcolor{ForestGreen!12}
            Parameters, $\text{WSS}$ net & $4.36\times10^{6}$          & $22.17\times10^{6}$ \\
            \bottomrule
        \end{tabular}
        \caption{Main architectural hyperparameters and number of trainable parameters of the Graph U-Net and the Transformer U-Net.
        One independent network is trained per target field ($\ve{u}$, $p$, $\text{WSS}$). The velocity and WSS networks have output dimension $3$, while the pressure network has output dimension $1$. Details of the parameters are provided in Table~\ref{tab:graph_breakdown}.}
        \label{tab:arch_params}
    \end{table}

    \begin{table}[!htb]
        \centering
        \renewcommand{\arraystretch}{1.2}
        \rowcolors{2}{RoyalBlue!7}{white}
        \begin{tabular}{lcc}
            \toprule
            \rowcolor{RoyalBlue!25}
            \textbf{} & \textbf{Graph U-Net} & \textbf{Transformer U-Net} \\
            \midrule
            Total training time [h]        & $38.05$  & $24.80$ \\
            Average time per iteration [s]  & $0.61$   & $0.40$ \\
            \bottomrule
        \end{tabular}
        \caption{Wall-clock training cost of the velocity network for each architecture, measured on a single NVIDIA A100-SXM4-40GB: total training time and average time per iteration. }
        \label{tab:arch_timings}
    \end{table}

    Table~\ref{tab:graph_breakdown} reports the layer-by-layer number of trainable
    parameters of the two architectures, taking the velocity network ($n_\text{out}=3$) as reference. The pressure
    network ($n_\text{out}=1$) only differs in the final projection, yielding $4\,362\,778$ parameters for the Graph
    U-Net and $22\,173\,121$ for the Transformer U-Net, against $4\,363\,558$ and $22\,173\,251$ for the velocity and
    WSS networks. The input RFF encoder contributes $8\,320$ parameters per stream
    (the Graph U-Net uses a single fused stream, whereas the Transformer U-Net uses two, resulting in $16\,640$ parameters).

\begin{table}%
  \centering
  \renewcommand{\arraystretch}{1.2}
      \rowcolors{2}{RoyalBlue!7}{white}
  \begin{subtable}[t]{0.49\textwidth}
    \centering
     {\small \begin{tabular}[t]{p{4.4cm}r}
        \toprule
        \rowcolor{RoyalBlue!25}
        \textbf{Graph U-Net} & \textbf{Parameters} \\
        \midrule
        Encoder (3 ConvBlocks $+$ pooling) & $1\,523\,520$ \\
        Bottleneck block \texttt{mid\_1}    & $1\,186\,560$ \\
        Bottleneck block \texttt{mid\_2}    & $1\,186\,560$ \\
        Decoder (3 ConvBlocks $+$ unpooling)& $457\,632$ \\
        Final projection block              & $966$ \\
        RFF encoder                        & $8\,320$ \\
        \midrule
        \rowcolor{ForestGreen!12}
        \textbf{Total ($\ve{u}$ net)}       & $\mathbf{4\,363\,558}$ \\
        \bottomrule
      \end{tabular}
      }
  \end{subtable}
  \hfill 
  \begin{subtable}[t]{0.49\textwidth}
    \centering
    {\small \begin{tabular}[t]{p{4.4cm}r}
        \toprule
        \rowcolor{RoyalBlue!25}
        \textbf{Transformer U-Net} & \textbf{Parameters} \\
        \midrule
        Input embedding                & $4\,160$ \\
        Conditioning projection        & $4\,160$ \\
        Encoder (3 stages)             & $3\,565\,760$ \\
        Bottleneck projection          & $131\,584$ \\
        Bottleneck (3 transformer blocks) & $12\,602\,880$ \\
        Decoder (3 stages)            & $5\,847\,744$ \\
        Output norm $+$ projection     & $323$ \\
        RFF encoders ($\times2$) & $16\,640$ \\
        \midrule
        \rowcolor{ForestGreen!12}
        \textbf{Total ($\ve{u}$ net)}  & $\mathbf{22\,173\,251}$ \\
        \bottomrule
      \end{tabular}
      }
  \end{subtable}
\caption{
  Left: Per-component trainable parameter count for the Graph U-Net velocity
  network. Pooling and unpooling layers are parameter-free
  $k$-nearest-neighbor linear interpolators.
  Right: Per-component trainable parameter count for the Transformer U-Net
  velocity network. Each transformer block contains a cross-attention layer and
  a $\times4$ convolutional feed-forward expansion. The latter dominates
  parameter allocation at the $512$-channel bottleneck.
  }
      \label{tab:graph_breakdown}
\end{table}

    \subsubsection{Inference results}
\begin{figure}[ht!]
        \centering
        \includegraphics[width=1\textwidth]{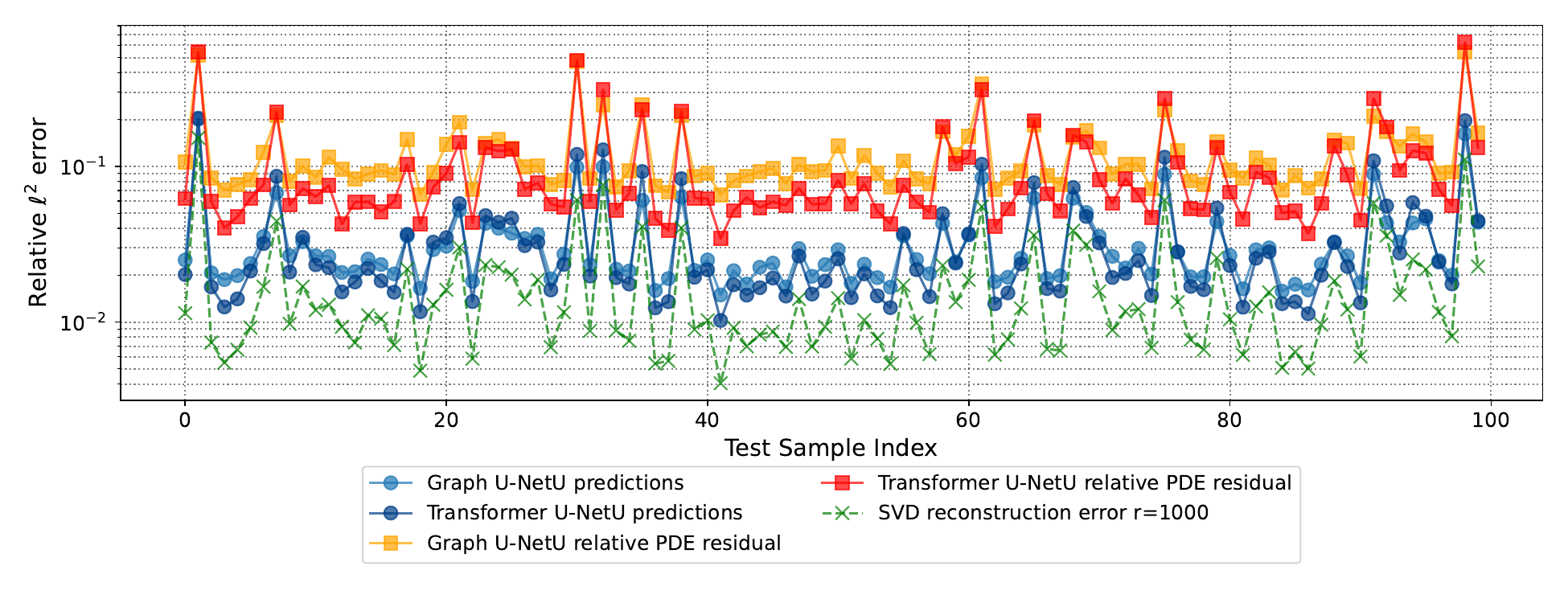}
        \caption{Results on a test set of $100$ random geometries. From values of the velocity field at
            inlet and outlet boundaries and the geometrical encoding, the Transformer U-Net and Graph U-Net predict the velocity field at peak systole on the whole domain.
            }
        \label{fig:inference}
    \end{figure}

\begin{figure}
        \centering
        \includegraphics[width=0.9\linewidth]{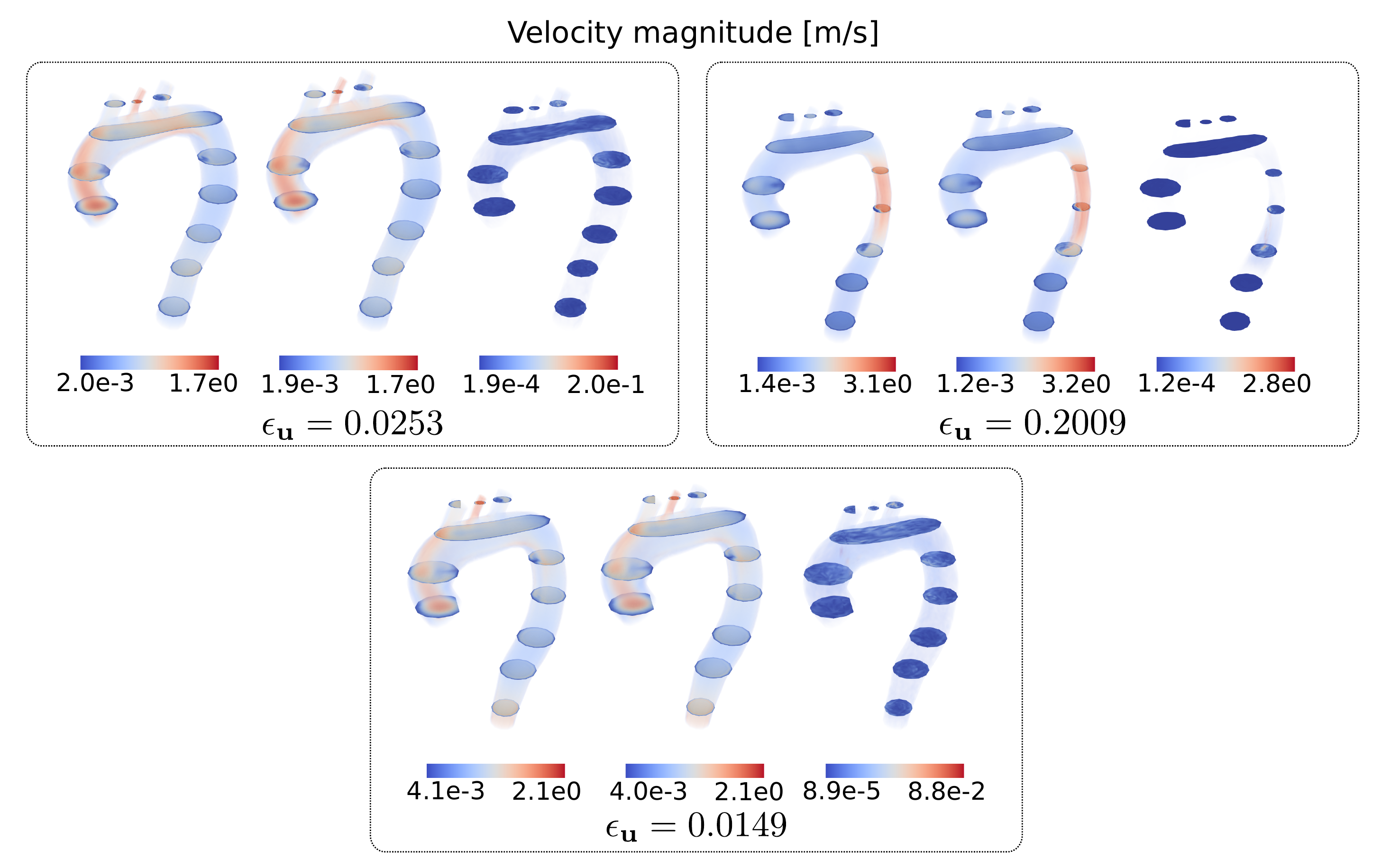}
        \caption{Median, maximum, and minimum relative $\ell^2$ errors of the Graph U-Net velocity prediction.}
        \label{fig:unetu_snaps}
    \end{figure}

    We use the Graph and Transformer U-Nets to infer the different QoIs
    (velocity, pressure, and WSS), focusing on the results at peak systole,
    given that the high-fidelity model's response is most complex here.
    To benchmark the performance of the surrogate model, we compare the
    inference results against the rSVD reconstruction error using $r=1000$ modes
    in terms of relative $\ell^2$ error.
    Figure~\ref{fig:inference} displays the results for the velocity inference.
    The error in the velocity is marginally higher than the rSVD reconstruction
    error, while large values in the reconstruction error coincide with cases
    featuring a large loss. Figure~\ref{fig:inference} also shows the relative
    PDE residual for both architectures, which also aligns well with the cases
    showing the largest losses and reconstruction errors. From these results,
    we conclude that the test cases featuring large losses are
    those in which both the reduced approximation and the inference are
    challenging due to the specific set of geometric hyperparameters.
    Overall, the errors range from 1\% to 20\%, with a mean loss of
    $\approx2.5\%$. It is also observed that, although the
    overall performance of the two methods is comparable, the Transformer
    U-Net slightly yet consistently outperforms the Graph U-Net over all cases.
    The cases leading to the maximum, median, and minimum relative
    $\ell^2$ error for the velocity field predicted by the Graph U-Net are depicted in
    Figure~\ref{fig:unetu_snaps}, which shows that geometric hyperparameters
    leading to a narrowing of the descending aorta correspond to the cases
    with the largest errors.

\begin{figure}[ht!]
        \centering
        \includegraphics[width=1\textwidth]{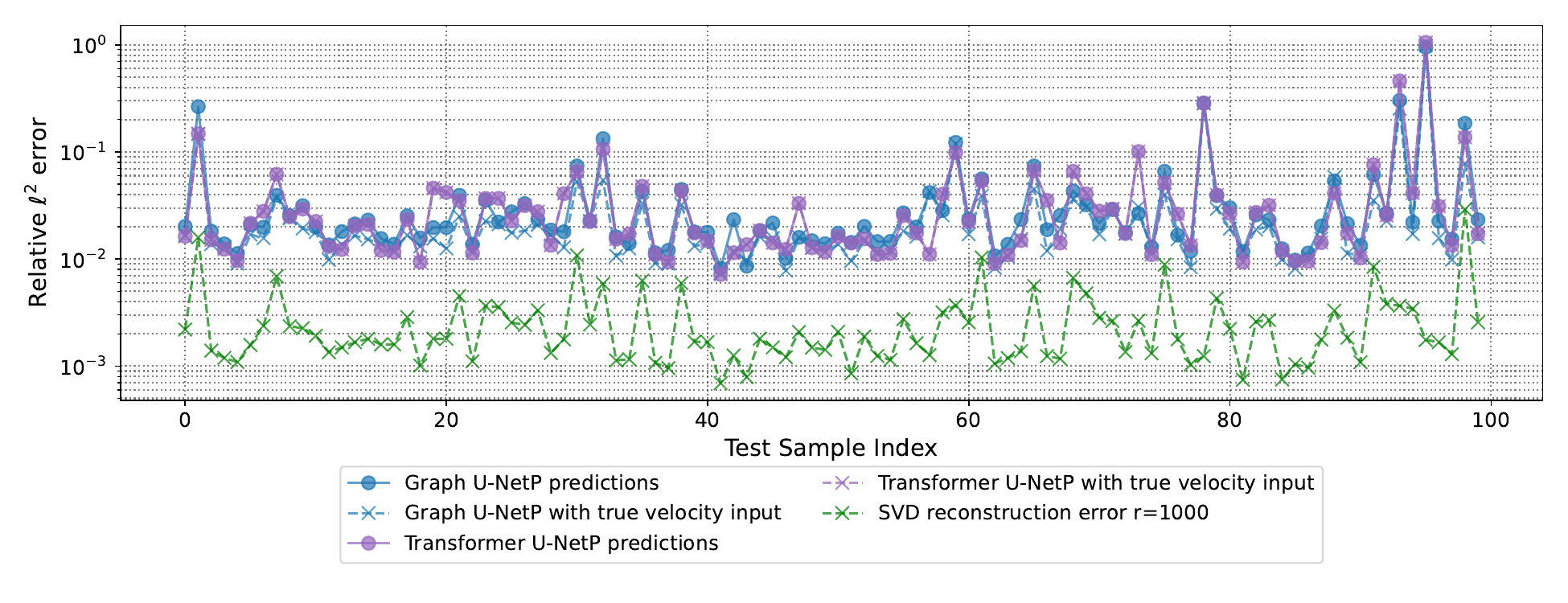}
        \caption{Results on a test set of $100$ random geometries. From values of the velocity field on the whole domain and the
            geometrical encoding, the Graph and Transformer U-Nets predict the pressure field at peak systole on the whole domain.
            The velocity field passed as input can be the full-order velocity field (\textit{U-NetP with full-order velocity input})
            or the one predicted by \textit{U-NetU} (\textit{U-NetP predictions}).
            }
        \label{fig:inferencep}
    \end{figure}

    The errors for pressure inference are reported in Figure~\ref{fig:inferencep}.
    One can draw the following conclusions:
    (i) the prediction errors using the Graph U-Net and the Transformer U-Net are similar,
    (ii) the cases with increased inference error show large rSVD reconstruction errors,
    (iii) the two considered architectures have a very similar performance
    across the dataset, and
    (iv) compared with the velocity inference, the maximum test error is greatly
    increased to $\approx 100\%$, while the mean error lies around $2\%$.

\begin{figure}[ht!]
        \centering
        \includegraphics[width=1\textwidth]{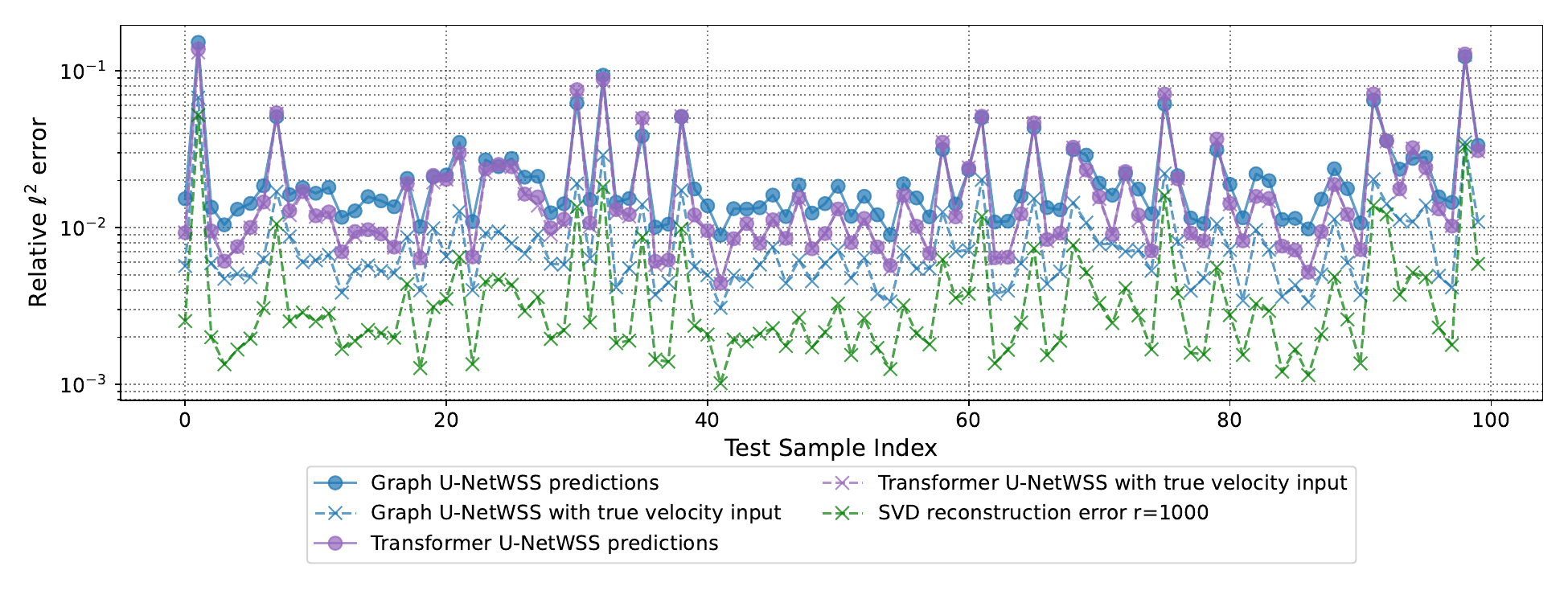}
        \caption{Results on a test set of $100$ random geometries. From values of the velocity field on the whole domain and the
            geometrical encoding, the Graph and Transformer U-Nets predict the WSS field at peak systole on the whole domain.
            The velocity field passed as input can be the full-order velocity field (\textit{U-NetWSS with full-order velocity input})
            or the one predicted by \textit{U-NetU} (\textit{U-NetWSS predictions}).
            }
        \label{fig:inferencewss}
    \end{figure}

    The WSS U-Net results are given in Figure~\ref{fig:inferencewss}.
    In this case as well, we conclude that test cases performing poorly
    (high WSS losses) are characterized by high reconstruction errors.
    The WSS errors show similar trends to the velocity errors, and overall
    error levels lie around $2\%$, while the maximum observed error lies at
    $\approx10.5\%$.
    In the same figure, we also compare the results with those obtained
    by replacing the input velocity field taken from the respective velocity U-Net
    with the full-order solution. This yields a further improvement in the results.
    On the other hand, if the full-order solution is available,
    an approximation of the WSS can be readily obtained directly by postprocessing
    the field, yielding the appropriate standard numerical approximation,
    which serves as reference within the current work.

   From these results, we conclude that Graph and Transformer U-Nets are valuable frameworks for surrogate modeling of aortic flow across a parametrized set of shapes.
   However, the accuracy of the results strongly depends on the type of geometric parametrization.
    Increasing the numerical resolution might be beneficial for the overall framework, removing noise from the snapshot data, but it is currently not computationally feasible.
    The rSVD reconstruction errors and high inference errors in all the quantities are aligned, indicating that certain parameter sets lead to challenging configurations, where the nonlinear physics cannot be captured as easily.

    \subsection{Inter-patient active learning}
    \label{sec:active_learning}

    One of the crucial challenges in inter-patient surrogate modeling is
    sampling efficiency due to the high dimensionality of the space of possible
    aortic shapes. AL is specifically devised to address this
    problem by extending the
    training set with the most informative samples. In this section, we
    analyze the performance of different AL strategies for the considered
    problem.

    \subsubsection{Active learning strategies}
    \label{subsubsec:al_strategies}

    Suppose that an initial training dataset and a pool $\mathcal{P}$ of
    candidate geometries not included in it are available. For each geometry
    $i$, we denote by $\hat{\ve{u}}_i$ the network prediction
    (Graph U-Net or Transformer U-Net) and by $(\ve{u}_i,\,p_i)$ the corresponding
    CFD solution.
    The AL strategies are based on ranking each candidate $i \in \mathcal{P}$
    according to a certain \emph{acquisition score}
    \[
    a: \mathcal P \to \mathbb R\,,
    \]
    and then selecting, for a given $N = |\mathcal{S}|$, the batch
    $\mathcal{S}\subset\mathcal{P}$ achieving the highest score.
    We benchmark several strategies, considering scoring functions based on
    worst error, residual, Fisher information, divergence, active subspaces,
    query by committee, geometry, and random sampling.
    Note that obtaining $(\ve{u}_i,\,p_i)$ by evaluating the CFD model and
    retraining the network with the newly selected snapshot data are both costly
    operations. Therefore, selection criteria based solely on the network
    prediction $\hat{\ve{u}}_i$ can significantly reduce the computational
    cost. The considered acquisition scores are detailed below.

    \paragraph{Worst error} In the first approach, one selects the geometries
    with the highest errors, which are the least well represented in the
    current training set. In this case, the acquisition score depends only on
    the relative error of the predicted velocity with respect to the CFD solution:
    \begin{equation}
        a(i)
        :=
          \frac{%
            \lVert
              \hat{\ve{u}}_i-\ve{u}_i
            \rVert
          }{%
            \lVert
              \ve{u}_i
            \rVert
          }
          \,.
    \end{equation}

    \paragraph{Residual} Assuming that the full-order solution is accurate enough,
    the discrete residual associated with the spatial discretization might be
    close to optimal and hence considered a target.
    Based on this consideration, the idea of the second approach is that the
    relative error in the momentum residual, i.e.,
    \[
    \ve{r}(\ve{u},p) = \frac{\partial}{\partial t}\ve{u}
        +
        \ve{u} \cdot \nabla \ve{u}
        -
        2 \nabla \cdot \left(\nu(\ve{x}) \nabla^\mathrm{S}\ve{u} \right)
        +
        \nabla p
        \,,
    \]
    might be a useful selection criterion. For a given solution $(\ve{u},p)$,
    the residual $\ve{r}(\ve{u},p)$ is assembled on the corresponding domain,
    i.e., the mapped geometry, and the score is defined
    as the associated momentum equation relative error:
    \begin{equation}
        a(i)
        :=
          \frac{%
            \lVert
              \ve{r}(\hat{\ve{u}}_i)
              -
              \ve{r}(\ve{u}_i)
            \rVert
          }{%
            \lVert
              \ve{r}(\ve{u}_i)
            \rVert
          }\,.
    \end{equation}

    \paragraph{Fisher information}
      This approach considers the amount of information of each training sample
      with respect to the model parameters. To this end,
      we consider the normalized energy of the loss gradient.
      Let
      $L_i=\lVert\hat{\ve{u}}_i-\ve{u}_i\rVert$
      denote the prediction error for sample $i$.
      The information score is then defined as
      \begin{equation}
        a(i)
        =
        \sum_{l}
          \frac{%
             \lVert
              \nabla_{\theta_l}L_i
            \rVert^2
          }{%
            \lVert
              \theta_l
            \rVert^2
          }\,,
      \end{equation}
      where the sum extends over all parameter tensors $\theta_l$ of the neural
      network (NN). The numerator measures the sensitivity of the loss to
      perturbations of the corresponding parameters, while the normalization
      by $\lVert\theta_l\rVert^2$ removes the dependence on the parameter scale.
      Consequently, samples that induce large parameter sensitivities receive a
      higher information score, indicating that they carry more information for
      updating the model during training.

    \paragraph{Divergence} This approach is similar to residual scoring
    and considers the magnitude of the discrete divergence of the predicted
    velocity, i.e.,
    \begin{equation}
        a(i)
        =
        \lVert
          \nabla\!\cdot\hat{\ve{u}}_i
        \rVert\,.
    \end{equation}
    This score measures the physical conformity of the network output based
    solely on the predicted velocity and, therefore, entails a low computational
    cost.

    \paragraph{Active subspaces}
      The goal of this approach is to select the best candidates considering the
      geometric variations that have the strongest influence on the prediction
      error. The identification of those geometries relies on an active subspace
      analysis~\cite{doi:10.1137/1.9781611973860}.
      Each geometry is represented in a lower-dimensional space by a parameter
      vector $\ve{x}_i\in\mathbb{R}^{20}$, corresponding to the leading rSVD
      coefficients of the nodal displacement field.

      Based on this dataset, a Gaussian-process surrogate
      $a(i)=\hat e(\ve{x}_i)$ is trained to approximate the velocity prediction
      error as a function of these geometric parameters. The active subspace is
      derived from the covariance matrix of the surrogate gradients,
      \begin{equation}
        C
        =
        \frac{1}{M}
        \sum_{j=1}^{M}
        \nabla_{\ve{x}}\hat e(\ve{x}_j)
        {\nabla_{\ve{x}}\hat e(\ve{x}_j)}^{\!\top}
         =
        W\Lambda W^{\top}
        \,,
      \end{equation}
      where $M$ denotes the number of sampled geometries. The dominant
      eigenvector $\ve{w}_1$ identifies the direction in the parameter space
      along which the predicted error varies most strongly. Each candidate
      geometry is subsequently projected onto this one-dimensional active
      coordinate, $z_i=\ve{w}_1^{\!\top}\ve{x}_i$. The projected samples are
      partitioned using $k$-means clustering, with the number of clusters
      selected according to the silhouette criterion. Finally, the desired
      number of samples $N$ is obtained through uniform sampling of all
      clusters, ensuring that the selected geometries cover the dominant
      error-sensitive directions in the parameter space.

    \paragraph{Query by committee}
    In this approach, the selection is based on an estimate of the epistemic
    uncertainty of the surrogate model. We employ a query-by-committee strategy
    based on Monte Carlo DropConnect. For each candidate geometry $i$, $S$
    stochastic forward passes are performed, yielding predictions
    $\hat{\ve{u}}_i^{(s)}$, $s=1,\ldots,S$, corresponding to different random
    realizations of the network weights. The acquisition score is defined as the
    mean predictive variance over all mesh nodes, i.e.,
      \begin{equation}
        a(i)
        =
        \frac{1}{|V_i|}
        \sum_{v\in V_i}
        \operatorname{Var}_s
        \!\left[
        \hat{\ve{u}}_i^{(s)}(v)
        \right]\,,
      \end{equation}
      where $V_i$ denotes the set of mesh nodes of sample $i$, and
      $\operatorname{Var}_s$ denotes the sample variance over the $S$
      stochastic forward passes. A large predictive
      variance indicates strong disagreement among the committee members and,
      consequently, high model uncertainty. Samples with the largest acquisition
      scores are therefore considered the most informative and are preferentially
      selected for inclusion in the training set.

    \paragraph{Geometry}
      To increase geometric diversity within the training set, in this method,
      candidate geometries are ranked according to their distance from the
      geometries already contained in the current training set $\mathcal{T}$.
      Each candidate geometry $i$ is represented by its nodal displacement
      vector $\ve{d}_i$ with respect to the reference configuration. The
      acquisition score is defined as the mean Euclidean distance to all
      training samples,
      \begin{equation}
        a(i)
        =
        \frac{1}{|\mathcal{T}|}
        \sum_{j\in\mathcal{T}}
        \lVert
        \ve{d}_i-\ve{d}_j
        \rVert\,.
      \end{equation}
      High acquisition scores correspond to geometries distinct from those
      already present in the training set. Therefore, selecting samples with the
      largest scores promotes broad coverage of the geometric
      parameter space and reduces redundancy among the selected geometries.

    \paragraph{Random}
      As a baseline strategy, additional geometries are randomly selected from
      the pool of available samples $\mathcal{P}$, i.e.,
      $
        \mathcal{S}\sim\mathcal{U}(\mathcal{P}),
      $
      such that every candidate geometry has the same probability of being
      selected. This strategy serves as a reference for evaluating the
      effectiveness of the other acquisition methods.

    \subsubsection{Numerical results}
    \label{subsubsec:al_numerical_results}

\begin{figure}[htp!]
      \centering
      \includegraphics[width=\linewidth]{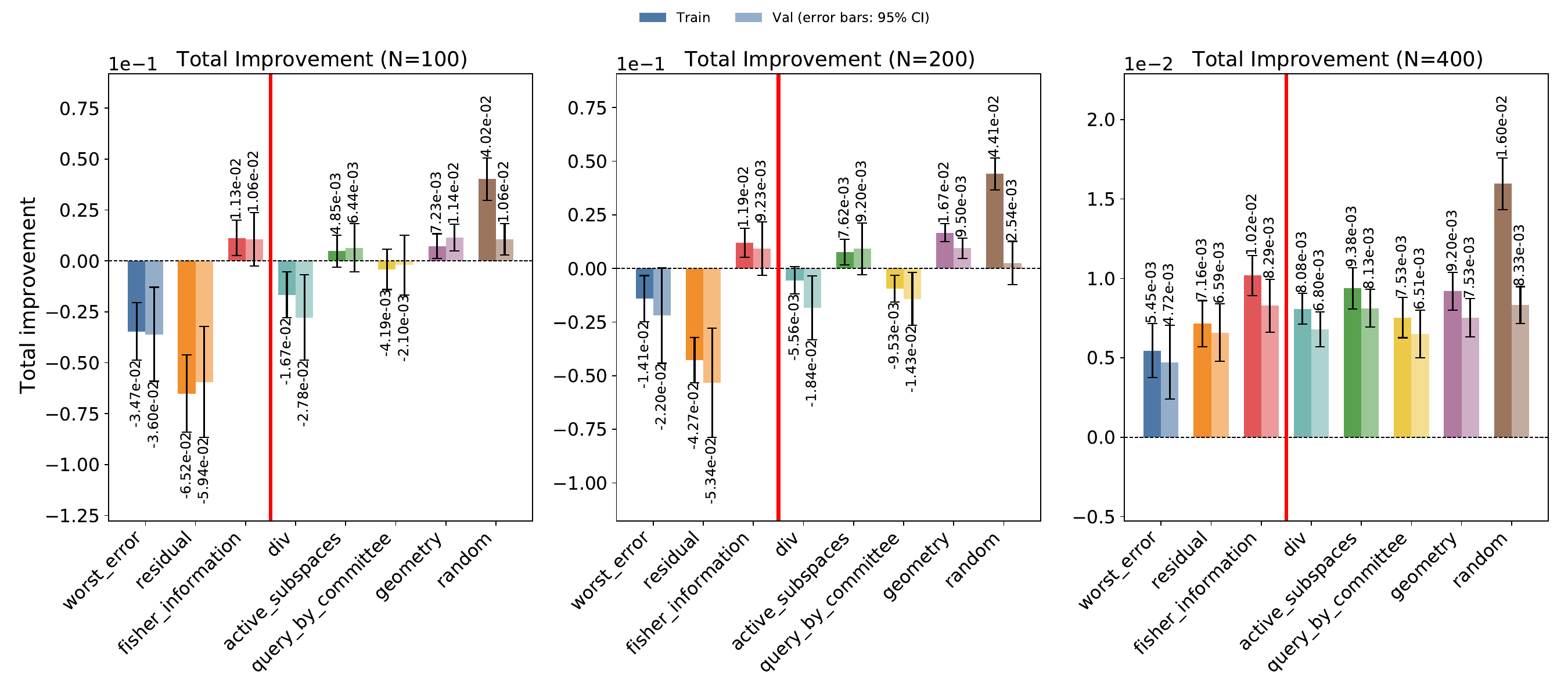}
      \caption{Warm-up regime: improvement in training and validation of the
      velocity field for $n_{\text{train}}^{\text{before}}\in\{100, 200, 400\}$
      and increasing batch size $N=n_{\text{train}}^{\text{before}}$.
      Error bars denote the $95\%$ confidence interval of the mean
      ($1.96\,\sigma/\sqrt{N}$). Positive values indicate lower errors after
      adding the selected batch to the surrogate model. The red line separates
      label-dependent strategies (worst error, residual, Fisher information)
      from label-free ones (divergence, active subspaces, query by committee,
      geometry, random).}
      \label{fig:al_sum}
    \end{figure}

    To evaluate the different strategies, we define the
    \emph{total improvement} as the difference of the relative errors for the
    velocity field before and after the AL step:
    \begin{equation}
        \text{Total Improvement} 
        = 
        \frac{1}{n}\sum_{i=1}^{n}
        \left(
          \epsilon_{\ve{u}, i}^{\text{before}} - \epsilon_{\ve{u}, i}^{\text{after}}
        \right)
        \,,
    \end{equation}
    where the superscripts before and after denote the results without and with the additional
    sample set $\mathcal{S}$. Here, only the Graph U-Net is considered.
    A positive value of the total improvement denotes an error reduction.
    The average is taken over the $n=n_{\text{test}}=100$ held-out geometries
    introduced in Section~\ref{sec:filtering} (and also over the
    $n=n_{\text{train}}^{\text{after}}$ training cases in
    Figure~\ref{fig:al_sum}), i.e., over the same set used to report the
    inference results in Section~\ref{sec:inference}: no separate validation
    split is employed, and the terms \textit{validation} and \textit{test}
    are used interchangeably in the following. Both the initial training sets of
    size $n_{\text{train}}^{\text{before}}$ and the candidate pool
    $\mathcal{P}$ are selected from the remaining $n_{\text{train}}=4560$
    training geometries, ensuring that no candidate is ever part of the evaluation set.

    Notice that the first three strategies (worst error, residual,
    Fisher information) are label-dependent, i.e., they require the calculation
    of the corresponding CFD solution, and will act in the following study as
    an oracle reference.
    The label-free ones (divergence, active subspaces, query by committee, geometry, random)
    only require an approximation with the surrogate, and are thus usable in deployment.

    \paragraph{Warm-up regime} In the first set of tests, the AL
    strategies are applied to small training sets,
    $n_{\text{train}}^{\text{before}}=\{100, 200, 400\}$, with a batch of size
    $N=n_{\text{train}}^{\text{before}}$ that doubles the set, i.e.,
    $n_{\text{train}}^{\text{after}}=\{200, 400, 800\}$.
    Figure~\ref{fig:al_sum} reports the improvement in training and validation
    compared to the model trained without the additional batch. Error- and
    residual-driven acquisition (worst error, residual, divergence)
    \emph{degrades} validation at the smallest batches ($\epsilon_{\ve{u}}$
    improvement $-0.036$ for worst error and $-0.059$ for residual at $N=100$),
    as it oversamples geometric outliers, and recovers to marginally positive values
    only for the largest batch size $N=400$.
    The diversity/uncertainty strategies (geometry, Fisher information, active
    subspaces) yield small but consistently positive gains, yet none of the AL
    strategies beats random selection on validation.
    For the smallest training set and batch size ($N=100$), Fisher information
    and random are comparable ($\approx+0.01$), and for $N=400$ all strategies
    collapse to parity ($+0.005$ to $+0.008$).
    The results therefore suggest that the additional effort associated with the
    AL strategies does not pay off in a scenario with low data
    availability.

\begin{figure}[htp!]
        \centering
        \includegraphics[width=\linewidth]{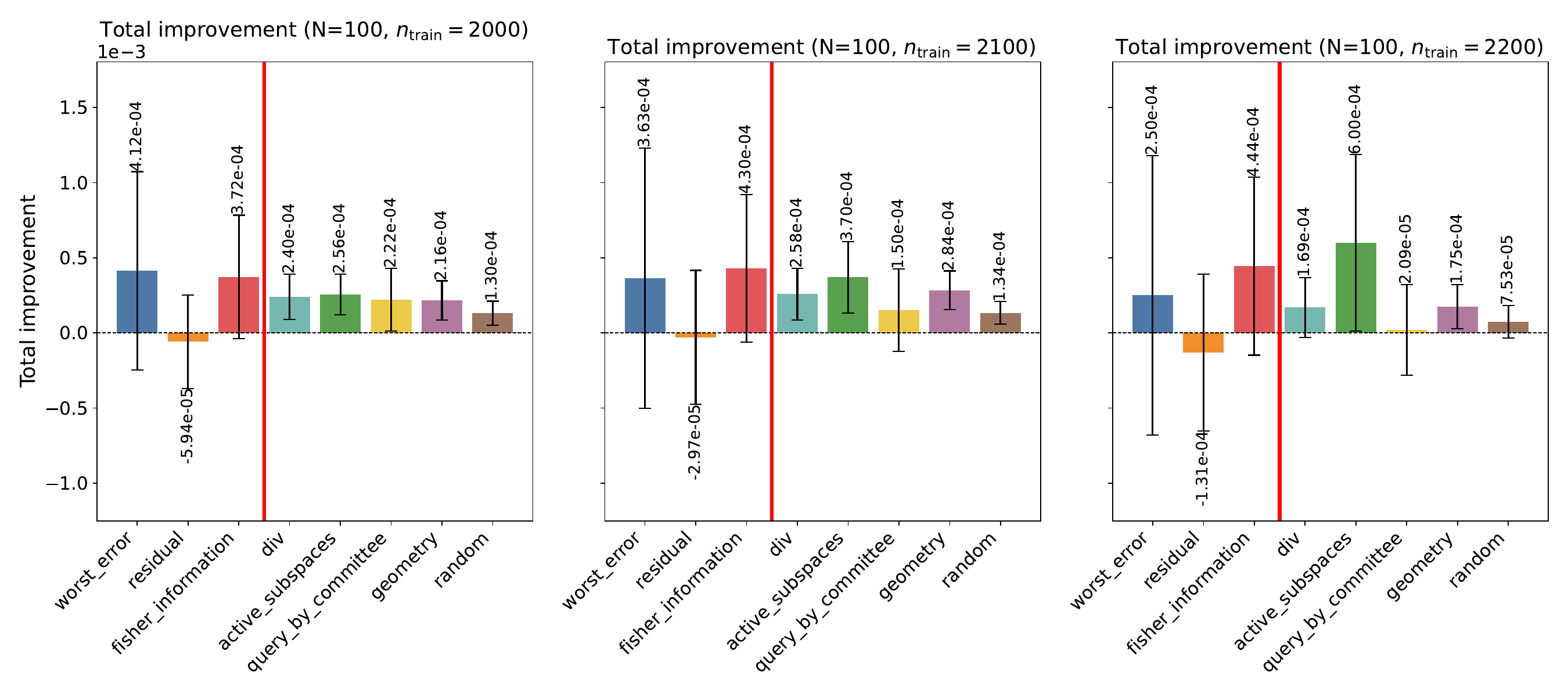}
        \caption{
          Fine-tuning regime: validation total improvement when AL
          is applied from $n_{\text{train}}=2000$ consecutively for $3$ rounds
          with the same batch size $N=100$. Error bars denote the $95\%$
          confidence interval of the mean ($1.96\,\sigma/\sqrt{N}$). Positive
          values indicate lower errors after adding the selected batch to the
          NN model. The red line separates label-dependent strategies
          (worst error, residual, Fisher information) from label-free ones
          (divergence, active subspaces, query by committee, geometry, random).
          The three panels share the same vertical scale.
        }
        \label{fig:al_late}
    \end{figure}

    \paragraph{Fine-tuning regime} In the second set of tests, we start from a
    data-rich model, $n_{\text{train}}^{\text{before}}=2000$, and add three
    consecutive batches of size $N=100$
    (i.e., $n_{\text{train}}^{\text{after}} 
    = 2000\!\to\!2100\!\to\!2200\!\to\!2300$).
    Figure~\ref{fig:al_late} reports the validation total improvement after
    each round. Gains are two orders of magnitude smaller ($\sim\!10^{-4}$)
    than in the warm-up regime, reflecting saturation, but targeting now clearly
    outperforms random selection (the weakest positive baseline, $\sim\!10^{-4}$).
    Residual is the only strategy that consistently \emph{worsens} the model
    (negative total improvement at every round), while query by committee
    decays to nearly zero. Worst error dominates the first round ($4.1\times10^{-4}$)
    but fades out in the following stages, whereas Fisher information and active
    subspaces give the most consistent and growing gains, the latter becoming
    best by the third round ($6.0\times10^{-4}$). Thus, in the data-rich regime,
    gradient- and subspace-based acquisition is preferable.

    These results indicate that the benefit of AL in the 
    application at hand depends heavily on the specific regime.
    During the data-scarce warm-up phase, sophisticated acquisition strategies do
    not outperform random sampling, whereas in the data-rich fine-tuning phase,
    gradient- and subspace-based methods provide the largest improvements.

    \subsection{Physical consistency as a soft constraint}
    \label{sec:physics_consistency}
    The NN surrogate models analyzed in the previous sections are trained to
    minimize the data misfit, but do not explicitly enforce the governing equations.
    To incorporate physical consistency during optimization, we augment the
    training objective with a differentiable residual-based penalty obtained by
    evaluating the weak form of the incompressible Navier--Stokes equations.
    The NN predictions are interpreted as finite element DoFs, allowing the
    residual and its gradients with respect to the predicted fields to be
    assembled efficiently via automatic differentiation. Within this section,
    we consider only the Graph U-Net for the sake of brevity, noting that no
    substantial differences in the results are expected when using the
    Transformer U-Net.

  \subsubsection{Velocity-dependent residual}

    For the physical consistency studies, we restrict ourselves to the surrogate
    model that predicts the velocity field. To compute the residual for a given
    solution $\ve{u}$, we consider the velocity-dependent part of the weak
    formulation in the interior domain, i.e.,
    \begin{equation}
        \label{eq:weakform}
        \mathcal{R}(\ve{u}; \ve{v}, q)
        =
        \int_\Omega
          \left(
            \ve{u}\cdot\nabla\ve{u}
          \right)
          \cdot \ve{v}
          +
          2
          \nu(\dot\gamma)\, \grads \ve{u} : \nabla \ve{v}
          +
          q \nabla\cdot\ve{u}~\mathrm{d}\ve{x},
          \quad
          \mbox{for}\; (\ve{v}, q) \in V\times Q\,,
    \end{equation}
    where $V$ and $Q$ denote the function spaces for velocity and pressure,
    respectively. In particular, we consider $V\subset H^{1}_0(\Omega)$,
    hence restricting to test functions vanishing on the boundary.

    The contributions to the residual of the pressure and time-derivative terms
    are not computed as a function of $\ve{u}$, but are collected in a
    precomputed right-hand side $\ve{b}$ assembled from the full-order velocity
    and pressure fields $(\ve{u}_h^\star, p_h^\star)$, i.e.,
     \begin{equation}
        \label{eq:time_remainder}
        \int_\Omega
        \partial_t\ve{u}_h^\star \cdot \ve{v}_h~\mathrm{d}\ve{x}
        =
        -
        \int_\Omega
        \left(
        \ve{u}_h^\star\cdot\nabla\ve{u}_h^\star
        \right)
        \cdot \ve{v}_h
        +
        2
        \nu(\dot\gamma^\star)\, \grads \ve{u}_h^\star :\nabla\ve{v}_h
        - p_h^\star \nabla\cdot\ve{v}_h~\mathrm{d}\ve{x}\,,
    \end{equation}
    and
    \begin{equation}
        \label{eq:rhs}
        \ve{b}(\ve{u}_h^\star, p_h^\star; \ve{v}_h)
        =
        \int_\Omega
          p_h^\star\, \nabla\cdot\ve{v}_h
          -
          \partial_t\ve{u}_h^\star \cdot \ve{v}_h~\mathrm{d}\ve{x}\,.
    \end{equation}
    Combining Equations~\eqref{eq:weakform}, \eqref{eq:time_remainder}, and \eqref{eq:rhs},
    the physical consistency penalty is defined as the squared norm of the
    assembled residual. Thus
    \begin{equation}
        \label{eq:residual}
        r(\ve{u}_h)
        =
        \lVert
          \mathcal{R}(\ve{u}_h; \ve{v}_h, q_h)
          -
          \ve{b}(\ve{u}_h^\star, p_h^\star; \ve{v}_h)
        \rVert^2
        \,,
    \end{equation}
    where $\ve{v}_h$ and $q_h$ are the standard discrete test functions from the
    finite element space living on $\Omega_h$. Note that the physical
    consistency penalty vanishes when the predicted velocity $\ve{u}_h$
    coincides with the full-order velocity $\ve{u}_h^\star$.
    Altogether, the loss function takes the form
    \begin{align}
      \label{eq:piloss}
      \mathcal{L}(\theta)
      =
      \frac{1}{n_{\text{batch}}}
      \sum_{i=1}^{n_{\text{batch}}}
      \Bigg[
        \underbrace{
          \frac{
            \big\lVert
              \hat{\ve{u}}_{h,i}(\theta) - \ve{u}^\star_{h,i}
            \big\rVert_{\ell^2}
          }{
            \big\lVert
              \ve{u}^\star_{h,i}
            \big\rVert_{\ell^2}
          }
        }_{\epsilon_{\ve{u}, i}(\theta)}
        +
        \beta
        \underbrace{
          \frac{
            r\big(\hat{\ve{u}}_{h,i}(\theta)\big)
          }{
            \big\lVert
              \ve{b}(\ve{u}^\star_{h,i}, p^\star_{h,i}; \ve{v}_h)
            \big\rVert^2
          }
        }_{\text{physical consistency}}
      \Bigg]
      \,,
    \end{align}
    where $n_{\text{batch}}$ is the batch size, $\theta$ collects the trainable
    network parameters,
    $\hat{\ve{u}}_{h,i}(\theta)$ denotes the predicted velocity for the $i$-th
    training geometry, interpreted as a finite element function, and
    $(\ve{u}^\star_{h,i}, p^\star_{h,i})$ is the corresponding full-order solution.
    The first term is the relative $\ell^2$ misfit $\epsilon_{\ve{u}, i}$ defined
    in Equation~\eqref{eq:err}, here evaluated for the surrogate prediction rather than for
    the rSVD reconstruction, while the second term is the physical consistency
    penalty in Equation~\eqref{eq:residual}, normalized by the precomputed right-hand side.
    Both contributions are therefore dimensionless and of comparable magnitude.
    The weight $\beta\geq0$ balances the deviation from the data against physical
    consistency, with $\beta=0$ recovering the purely data-driven reference case.
    We employ $\beta=1$; for other possible approaches, see~\cite{Liu2025ConFIG}.

    \begin{remark}
      The right-hand side $\ve{b}$ in Equation~\eqref{eq:rhs} is assembled from the full-order
      velocity and pressure fields, so that the penalty in Equation~\eqref{eq:residual} is, in the
      terminology of Section~\ref{subsubsec:al_numerical_results},
      \emph{label-dependent}: it can be evaluated only on samples for which the CFD
      solution is available, and it vanishes exactly at the full-order solution.
      This is a consequence of the design of the surrogate models rather than of the
      formulation itself. The networks introduced in Section~\ref{sec:inference}
      predict the velocity field at a single time instant (peak systole), so that
      neither the temporal derivative $\partial_t\ve{u}$ nor the pressure $p$ can be
      recovered from the prediction alone; both are therefore supplied by the reference
      solution through $\ve{b}$.
    \end{remark}

    The surrogate models introduced in Section~\ref{sec:inference} predict nodal
    velocity and pressure values on the shared reference grid. To compute the
    penalty term, we interpret these fields as the DoFs
    of a piecewise linear finite element function, i.e., the coefficients of the
    finite element fields with respect to basis functions with support points
    on the mesh vertices.
    In practice, the reference mesh is first deformed onto the particular
    patient-specific geometry using the nodal displacements predicted alongside
    the flow fields, and the resulting coefficient vector is passed as the
    evaluation point to an assembly routine.

    Specifically, for each cell, we compute the interpolation of $\ve{u}_h$ and
    $\nabla\ve{u}_h$ at the quadrature points from the nodal values via the
    finite element basis and accumulate the discrete weak residual defined
    in Equation~\eqref{eq:residual}.
    The element residual is registered with respect to the local nodal velocity
    values and differentiated by forward-mode automatic differentiation via
    \texttt{Sacado}~\cite{Phipps2021}. The same pass returns both the global
    residual vector and its Jacobian, such that the squared residual norm
    enters the loss as a soft penalty for physical consistency, while the
    transpose Jacobian-vector product propagates its gradient back into the
    network. The implementation also takes into account batched predictions
    from the NN surrogate models.

    \subsubsection{Verification of convergence in a single geometry}
    \label{subsubsec:physics_consistency_single_geometry}
\begin{figure}[!htbp]
      \centering
      \begin{minipage}[c]{0.66\linewidth}
          \centering
          \includegraphics[width=1\linewidth]{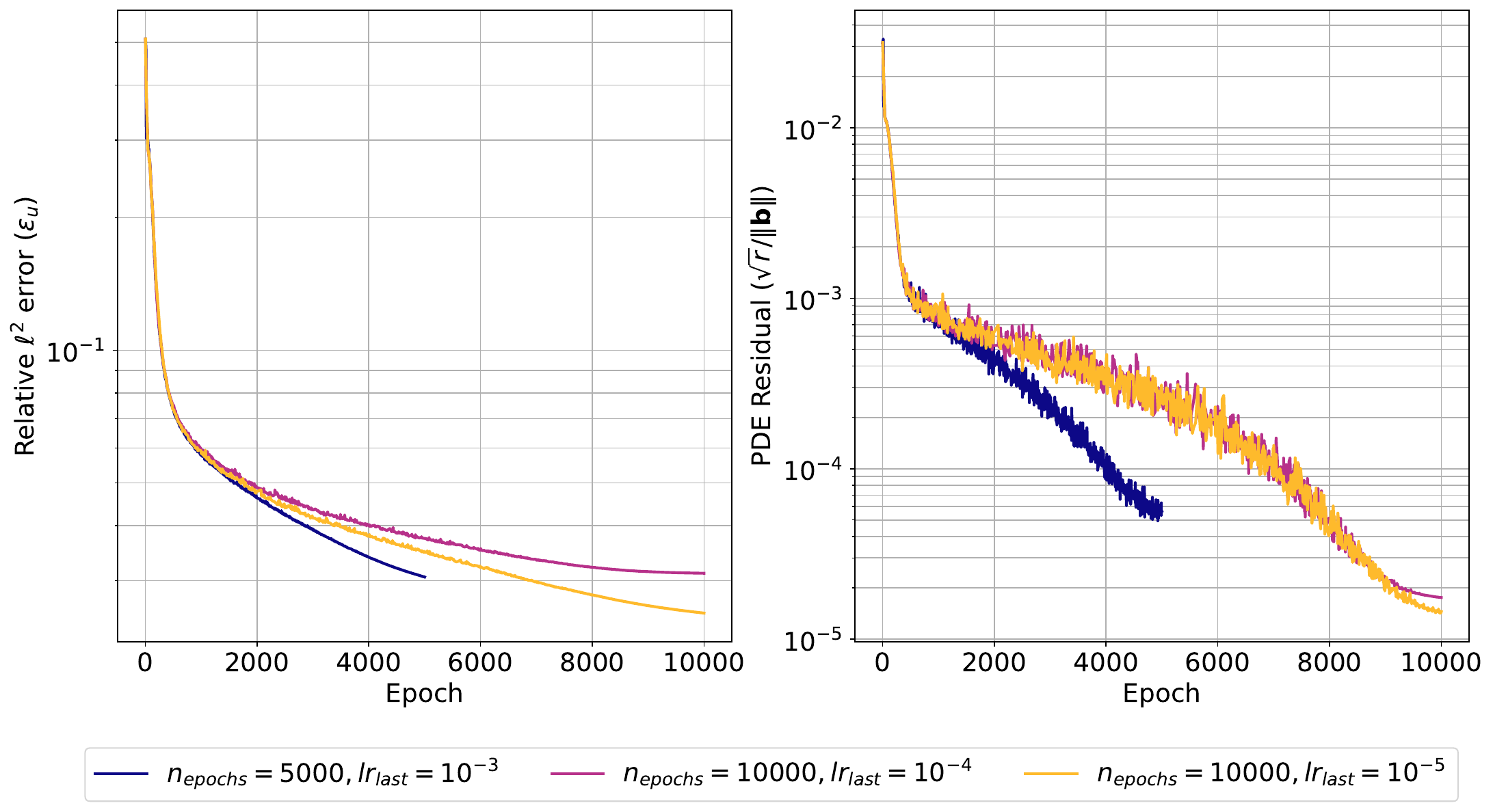}
      \end{minipage}\hfill
      \begin{minipage}[c]{0.32\linewidth}
          \centering
          \footnotesize
          \setlength{\tabcolsep}{4pt}
          \begin{tabular}{lrrr}
              \toprule
              \multirow{2}{*}{$\mathit{lr}_{\text{last}}$} & time  & time/iter. & epochs \\
                                                  & in min & (avg.) in s & \\
              \midrule
              $10^{-3}$ & $55$ & $0.662$ & $5000$\\
              $10^{-4}$ & $98$ & $0.592$ & $10000$\\
              $10^{-5}$ & $88$ & $0.529$ & $10000$\\
              \bottomrule
          \end{tabular}
      \end{minipage}
      \caption{
        Physical consistency loss convergence for a single geometry. The SOAP
        optimizer and the cosine annealing scheduler with initial and final
        learning rate $\mathit{lr}_{\text{start}}=10^{-2}$,
        $\mathit{lr}_{\text{last}}\in\{10^{-3}, 10^{-4}, 10^{-5}\}$, and
        number of epochs $n_{\text{epochs}}\in\{5000, 10000, 10000\}$ are
        employed. The table (right) reports the total wall-clock time and
        the average wall-clock time per iteration for each run.
      }
      \label{fig:pinn_single_geo}
  \end{figure}

As a preliminary step, we validate the physical consistency loss for a fixed single geometry:
\begin{align}
  \label{eq:resconv}
      \mathcal{L}(\theta)
      =
        \frac{
          r\big(\hat{\ve{u}}_{h}(\theta)\big)
        }{
          \big\lVert
            \ve{b}(\ve{u}^\star_{h}, p^\star_{h}; \ve{v}_h)
          \big\rVert^2
        }
      \,.
    \end{align}
For this convergence study, instead of an NN surrogate model for the velocity,
we solve for the
velocity field $\ve{u}_h$ by minimizing the residual in
Equation~\eqref{eq:resconv}. Figure~\ref{fig:pinn_single_geo} reports the
results for different numbers of training epochs and final learning rates
$\mathit{lr}$ within the cosine annealing scheduler of the SOAP
optimizer~\cite{Vyas2025SOAP}, showing
that the relative $\ell^2$ error and the PDE residual converge.
The accuracy could be improved by multi-stage training~\cite{WANG2024112865}
or with other second-order optimizers.
 The optimization converges up to an accuracy of $2\%$ in terms of
 $\epsilon_{\ve{u}}$, which validates the correctness of the implementation of
 the physical consistency loss and its gradient.

    \subsubsection{Effectiveness in training and fine-tuning}
    \label{subsubsec:physics_consistency_training_finetuning}

    The physical consistency loss can either be enforced throughout training or
    introduced only after a purely data-driven model has converged. We therefore
    investigate its effectiveness in both scenarios, considering the
    Graph U-NetU trained from scratch in low- and high-data regimes
    ($n_{\text{train}}=50$ and $n_{\text{train}}=500$), as well as the
    fine-tuning of the best pretrained model.

    \paragraph{Physical consistency during training}
    We compare the accuracy of the Graph U-NetU trained including the physical
    consistency term with the baseline model trained only with the data term.
    In the physically consistent variant, the loss to be minimized has two
    terms, the relative $\ell^2$ error $\epsilon_{\ve{u}}$ and the relative
    residual $r(\ve{u}_h)/\lVert\ve{b}\rVert^2$, as in
    Equation~\eqref{eq:piloss}.

\begin{figure}[!htb]
        \centering
        \includegraphics[width=0.49\linewidth]{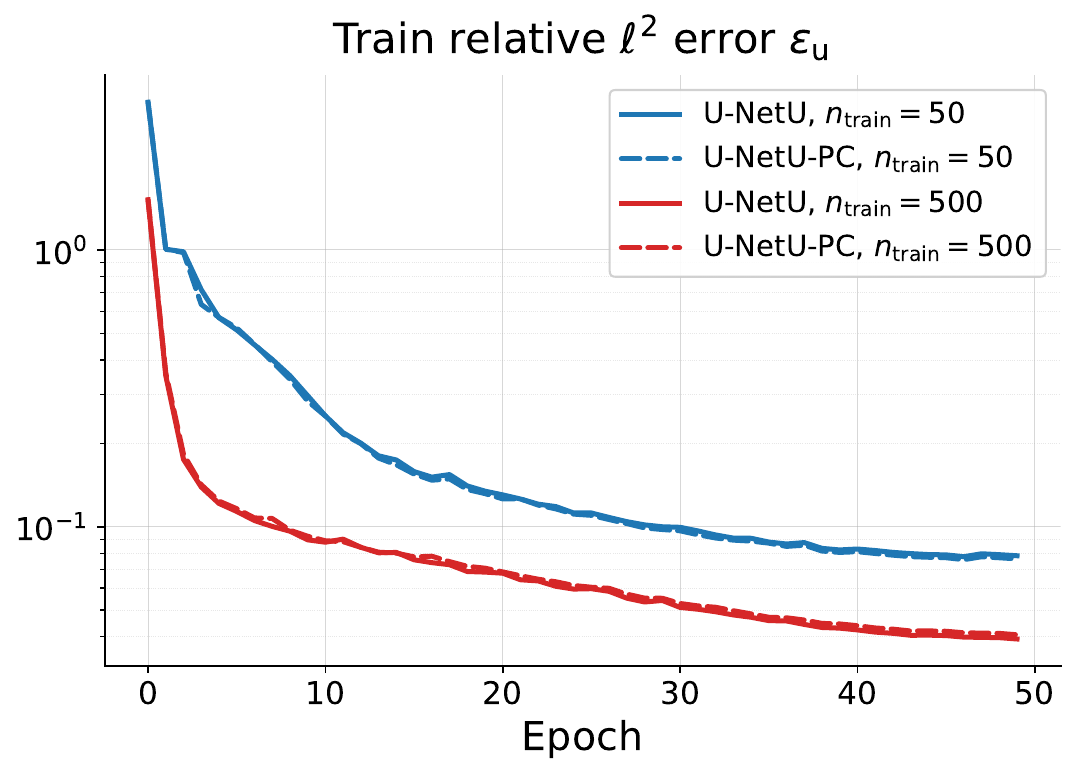}
        \includegraphics[width=0.49\linewidth]{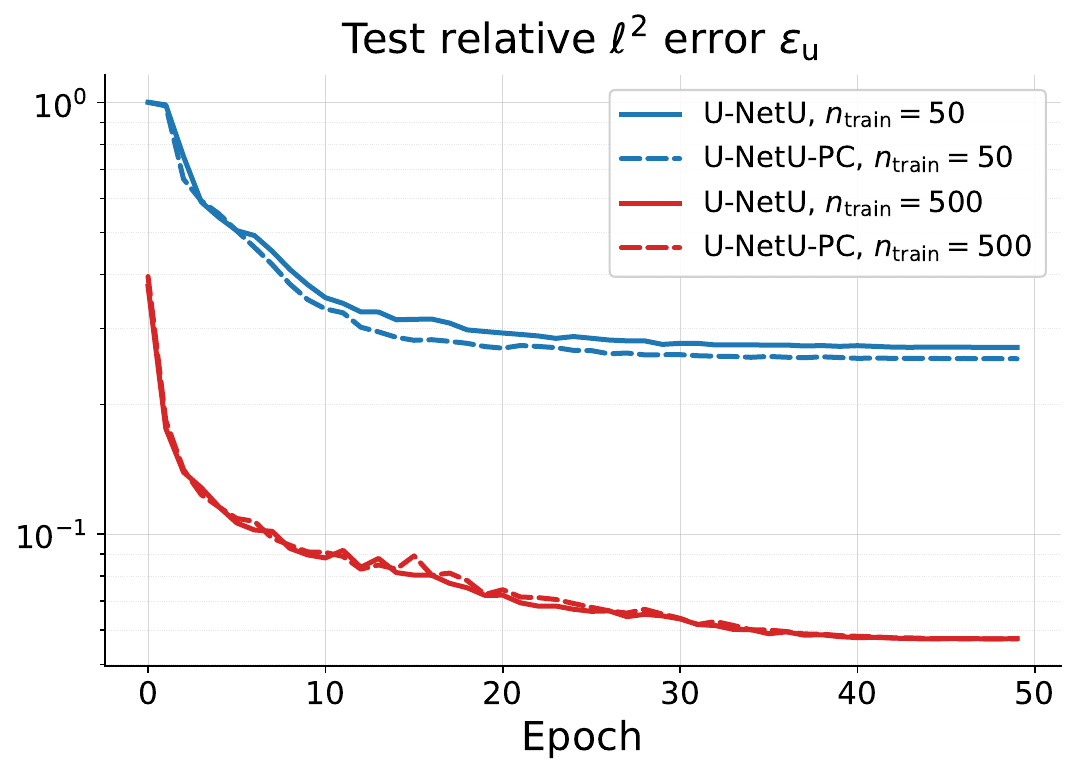}\\[1ex]
        \begin{minipage}[c]{0.49\linewidth}
            \centering
            \includegraphics[width=\linewidth]{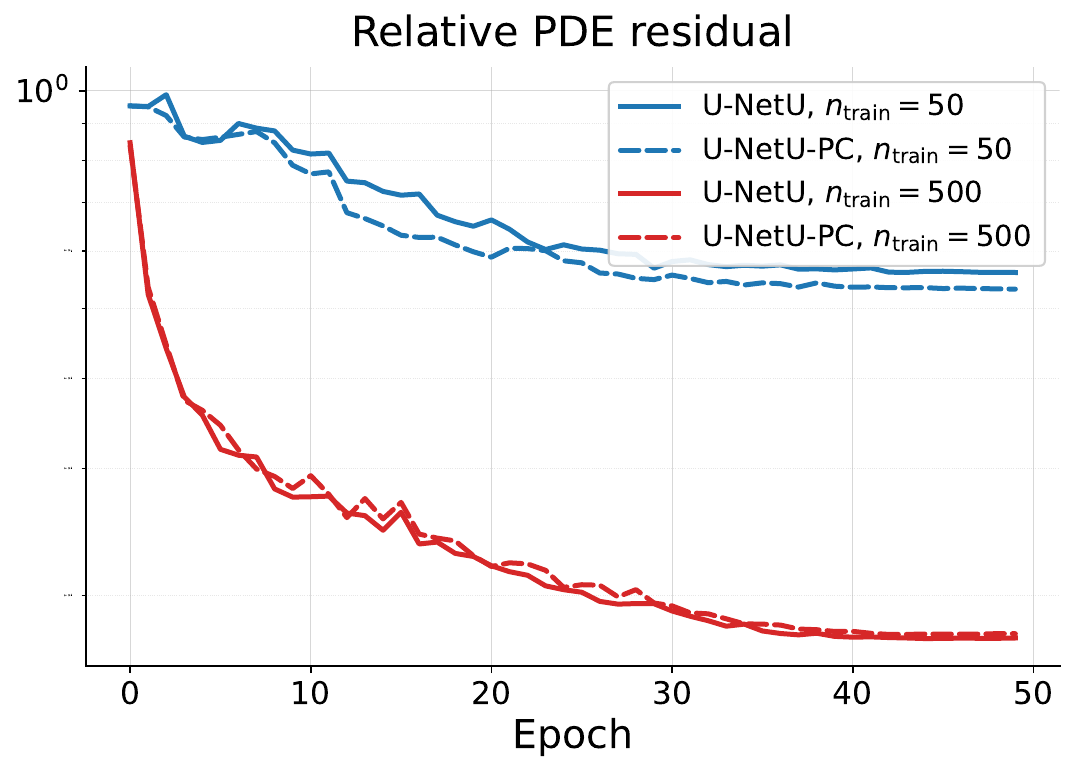}
        \end{minipage}\hfill
        \begin{minipage}[c]{0.49\linewidth}
            \centering
            \begin{tabular}{lrrr}
                \toprule
                          & \multirow{2}{*}{$n_\text{train}$}
                                  & time    & time/iter. \\
                          &       & in h    & (avg.) in s \\
                \midrule
                U-NetU    & $50$  & $2.03$  & $1.46$ \\
                U-NetU-PC & $50$  & $5.60$  & $4.04$ \\
                U-NetU    & $500$ & $8.44$  & $0.61$ \\
                U-NetU-PC & $500$ & $40.40$ & $2.91$ \\
                \bottomrule
            \end{tabular}
        \end{minipage}

        \caption{
          Physical consistency (PC) during Graph U-NetU training, for the
          low-data ($n_{\text{train}}=50$, blue) and high-data
          ($n_{\text{train}}=500$, red) regimes. Each curve compares the
          baseline U-NetU (solid) against its physics-informed variant
          U-NetU-PC (dashed) at the same data regime (same color).
          Top: training and test relative $\ell^2$ velocity errors.
          Bottom-left: relative PDE residual.
          Bottom-right: wall-clock times.
        }
        \label{fig:pinn_training}
    \end{figure}

    The results are reported in Figure~\ref{fig:pinn_training}. The physical
    consistency loss yields a mild but consistent gain only in the low-data
    regime, while in the high-data regime the data term already dominates and
    the two variants are essentially indistinguishable. In terms of wall-clock
    time,
    enforcing the physical consistency loss increases the per-iteration cost by
    roughly $3$--$5\times$ over the purely data-driven baseline. Although the
    evaluation of the finite element residual inevitably incurs additional
    computational cost, it should be noted that the timings reported here are
    representative only of the present implementation, which has not yet been
    optimized for assembly and automatic differentiation.

    \begin{remark}
    Recent work has proposed adaptive weighting and staged optimization
    strategies to mitigate conflicts between the data-fitting and physical
    consistency losses~\cite{Liu2025ConFIG}. Such approaches, however, address
    optimization rather than computational efficiency, since they continue to
    require an evaluation of the residual at each training step. Furthermore,
    near convergence the two objectives become largely compatible, as a
    reduction in the prediction error also reduces the residuals of the
    governing equations.
    \end{remark}

    \paragraph{Physical consistency during fine-tuning}
    Next, we test the soft physics constraint in the loss of the
    Graph U-NetU model as a fine-tuning procedure. Starting from the best
    data-driven Graph U-Net (whose weights are frozen), a lightweight
    feed-forward correction network is trained with the physical consistency
    loss to predict a nodal velocity correction that is added to the frozen
    prediction on the interior nodes, while the boundary and wall velocities
    are kept at their prescribed values. Thus
    \begin{align*}
      \mathcal{L}(\theta)
      =
      \frac{1}{n_{\text{batch}}}
      \sum_{i=1}^{n_{\text{batch}}}
      \Bigg[
        \beta
        \underbrace{
          \frac{
            r\big(\hat{\ve{u}}_{h,i}(\theta)\big)
          }{
            \big\lVert
              \ve{b}(\ve{u}^\star_{h,i}, p^\star_{h,i}; \ve{v}_h)
            \big\rVert^2
          }
        }_{\text{physical consistency}}
      \Bigg]\,.
    \end{align*}
    For the fine-tuning studies, the full training dataset with
    $n_{\text{train}}=4560$ is employed.

    As in the two main architectures, the per-node geometric input
    ($8$ components: the $5$-dimensional centerline encoding and the
    $3$-dimensional deformed position) is first mapped by an RFF encoding
    (with $2$ features over $16$ scales with standard deviation $\sigma=1$) to a
    $64$-dimensional embedding and refined by a two-layer FFN ($64\!\to\!64$
    map with sigmoid linear unit activation function). The correction network
    is a four-layer multilayer perceptron with three hidden layers of width
    $512$ and sigmoid linear unit activation functions
    ($64\!\to\!512\!\to\!512\!\to\!512\!\to\!3$). The whole module is trained
    with the SOAP optimizer and a cosine annealing scheduler with an initial
    learning rate of $10^{-2}$ and a final learning rate of $10^{-4}$.
    \begin{figure}[!htb]
        \centering
        \begin{tabular}{lrrr}
            \toprule
                & \multirow{2}{*}{epochs}
                         & time & time/iter.\\
                &        & in h & (avg.) in s\\
            \midrule
            U-NetU-PC & $10$  & $12.38$ & $1.06$ \\
            U-NetU    & $100$ & $14.84$ & $0.59$ \\
            \bottomrule
        \end{tabular}\\[1ex]
        \includegraphics[width=1\linewidth]{
          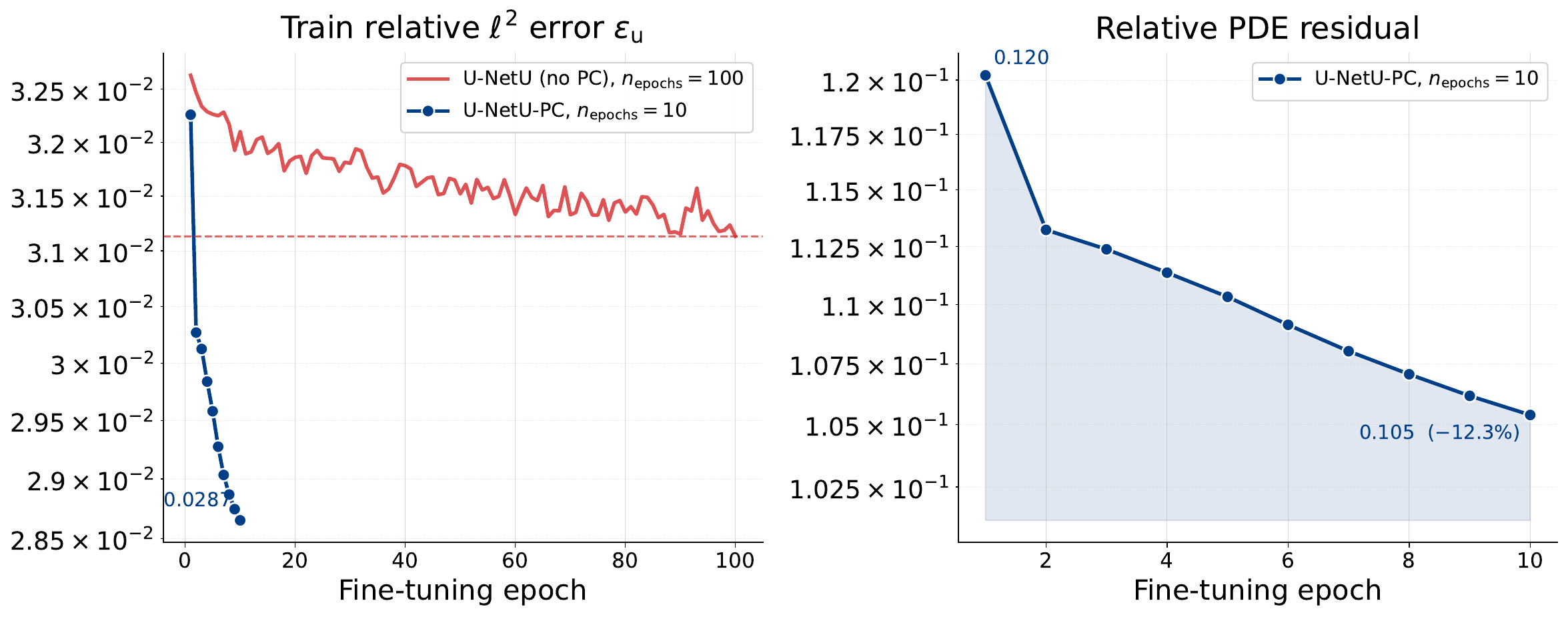
          }
          \\[1ex]
        \includegraphics[width=0.8\linewidth]{
          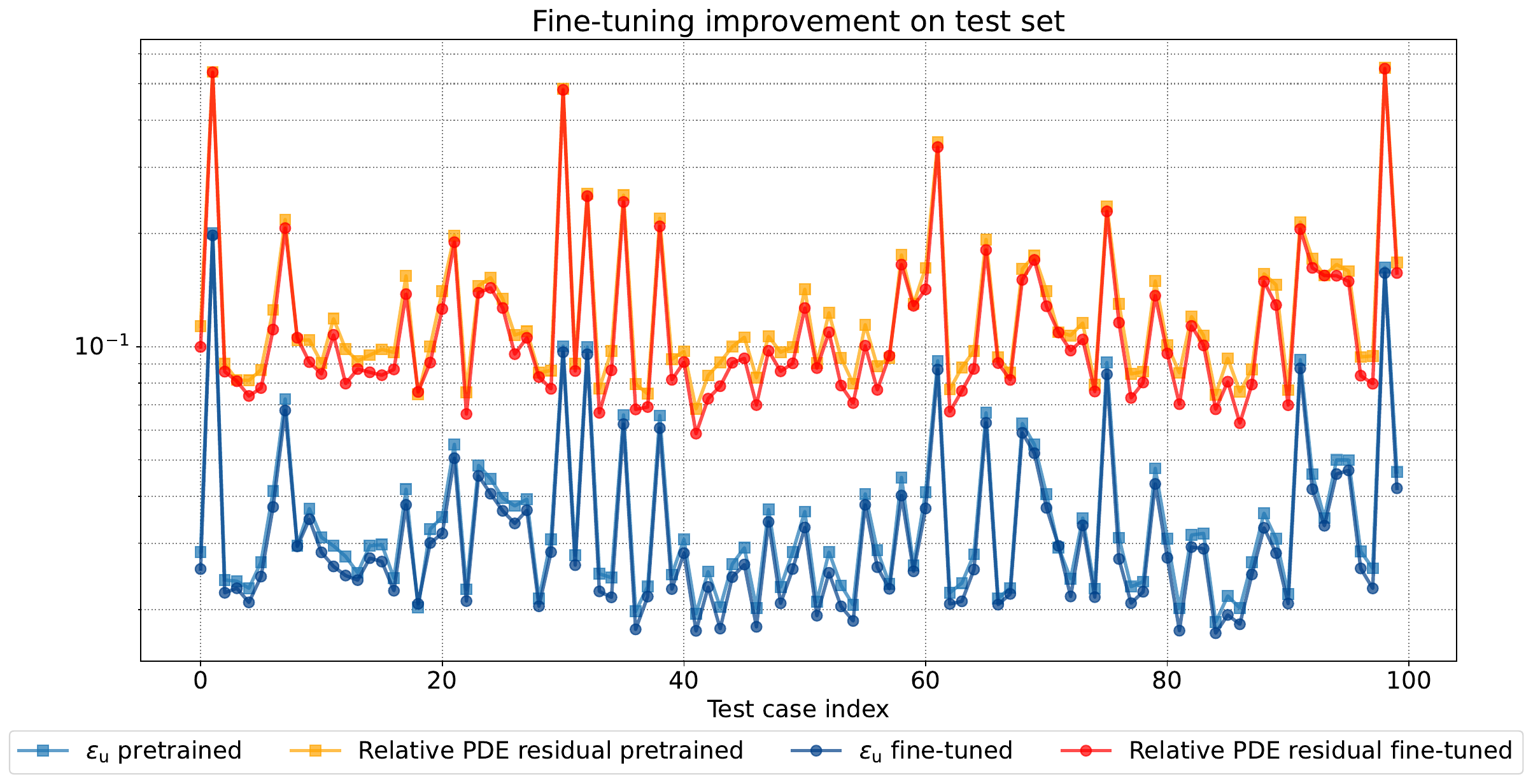
          }
        \caption{
          Fine-tuning with physical consistency loss $n_{\text{epochs}}=10$
          compared to fine-tuning without physical consistency loss
          $n_{\text{epochs}}=100$. Wall-clock times (top row),
          relative $\ell^2$ error and relative PDE residual (middle row),
          and improvement on the test set (bottom row).
        }
        \label{fig:pinn_finetuning}
    \end{figure}

    The results (Figure~\ref{fig:pinn_finetuning}) demonstrate that the physical
    consistency loss is considerably more effective for fine-tuning than as a
    regularizer applied throughout training. Although evaluating the finite
    element residual increases the per-iteration cost by approximately a factor
    of $1.8$ in fine-tuning, the physics-informed correction network reaches a
    lower validation error after only $10$ fine-tuning epochs than the purely
    data-driven baseline achieves after $100$ epochs, resulting in a lower
    overall wall-clock time. These findings suggest that the residual, once a
    surrogate has learned an accurate data-driven approximation, provides a
    complementary optimization signal that efficiently corrects remaining
    physical inconsistencies without requiring the entire network to be
    retrained.

    \begin{remark}
      A surrogate model resolving the full cardiac cycle and
      predicting the pressure jointly with the velocity would allow assembling the
      complete unsteady residual solely from the predicted fields. In this case,
      the penalty can be interpreted as
      an unsupervised regularizer that could also be evaluated on geometries for
      which the numerical solutions have not yet been computed.
      The conclusions drawn in
      Section~\ref{subsubsec:physics_consistency_training_finetuning} are unaffected by
      this distinction, as they concern the relative effectiveness of the residual as an
      optimization signal in training and fine-tuning,
      independent of whether $\ve{b}$ is precomputed or assembled at runtime.
    \end{remark}

    \subsection{Out-of-distribution generalization} 
    \label{sec:results_ood}

    The surrogate models presented and tested in the previous sections have been
    trained and evaluated on geometries sampled from the SynthAorta
    parametrization~\cite{Bosnjak2025SynthAorta}, i.e., the training and
    test sets share the same underlying parametric distribution. To assess their
    robustness and practical applicability in more general contexts, we now
    evaluate the performance of the surrogate models
    on pools of patient-specific aortic geometries that lie outside this distribution.
    This experiment aims at quantifying the extent to which the learned geometric
    representation generalizes beyond the parametrized shape space and
    highlights the limitations of the current training dataset. For the sake of
    brevity, only results obtained via the fully trained Graph U-NetU model are
    presented.

    To this end, we consider (i) a dataset of synthetic aortic shapes generated
    via statistical shape modeling from a set of aortic coarctation
    patients~\cite{Yevtushenko2022, Katz2026} and (ii) aortic shapes obtained
    from the VMR~\cite{Wilson2013}.
    The out-of-distribution geometries were registered to the current dataset's
    template via the multi-level residual network-based large deformation diffeomorphic
    metric mapping (ResNet-LDDMM) registration method introduced
    in~\cite{Romor2025Assimilation} to compute the corresponding displacement
    field from the reference shape.

\begin{figure}[!htb]
      \includegraphics[width=1\linewidth]{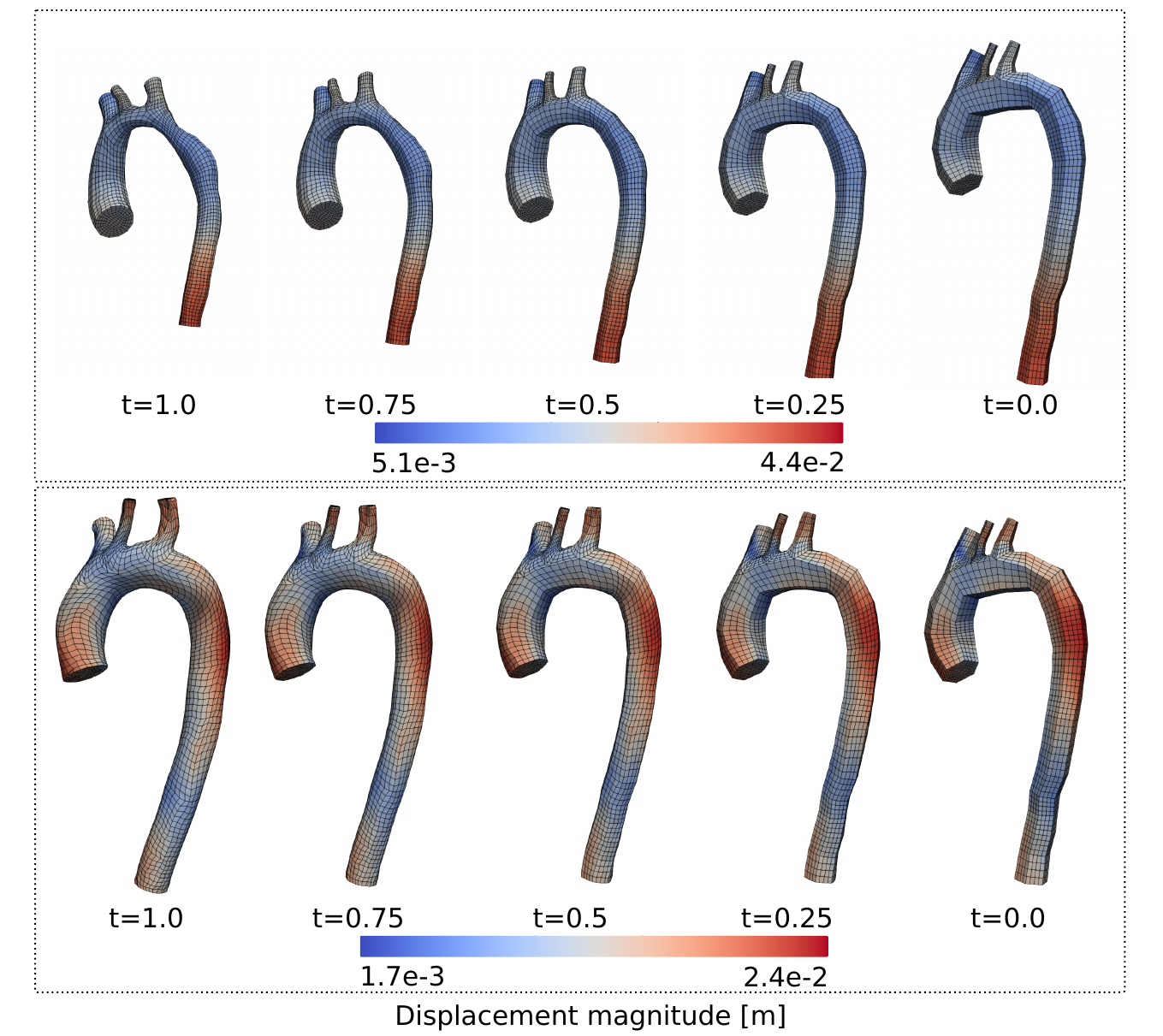}
      \caption{
        Intermediate steps of multi-level ResNet-LDDMM
        registration~\cite{Romor2025Assimilation} of the VMR and coarctation
        datasets to the SynthAorta template. The first row is a geometry (left)
        from the coarctation dataset registered onto the SynthAorta template
        (right); the second row is a geometry (left) from the VMR dataset
        registered onto the SynthAorta template (right).
      }
      \label{fig:registration}
    \end{figure}

    The registration proceeds from a coarse resolution and progressively refines
    at finer resolutions. The ResNet-LDDMM framework allows for the computation
    of diffeomorphic transformations that preserve the topology of the
    geometries, ensuring that the registered shapes remain anatomically
    plausible. For both datasets, Figure~\ref{fig:registration} illustrates the
    registration process, showing the initial geometry ($t=0.0$), the
    intermediate deformations ($t\in\{0.25, 0.5, 0.75\}$), and the final
    registered geometry ($t=1.0$).

\begin{figure}[!htb]
        \centering
        \includegraphics[width=1\linewidth]{
          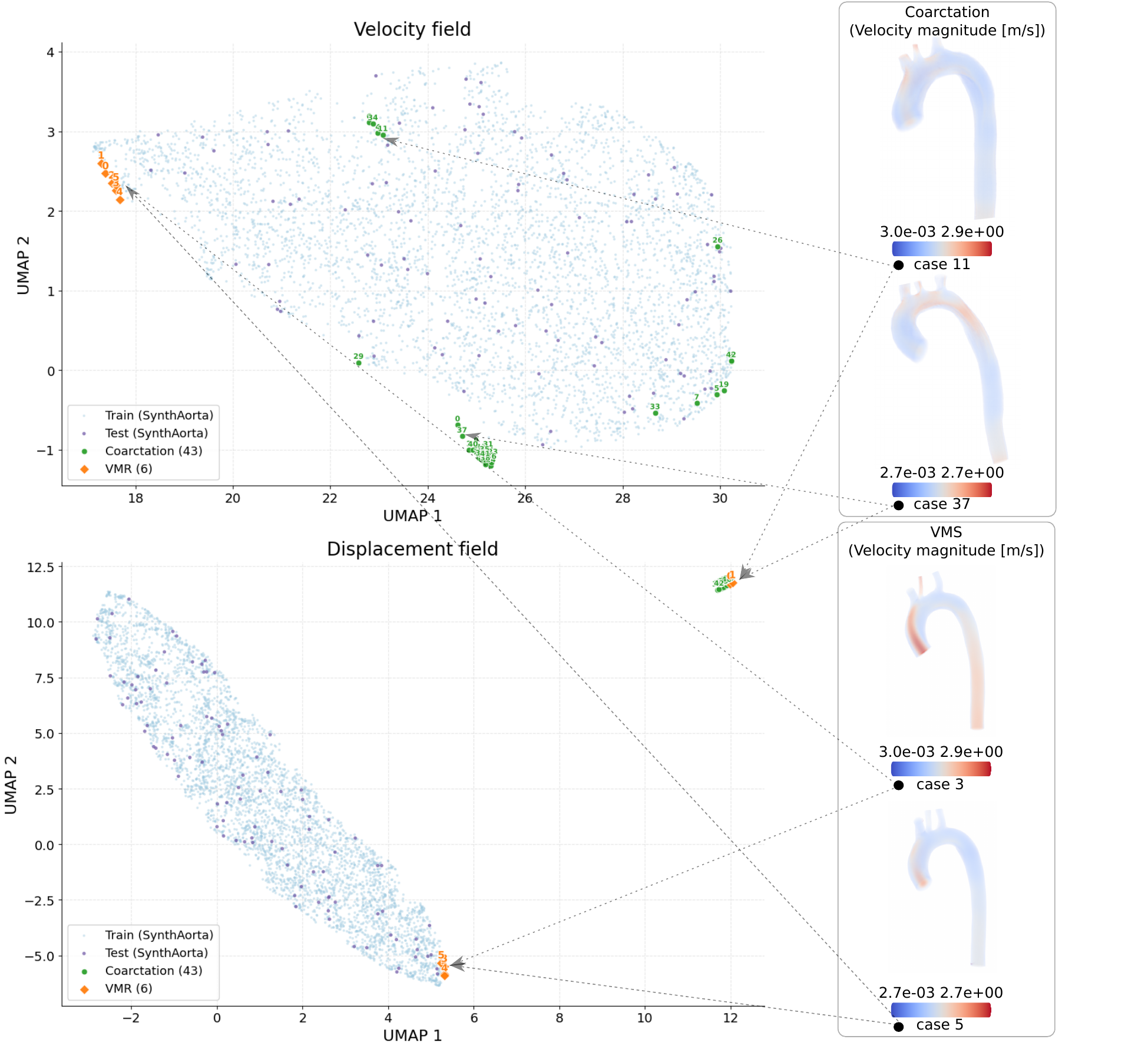
        }
        \caption{
          UMAP $2$D embedding of the reduced velocity (top) and
          displacement (bottom) fields for the dataset including the
          out-of-distribution geometries. The training and test sets of the
          SynthAorta dataset are shown as a background point cloud, while the
          out-of-distribution coarctation (green) and VMR (orange) geometries
          are highlighted and labeled with their indices within the test set.
          The coarctation cases concentrate in a few sub-regions, whereas the
          VMR cases---particularly in the displacement embedding---appear as
          outliers away from the main manifold.
        }
        \label{fig:clustering_ood}
    \end{figure}

    As a preliminary step, Figure~\ref{fig:clustering_ood} shows the embedding
    of the reduced velocity and displacement fields of the new geometries in
    the two-dimensional (2D) UMAP space of the SynthAorta dataset (analogous to what
    has been shown in Figure~\ref{fig:umap_clustering} for the SynthAorta dataset).
    Cases of coarctation are concentrated in a few sub-regions, whereas the VMR
    cases---particularly in the displacement embedding---are outliers away from
    the main manifold. This confirms that the new geometries are
    out-of-distribution with respect to the training set.

\begin{figure}[!htb]
        \centering
        \includegraphics[width=1\linewidth]{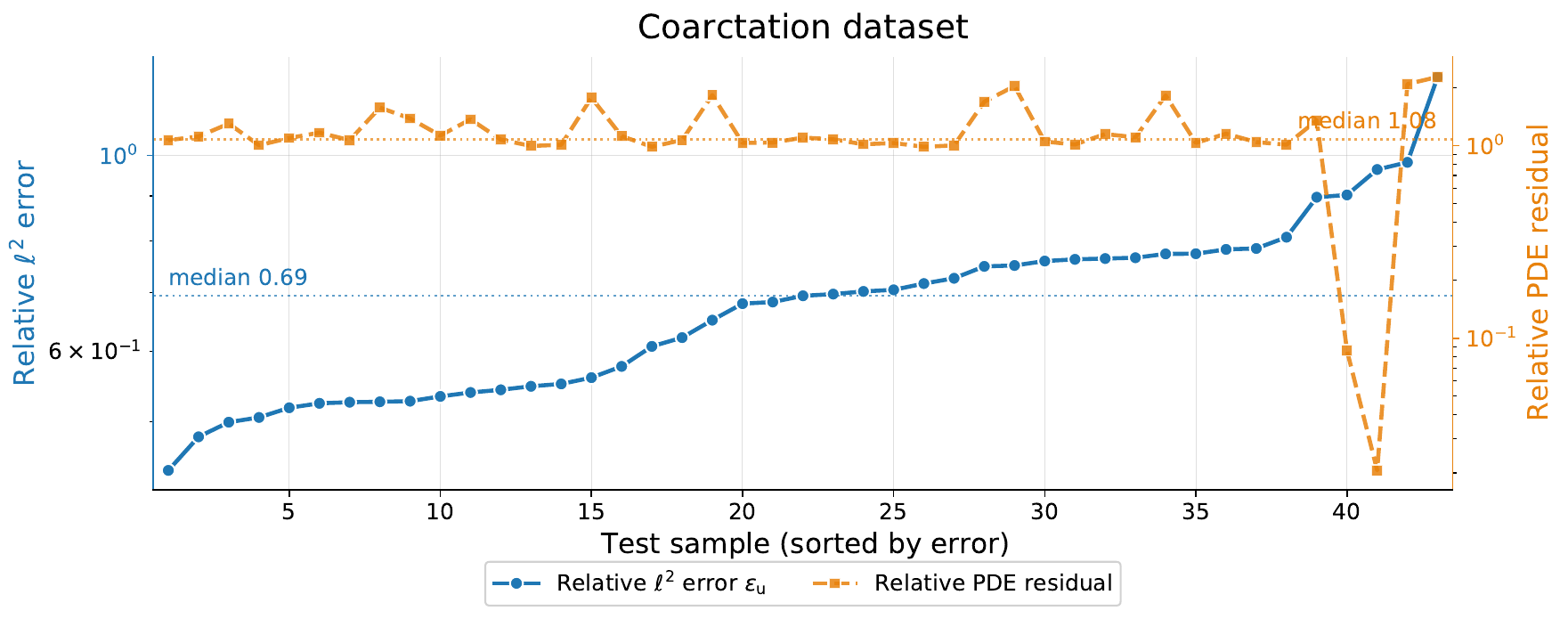}\\[1ex]
        \includegraphics[width=1\linewidth]{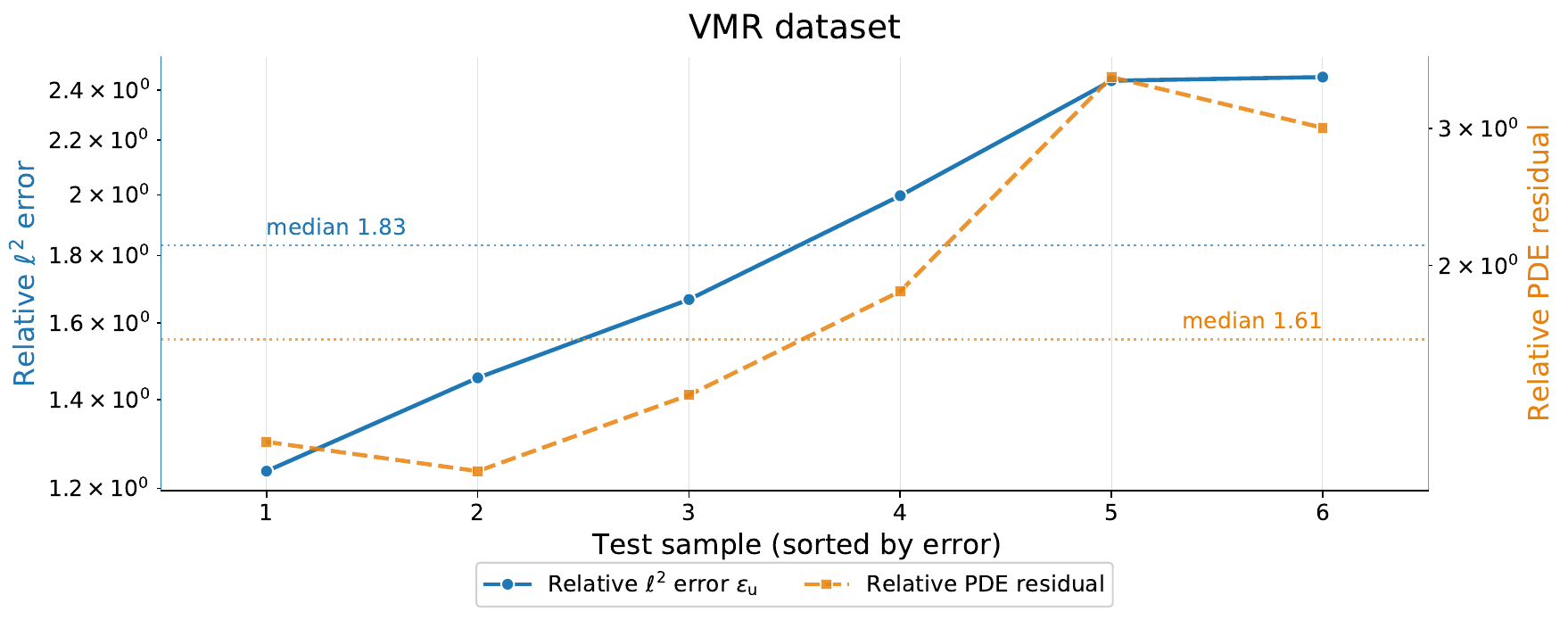}
        \caption{
          Out-of-distribution inference of the Graph U-NetU velocity
          surrogate on two external datasets: aortas affected by
          coarctation~\cite{Romor2025Assimilation} (top) and the
          VMR dataset~\cite{Wilson2013} (bottom). For each test sample, sorted
          by relative $\ell^2$ error, the relative $\ell^2$ velocity error
          (blue, left axis) and the relative PDE residual (orange, right axis)
          are reported, with the respective medians marked by dotted lines.
        }
        \label{fig:coarctation_inference}
    \end{figure}

    The results in Figure~\ref{fig:coarctation_inference} underline the restricted
    validity of the parametrization used to describe inter-patient geometries.
    For the VMR dataset, the median relative $\ell^2$ error lies at $183\%$, whereas
    the coarctation dataset yields a median relative $\ell^2$ error of $69\%$. Both
    indicate that the parametrization underlying the SynthAorta dataset
    does not sufficiently cover the full physiological spectrum---let alone
    pathological configurations.

    These findings should therefore be interpreted as a limitation of the
    available training data rather than of the proposed methodology itself.
    As larger collections of patient-specific anatomies become available, they
    can be registered to the common reference configuration and used to enrich
    the parametrized shape space through data augmentation, thereby extending
    the domain of validity of the surrogate models. The proposed framework is
    well suited for such incremental refinement, enabling continuous
    improvements in robustness and generalization without fundamental changes to
    the underlying methodology.

    \FloatBarrier

    \section{Conclusion}\label{sec:conclusion}

    This work investigated different aspects of inter-patient surrogate modeling
    for the simulation of blood flow in the aorta. Building upon the
    SynthAorta~\cite{Bosnjak2025SynthAorta} parametrization, we generated a
    large dataset of CFD solutions, which was then used to train graph- and
    transformer-based neural architectures for the approximation of relevant
    QoIs. Besides the predictive accuracy and the performance
    of the surrogate models, we tested several AL strategies for
    data augmentation, as well as the impact of including physical consistency
    constraints during the training stage. Geometric parametrization makes it
    possible to map functions defined on different geometries, such as numerical
    solutions, predictions, and residuals, onto a reference domain,
    thereby enabling the study of accuracy, AL, and physical
    consistency constraints in a common framework.

    Several conclusions can be drawn from the numerical results. First, the
    parametric definition of the dataset allows defining a common reference
    geometry and displacement-field representations of all considered shapes.
    These representations provide an effective basis for inter-patient
    surrogate modeling, which does not require an additional geometric
    registration of shapes and automatically preserves point-to-point
    correspondence across patients. The surrogate achieves a
    median relative $\ell^2$ error of $2\%$ for the velocity
    field: between $500$ and $1000$ rSVD modes would be needed to achieve a
    comparable reconstruction error, making the use of classical reduced basis
    methods too expensive.
    Second, the benefit of AL is strongly regime-dependent. Our
    results show that sophisticated acquisition strategies do not outperform
    random sampling during the warm-up phase---when only limited training data
    are available---whereas gradient- and subspace-based methods consistently
    improve performance when fine-tuning models that are already accurate.
    Third, incorporating the governing equations through a differentiable finite
    element residual is more effective as a fine-tuning objective than as
    a regularizer throughout training. Once the data-driven surrogate has
    converged, the residual provides a complementary optimization signal that
    efficiently corrects the remaining physically inconsistent errors at a lower
    overall computational cost than prolonged conventional training.
    Finally, the analysis of accuracy for out-of-distribution shapes
    demonstrates that the dominant limitation of the present framework is not
    the surrogate architecture itself, but the geometric support of the
    underlying parametrization and training dataset. Since these considerations
    apply to any dataset size---given that computational resources are always
    limited, problem sizes will be adjusted, and parametrizations
    extended---these results
    naturally lead to the question of how best to invest a given computational
    budget in extending snapshot data to minimize
    inference loss in geometrically parametrized aortic flow.

    The primary contribution of this work is the demonstration of an
    integrated workflow for shape-informed surrogate modeling of vascular
    flow, combining synthetic dataset generation, geometry-aware neural
    architectures, differentiable physics constraints, and data acquisition
    strategies within a unified framework. Beyond the individual algorithmic
    components, the study provides practical insight into when additional
    information is most valuable during surrogate construction: increasing
    dataset diversity is more beneficial than targeted sample selection during
    early training, whereas physics-based optimization and informed sample
    selection become increasingly valuable once an accurate data-driven
    approximation has been established.

    Nevertheless, the present study is subject to some limitations. The accuracy
    attainable with external cohorts depends strongly on the processing of the
    vascular surfaces, as the position of the planes truncating the inlet and
    the outlets, as well as the extent of the remaining vessel segments, condition the
    quality of the registration onto the common template. Moreover, the
    surrogates are trained at a single instant of the cardiac cycle and rely on
    rigid vessel walls and a single combination of boundary conditions per
    geometry. The presented framework is nevertheless readily extensible as
    larger collections of patient-specific anatomies become available, offering
    a scalable route towards more robust and clinically relevant surrogate
    models for cardiovascular flow simulations.

    Future work will focus on validating the surrogates against clinical
    measurements---such as \emph{in vivo} 4D-flow MRI
    acquisitions---representing a necessary step for their application
    in clinical practice.

    \section*{Acknowledgements}
    This work received support from the European Union’s Horizon 2024 research
    and innovation program under the Marie Skłodowska-Curie
    project ``VascularROM'', grant agreement no.~101209352.
    During preparation of this manuscript and underlying software development,
    the authors used Claude Code Opus 5.5 (Anthropic)
    and GPT-5.6/GPT-6 (OpenAI) to assist with style and language editing
    and coding. The authors take full responsibility for all content in the
    paper and the employed software.

    \FloatBarrier

    \bibliographystyle{unsrtnat}
    \bibliography{bibliography}

\end{document}